\documentclass[12pt,preprint]{aastex}

\shorttitle{X-ray stars in the NGC 1333}
\shortauthors{Getman et al.}

\slugcomment{To appear in the Astrophysical Journal; scheduled for
2002, ApJ, 575 (August 10th)}

\begin{document}

\title{Chandra Study of Young Stellar Objects in the NGC 1333 Star-forming Cloud}

\author{
Konstantin V. Getman\altaffilmark{1}, Eric D.
Feigelson\altaffilmark{1}, Leisa Townsley\altaffilmark{1}, John
Bally\altaffilmark{2}, Charles J. Lada\altaffilmark{3}, Bo
Reipurth\altaffilmark{4}}

\altaffiltext{1}{Department of Astronomy \& Astrophysics, 525
Davey Laboratory, Pennsylvania State University, University Park,
PA 16802} \altaffiltext{2}{Center for Astrophysics and Space
Astronomy, Campus Box 389, University of Colorado, Boulder, CO
80309} \altaffiltext{3}{Harvard-Smithsonian Center for
Astrophysics, Mail Stop 72, Cambridge, MA 02138}
\altaffiltext{4}{Institute for Astronomy, 2680 Woodlawn Drive,
University of Hawaii, Honolulu, HI 96822}

\begin{abstract}
NGC 1333, a highly active star formation region within the Perseus
molecular cloud complex, has been observed with the ACIS-I detector
 on board the {\it Chandra X-ray Observatory}. In our image with a sensitivity limit of
$\sim 10^{28}$~erg~s$^{-1}$, we detect 127 X-ray sources, most
with sub-arcsecond positional accuracy. While 32 of these sources
appear to be foreground stars and extragalactic background, 95
X-ray sources are identified with known cluster members. The X-ray
luminosity function of the
 discovered
YSO population spans a range of $\log L_x \simeq 28.0$ to 31.5
erg~s$^{-1}$ in the $0.5-8$ keV band, and absorption ranges from
$\log N_H \simeq 20$ to 23 cm$^{-2}$. Most of the sources have
plasma energies between 0.6 and 3 keV, but a few sources show
higher energies
 up to $\sim 7$ keV.
Comparison with K-band source counts indicates that
 we detect all of the known cluster members
 with $K < 12$, and about half of members with $K > 12$. $K \simeq 11$, the peak of the K-band
luminosity function, corresponds to $0.2-0.4$ M$_\odot$ stars for
a cluster age of $\sim 1$ Myr.

We detect seven of the twenty known YSOs in NGC 1333 producing
jets or molecular outflows as well as one deeply embedded object
without outflows. No evident difference in X-ray emission of young
stars with and without outflows is found. Based on the complete
subsample of T Tauri stars, we also find no difference in X-ray
properties and X-ray production mechanism of stars with and
without K-band excess disks.

Several other results are obtained. We suggest that the X-ray
emission from two late-B stars which illuminate the reflection
nebula originates from unresolved late-type companions. Two T
Tauri stars are discovered in the ACIS images as previously
unknown components of visual binaries. A good correlation $L_x
\propto J$ is seen which confirms the well-known relation $L_x
\propto L_{bol}$ found in many star forming regions. Based on
spectral analysis for the X-ray counterpart of SVS 16, we
establish that the column density $N_H$ is much lower than that
expected from near-IR photometry so that its X-ray luminosity,
$\log L_x \simeq 30.6$ erg~s$^{-1}$, is not unusually high.
\end{abstract}



\section{Introduction} \label{intro_sec}

NGC 1333 is a star formation region within the Perseus molecular
cloud complex noted for its large population of protostars and
young stellar outflows. Named for a reflection nebula illuminated
by B-type members of the Per OB2 association and lying in the
Lynds 1450 dark cloud (Barnard 205), near-infrared (IR) imaging
surveys revealed two embedded clusters of about 150 low- and
intermediate-mass young stars within the inner parsec
\citep{Aspin94, Lada96}. The northern cluster includes the
reflection nebula, while the southern cluster is associated with
dense molecular material surrounding HH 7-11. Both clusters are
associated with concentrations of dense molecular material seen in
H$_{2}$CO, CS, and CO molecular-line observations \citep{Lada74,
Warin96, Knee00}.

Members of NGC 1333 produce an astonishing profusion of outflows
with at least 17 young stellar objects (YSOs) ejecting Herbig-Haro
objects, H$_{2}$ shocks and/or CO bipolar flows \citep{Bally96}.
The kinetic energy of these outflows dominates the dynamics of the
cloud. At least five objects are extremely young Class 0
protostars \citep{Knee00, Looney00, Chini01}. The large population
of short-lived outflows, together with the K-band luminosity
function, indicate that the star clusters are extremely young with
a high star formation rate around $4 \times 10^{-5}$ M$_\odot$
yr$^{-1}$ over the past 1 Myr \citep{Lada96}.

Herbig-Haro (HH) objects are jets of shocked gas
\citep{Reipurth01} which trace high-velocity bipolar outflows that
all young stellar objects appear to produce shortly after their
birth. In NGC 1333, several HH objects are powered by optically
visible low-mass young stars that suffer relatively low extinction
and seem not to be embedded within opaque cloud cores. Three of
the irradiated jets exhibit unusual C-shaped symmetry, suggesting
that the source stars, still forming protostars, have been ejected
from the cluster core \citep{Bally01}. NGC 1333 thus provides a
unique opportunity to study protostellar X-ray emission with
relatively little intervening absorption.

We adopt 318~pc for the distance to NGC~1333 based on the {\it
Hipparcos} parallactic measurements of the Perseus OB2 association
\citep{de Zeeuw99}. Values around $310-318$ pc were adopted in
previous X-ray studies of NGC~1333
\citep{Preibisch97b,Preibisch98,Preibisch02}, substantially more
distant than the value of 220~pc \citep{Cernis90} used by other
recent authors \citep[e.g.][]{Looney00,Bally01,Sandell01}. IC~348,
a nearby star-forming region that is part of the Perseus molecular
cloud complex as well, has also been the subject of recent {\it
Chandra} studies \citep{Preibisch01,Preibisch02}. These authors
use the distance to IC~348 (310~pc) obtained by Herbig (1998), who
shows that the photometric technique used by \citet{Cernis93} to
derive a distance of $260 \pm 16$~pc to IC~348 infers earlier
spectral types than direct spectroscopic measurements.  Since
Herbig's arguments apply equally well to NGC~1333 and more precise
distance estimates are unavailable for this confused region, we
choose to adopt the {\it Hipparcos} distance to the OB
association.  This facilitates comparisons to the earlier X-ray
studies but must be taken into account in quantitative comparisons
with studies that use a smaller distance.

X-ray studies give a unique view of star forming regions
\citep{Feigelson99}. X-ray luminosities are elevated $10^1-10^4$
fold above main sequence levels during all phases of pre-main
sequence evolution. X-ray surveys are particularly effective for
identifying YSOs that no longer exhibit circumstellar disks and,
with recent telescopes with good sensitivity in the hard X-ray
band, can penetrate deep into molecular cloud cores.  X-rays are
tracers of YSO magnetic activity; the emission arises from plasma
heated by violent magnetic reconnection flaring.  The NGC 1333
clusters were observed in the soft X-ray band  with the {\it
ROSAT} High Resolution Imager by \citet{Preibisch97b}.  This 40 ks
observation detected 16 YSOs, most with $0 < A_V < 9$ and
intrinsic luminosities $\log L_x \simeq 29.3$ to 31.2 erg
s$^{-1}$. The YSO SVS 16 was a notable exception with $A_V \simeq
26$ and X-ray emission apparently constant at the unusually high
level of $\log L_x \simeq 32.3$ erg s$^{-1}$ \citep{Preibisch98}.

We present here a study of the region with the {\it Chandra X-ray
Observatory} and its Advanced CCD Imaging Spectrometer (ACIS)
detector.  With its excellent mirrors and detectors, {\it Chandra}
achieves an order of magnitude higher sensitivity and spatial
resolution compared to {\it ROSAT}. More than half of the known
cluster members are detected, penetrating deep into the X-ray
luminosity function of the population. We first present data
analysis methods (\S \ref{obs_sec}), provide source lists (\S
\ref{src_list_prop_subsec}), and  discriminate between cluster
members and background sources (\S \ref{nocounterpart_subsec}). We
then consider the X-ray stellar population (\S
\ref{stel_pop_subsection}), summarize the K-band survey
 and the ACIS survey (\S \ref{cluster_membership_section}), present the X-ray properties of the cluster as a whole (\S
\ref{global_properties_subsec}) and discuss in particular the
X-ray emission from late B stars (\S
\ref{intermediate_mass_section}), the X-ray properties of CTTs,
WTTs (\S \ref{ctt_vs_wtt_section}, \S \ref{spectra2_section}),
younger protostars, and outflow sources (\S
\ref{protostars_subsec}), and  a few other interesting sources (\S
\ref{intersting_sources_subsec}).

\section{Observations and data analysis} \label{obs_sec}

\subsection{ACIS Observation} \label{obs_subsec}

\begin{figure}
\centering
\includegraphics[width=4.5in]{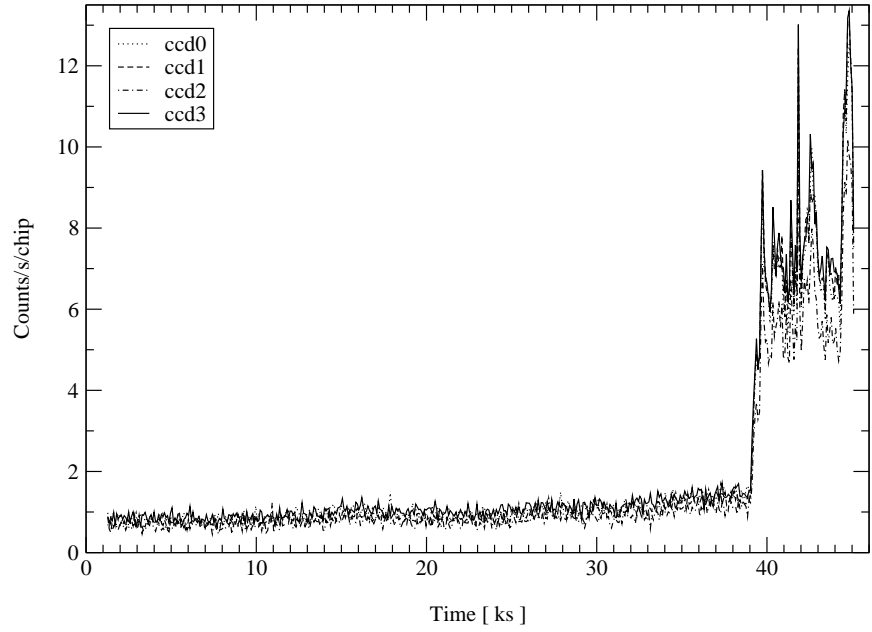}
\caption{Background flaring during the last 6 ks of our
observation due to solar energetic particles.  The four traces
show the count rate summed across each ACIS-I CCD chip.
\label{bg_flare_fig}}
\end{figure}

The {\it Chandra} observation of NGC 1333 was performed on
12.96-13.48 July 2000 utilizing the ACIS imaging array (ACIS-I)
consisting of four abutted front-side illuminated CCDs with a
field of view of about $17\arcmin \times 17\arcmin$.\/ The
aimpoint of the array is $3^h 29^m 06.1^s$, $+31^\circ 19\arcmin
38\arcsec\/$ (J2000) and the satellite roll angle (i.e.,
orientation of the CCD array relative to the north-south
direction) is 95.7$^{\circ}$. The focal plane temperature is
-120$^{\circ}$C. The total exposure time of our NGC 1333 image is
43.9 ks, but because of background flaring at the end of the
observation (Figure \ref{bg_flare_fig}), we utilize a reduced
exposure time of 37.8 ks. Two of the CCDs of the ACIS-S
spectroscopic array were also turned on, but these data will not
be discussed here as the mirror point spread function is
considerably degraded far off-axis. Detailed {\it Chandra} ACIS
instrumental information can be found in \citet{Weisskopf02} and
at \url{http://asc.harvard.edu}.

\subsection{Data selection and astrometric alignment} \label{data_selection_subsec}

We start data analysis with the Level 1 processed event list
provided by the pipeline processing at the {\it Chandra} X-ray
Center. The energy and grade of each data event are then corrected
for charge transfer inefficiency (CTI), applying the techniques
developed by \citet{Townsley00, Townsley01a}.  This provides a
more uniform gain and improved energy resolution, accounting for
the spatial redistribution of charge during the chip readout
process. A sequence of cleaning operations follows to remove
cosmic ray afterglows, hot columns/pixels, events arising from
charged particles, and detector noise.  These include filtering by
ASCA grades (0,2,3,4,6), status bits, timeline, and background
flaring due to solar activity. More extensive discussion of these
methods is provided by \citet{Feigelson02} [hereafter F02]. Note,
however, that our treatment of spectral calibration is improved
over the method described in that paper. The resulting ACIS-I
image is shown in Figure \ref{figure_acis_image}.

The effect of solar protons is illustrated in Figure
\ref{bg_flare_fig}, which shows that background flaring occurred
during the last 6 ks of the exposure.  This is attributed to
flares from solar active region 9077 which produced a rise in the
{\it GOES} satellite $>1$ MeV solar proton fluence from $10^1$ to
$10^3$ particles cm$^{-2}$ s$^{-1}$ sr$^{-1}$ during 13.1-13.5
July, peaking at $10^4-10^5$ particles cm$^{-2}$ s$^{-1}$
sr$^{-1}$ during 14.5-15.8 July.  The {\it Chandra} flare onset
coincided within two minutes of the geomagnetic Sudden Storm
Commencement at 0945UT 13 July. This was the strongest solar
proton event, and produced the worst geomagnetic storm, of the
decade. Detailed information about solar activity during this
period of time can be found on-line at
\url{http://www.sec.noaa.gov/weekly/weekly00.html}.

To correct for possible systematic errors in the {\it Chandra}
aspect system, which are typically around 1\arcsec, we compare the
X-ray positions of 2 ACIS sources with their optical counterparts
from the Tycho-2 catalog (BD +30$^\circ$549, BD +30$^\circ$547),
and 10 ACIS sources in the inner 5\arcmin\/ field with optical
counterparts from the USNO-A2.0 catalog. The resulting offset of
$-0.26\arcsec\/$ in right ascension and $-0.65\arcsec\/$ in
declination was applied to the ACIS field.

The ACIS image of NGC 1333 emerging from these data selection and
alignment procedures is shown in Figure \ref{figure_acis_image}a.
Sources are located with the {\it wavdetect} program within the
CIAO data analysis system that performs a Mexican hat wavelet
decomposition and reconstruction of the image \citep{Freeman02}.
Cosmic ray afterglow events were removed from the data and the
energy range was restricted to $0.5-8$ keV for source detection.
Wavelet scales ranging from 1 to 16 pixels in steps of $\sqrt{2}$
were used, where each pixel subtends $0.5\arcsec \times
0.5\arcsec$. To enhance selection sensitivity we run {\it
wavdetect} separately for a series of three $1024 \times 1024$\/
pixel concentric images at 1, 2 and 4 pixel resolutions for each
of the three energy bands (soft $0.5-2.0$, hard $2.0-8.0$, full
$0.5-8.0$ keV) with a source significance threshold of $1 \times
10^{-6}$. The merged catalog produced 117 sources. Eight sources
appeared to be spurious upon visual inspection of the image.
Repeating the detection procedure with a significance threshold of
$1 \times 10^{-5}$ did not reveal any convincing new sources.  Our
source list thus contains 109 sources, tabulated in Table
\ref{src_tbl} and displayed in Figure \ref{counterparts1_fig} and
Figure \ref{counterparts2_fig}.

We then performed a visual search for faint sources, looking for
concentrations of photons spatially coincident with known sources
from near-IR \citep{Lada96, Aspin94}, mm/submm \citep{Chini01,
Sandell01} and radio continuum \citep{Rodriguez99} catalogs.  We
were careful to avoid biasing these results; no other information
(e.g. IR magnitude) about the known sources was provided to the
person performing the matching. The results are shown in
Table~\ref{faint_src_tbl}, where $\theta$ is the off-axis angle.
Most of these candidate X-ray sources were matched to K-band
sources from \citet{Lada96}. No further tentative matches were
found to the radio sources of \citet{Rodriguez99} or the
protostars in \citet{Chini01} or \citet{Sandell01}. While some of
these concentrations of ACIS photons in Table~\ref{faint_src_tbl}
may be random coincidences, we believe most of them are real
sources. For small off-axis angles, assuming an average point
spread function size of roughly 10 arcsec$^{2}$ and the background
rate of 0.018 counts per arcsec$^{2}$, the likelihood of the
random occurrence of 3 or more events sampled from a Poisson
distribution of photons is $\leq 0.1\%$ on-axis.

\subsection{Optical imaging \label{zband_subsec}}

A deep broad-band image in the Sloan Digital Sky Survey I-band was
obtained with the MOSAIC CCD camera at the prime focus of the
Mayall KPNO 4 meter reflector on the night of 13 October 2001.
This camera uses eight 2048 $\times$ 4096 pixel SITe CCDs, has a
36\arcmin\ field of view and a 0.26\arcsec/pixel image scale. Five
individual 180 second exposures, obtained using a standard dither
pattern to remove the intra-CCD gaps, were combined to form an
image with a total exposure time of 900 sec in 0.8\arcsec\ seeing.
The IRAF package MSCRED was used for standard reductions. The USNO
catalog was used to establish an astrometric solution accurate to
about 0.6\arcsec\ and to compare with the {\it Chandra} source
positions. The limiting magnitude is about $m_I \sim 23$.

\subsection{Stellar counterparts and source positions} \label{source_positions_subsec}

Likely stellar identifications within 3\arcsec\/ of each ACIS
source in Table \ref{src_tbl} were selected from multiwavelength
studies of the NGC 1333 cloud. We found 6 compact radio continuum
counterparts \citep{Rodriguez99}, 1 mm/submm continuum source
\citep{Looney00, Chini01, Sandell01}, 3 far-IR  sources from the
Infrared Astronomical Satellite \citep{Jennings87}, 43 mid-IR
sources from the Infrared Space Observatory\footnote{We used the
ISO public data archive located on-line at
\url{http://www.iso.vilspa.esa.es/ISO/}.  These data are not
cleaned of instrumental artifacts, so additional faint ISO
counterparts may be present.}, 73 near-IR $JHK$ sources
\citep{Aspin94, Aspin97, Lada96}, and 70 optical band stars from
our I-band image, from the Hipparcos/Tycho, USNO-A2.0, HST Guide
Star catalogs, and studies of the region by \citet{Herbig83},
\citet{Bally96}, and \citet{Bally01}. We also searched the SIMBAD
and NED databases for additional counterparts without success, and
the 2MASS survey catalog of the region has not been released at
the time of writing. All 16 soft X-ray sources detected with {\it
ROSAT} \citep{Preibisch97b} in the field of view are recovered.
{\it ROSAT} source \#15 is resolved into two sources separated by
4\arcsec (ACIS \#64 and \#65; see Figure
\ref{figure_acis_image}e). These results are presented in Table
\ref{src_tbl} and Figures
\ref{counterparts1_fig}-\ref{counterparts2_fig}.

Altogether, 80 of the 109 ACIS sources (73\%) have counterparts at
a non-X-ray band.  We believe that these identifications are
highly reliable in the sense that they are not chance
coincidences.  Of the 38 sources with astrometric counterparts (2
from Tycho and 36 from USNO-A2.0), two-thirds have X-ray/optical
offsets less than 0.5\arcsec\/ with the remainder within
1.7\arcsec\/ (Table \ref{src_tbl}, column 9), well within
uncertainties of the {\it Chandra} ACIS instrument. We believe the
main uncertainty in identification lies in binarity and
multiplicity of most YSO stellar systems \citep{Mathieu00,
Looney00}. Figure \ref{figure_acis_image}b-e show four cases where
ACIS resolved close ($2\arcsec - 5\arcsec$) double sources that
were unresolved in the K-band image of \citet{Lada96}.  Many
closer multiple systems are likely present for which we cannot
identify the X-ray active component. Except for the early B-type
stars, we assume that most of the X-rays are produced by the
component that produces most of the bolometric emission.

\subsection{Source extraction and limiting sensitivity} \label{extraction_subsec}

Extraction of counts generally follows the procedures described in
F02. Briefly, we extract counts $C_{xtr}$ in the total ($0.5-8.0$
keV) band from within radius $R_{xtr}$, the radius encircling
about 90\% (99\% for very strong sources with $>1000$ counts) of
the energy, which increases with off-axis distance due to
deterioration of the point spread function. $B_{xtr}$ background
events are extracted from four circular regions with radii
$R_{xtr}/2$ surrounding the source. Background extraction circles
are manually adjusted for  crowded regions of the field. The mean
background level is $\simeq 0.018$ counts per arcsec$^{2}$. The
count rate $CR$ for each source is then given by $CR {\rm
(ct~ks^{-1})} = (C_{xtr}-B_{xtr}) / (f_{PSF} E_{net})$ where the
net exposure time $E_{net} = 37.8$ ks (\S \ref{obs_subsec}) and
the fraction of energy extracted is typically $f_{PSF} \simeq
0.9$. These extracted quantities are tabulated for each source in
Table \ref{properties_tbl}.

The faintest on-axis source emerging from the wavelet source
detection procedure (\S \ref{data_selection_subsec}) has 5
extracted counts. The corresponding minimum detectable X-ray
luminosity in the total ($0.5-8$ keV) band is $\log L_c \simeq
28.0$ erg s$^{-1}$ for a source with negligible interstellar
absorption ($A_V \sim 1$) and a typical source spectrum of a $kT
\sim 1$ keV thermal plasma\footnote{The determination of the
transformation factor from detector count rates to X-ray flux was
done using PIMMS. PIMMS is the Portable, Interactive Multi-Mission
Simulator located on-line at
\url{http://asc.harvard.edu/toolkit/pimms.jsp}}. This limit
increases to 28.6 and 29.3 erg s$^{-1}$ if the absorption is
increased to $\log N_H = 22.0$ ($A_V \sim 5$) and 22.6 cm$^{-2}$
($A_V \sim 20$), respectively. The sensitivity decreases by a
factor of 4 at the edge of the field as described in F02.

\subsection{Variability analysis} \label{var_analysis_subsec}

For each X-ray source, lightcurves in the total ($0.5-8.0$ keV)
band were created. The lightcurves are displayed with total number
of bins: 20 for strong sources and 10 for weaker sources. The
existence of variability was tested by comparing the observed
photon arrival times to that of a constant source using the
Kolmogorov-Smirnov one-sample test\footnote{Note that this
classification differs from that in F02, which used mainly
qualitative rather than quantitative classification criteria.}.
Sources were classified as `Constant' if the probability of
accepting the null hypothesis $P_{KS} > 0.05$, as `Possible flare'
if $0.005 \leq P_{KS} \leq 0.05$, and as `Flare' if $P_{KS} <
0.005$. Examples of these variability classes are shown in Figure
\ref{lightcurve_fig}.  The lightcurves in panels (c)-(f) are very
similar to the X-ray behavior of solar long decay flares with a
rise time of 1-2 hours and a decay time of a few hours up to 10
hours \citep{Getman00}. In contrast to short impulsive flares,
these often originate during posteruptive processes resulting in
magnetic reconnection in a vertical current sheet with the
subsequent formation and evolution of new giant coronal loops.

\subsection{Spectral analysis \label{spectral_analysis_subsec}}

Our spectral analysis utilizes new techniques to improve treatment
of the ACIS CCD charge transfer inefficiency (CTI) induced by
particle irradiation during the first weeks of the {\it Chandra}
mission. The CTI problem causes degradation of detector gain,
spectral resolution and high energy sensitivity as a function of
location on the chip.  Spectral analysis implemented in the CIAO
version 1 and 2 software fits the gain drop but does not seek to
improve the spectral resolution or sensitivity.  In our previous
study (F02), we corrected the data for all three effects using the
procedures described by \citet{Townsley00}, but did not have
appropriately corrected calibration auxiliary response files ({\it
arf}) and response matrix files ({\it rmf}) for scaling the
observed event energy distribution to the true spectrum incident
on the detector. Here, we both apply the CTI correction to the
data and use newly derived self-consistent {\it arf} and {\it rmf}
files \citep{Townsley01a,Townsley01b}. The spectral analysis here
minimizes the limitations or biases of previous studies as much as
possible.

Source and background pulse height distributions in the total band
$(0.5-8.0$ keV) were constructed for each X-ray object and grouped
into energy bins. Model fitting of background subtracted spectra
were performed using XSPEC 11.0.1 using the new {\it arf} and {\it
rmf} files. For nearly all sources, a one-temperature optically
thin thermal plasma MEKAL model \citep{Mewe86, Liedahl95}
successfully fits continuum and emission line strength, assuming a
uniform density plasma with 0.3 times solar elemental abundances.
In a few cases, multi-temperature or variable abundance models
were needed to obtain an adequate fit. X-ray absorption was
modelled using atomic cross-sections of \citet{Morrison83}.

The derived spectral parameters inherit a statistical uncertainty,
and a bias, from the $\chi^{2}$ fitting process. We used the {\it
fakeit} utility in the XSPEC package to perform a series of
simulations with sources of known spectra, considering the
instrumental model based on the {\it rmf} and {\it arf} files of
our observation, to estimate the statistical uncertainties of
$\log N_H$, $kT$ and broad-band luminosities.  Simulations were
carried out for a wide ranges of parameters: $20 \leq \log N_H
\leq 22.5$ cm$^{-2}$, $1 \leq kT \leq 10$ keV, and $15 \leq
C_{xtr} \leq 1000$ counts. The results show a tendency to
systematically underestimate the plasma energies of $kT = 10$ keV
sources by $40-50$\% ($15-30$ counts), $15-20$\% (100 counts), and
$5-15$\% (1000 counts). This can be explained by the rapid decline
in telescope effective area at high energies. The column density
can be over-estimated in weak unabsorbed sources because no data
below 0.5 keV is considered.

The simulations show that the standard deviations of derived
plasma energy values range from $\Delta(kT)/kT \simeq 50-90$\%
($15-30$ counts) to $30-50$\% (100 counts) and 10-15\% (1000
counts). Column density uncertainties  $\Delta(\log N_H)$ are 0.5
to 0.2 ($15-30$ counts), 0.4 to 0.1 (100 counts), and 0.3 to 0.02
(1000 counts) for unabsorbed and highly absorbed sources,
respectively. Log$~N_H$ values below $\sim 20.5$ cm$^{-2}$ are
often ill-determined because we consider data only above 0.5 keV.
  Due to the nonlinearity of the models and data,
correlated errors are naturally present. Broad-band luminosity
values exhibit standard deviations ranging from $20-35$\% ($15-30$
cts) to 15\% (100 cts) and 4\% (1000 cts), only somewhat larger
than the optimal $\sqrt{C_{xtr}}$.  All of these results are not
substantially affected by use of the likelihood ratio ({\it C})
statistic instead of the $\chi^2$ statistic.

We conclude from this simulation analysis that all broad-band
luminosities derived here are reliable, but that the individual
spectral parameters ({\it kT} and $\log N_H$) are unreliable for
the faintest sources.  We thus confine discussion of spectral
properties to the sources with $C_{xtr} > 30$ cts for which
estimated errors are smaller than the parameter ranges.

Figure \ref{spectra_fig} shows several examples of YSO X-ray
spectra. Panels (a) and (b) show typical hard spectra of highly
absorbed sources. Panel (a) is source \#71 with a heavily reddened
near-IR counterpart.  Its 150 counts are  fitted with $kT \simeq
3.3^{+2.7}_{-1.0}$ keV  (internal errors, evaluated from the XSPEC
$\chi^2$ statistic, indicate 1$\sigma$ confidence limit) and $\log
N_H \simeq 22.6^{+0.1}_{-0.1}$ cm$^{-2}$ corresponding to $A_V
\simeq 25$. Panel (b) is source \#49 with only mm/submm and radio
counterparts; it is probably a deeply embedded protostar. The 16
ACIS counts are fitted by a $kT \simeq 1.9^{+3.0}_{-1.2}$ keV
plasma and $\log N_H \simeq 22.9^{+0.4}_{-0.4}$ cm$^{-2}$. Panels
(c) and (d) show typical sources with moderate absorption. Panel
(c) is source \#25 with a near-IR counterpart with K-band excess,
probably a classical T Tauri star. Its 35 counts are fitted by a
$1.7^{+0.7}_{-0.3}$ keV plasma with $\log N_H \simeq
21.9^{+0.2}_{-0.3}$ cm$^{-2}$. Source \#52 in panel (d), also
probably a classical T Tauri star, exhibited a strong flare.  Its
478 counts are not well fit by the standard plasma model.  The
model shown here has $\log N_H \simeq 21.8^{+0.06}_{-0.07}$
cm$^{-2}$ and $kT \simeq 2.3^{+0.5}_{-0.2}$ keV. Excess abundance
of sulfur (2.5 keV) and a non-solar neon-to-iron ratio ($\sim 1.0$
keV) may be present. Panels (e) (source \#81) and (f) (source
\#91) show relatively soft spectra with little absorption and
optical counterparts. Panel (e), a likely weak-lined T Tauri star,
has 234 counts fitted by $kT \simeq 1.0^{+0.07}_{-0.08}$ keV
plasma and $\log N_H \simeq 21.4^{+0.2}_{-0.2}$ cm$^{-2}$. Its
spectrum may show excess oxygen (0.82 keV). Panel (f), also
probably a weak-lined T Tauri star, has 85 counts and is fitted by
a model with $\log N_H \simeq 21.1^{+0.4}_{-1.0}$ cm$^{-2}$ and
$kT \simeq 1.3^{+0.3}_{-0.2}$ keV.


\section{Source list and properties} \label{src_list_prop_subsec}

The database of sources found in the NGC 1333 ACIS field is
provided in Tables \ref{src_tbl} and \ref{properties_tbl}. The
first of these tables gives source locations and multiwavelength
properties while the second table gives X-ray properties.
Specifics regarding table entries follow.

\begin{description}

\item [Table \ref{src_tbl}, Column 1] Source number. Sources are sorted by their right ascension.

\item [Column 2] Source name in IAU format Jhhmmss.s-ddmmss with the IAU designation\\ CXONGC1333.

\item [Columns 3-4] Source position for epoch J2000 in sexagesimal units.

\item [Column 5] Distance from the aimpoint of ACIS field (\S \ref{obs_subsec}) in arcminutes.
This quantity is useful for evaluating extraction radii and point
spread functions (\S \ref{extraction_subsec}).

\item [Column 6] Source number in X-ray study of the NGC 1333 star forming region
 using the {\it ROSAT} High Resolution Imager \citep{Preibisch97b}.

\item [Columns 7-9] Optical counterparts: HJ = source number from the catalog derived from R-band
 photographic plates by  \citet{Herbig83}; BR = visible on H$\alpha$/[SII] CCD image
 by \citet[][see Figure \ref{counterparts2_fig}]{Bally01}; USNO = listed in USNO-A2.0 astrometric
 catalog \citep{Monet98}.  For USNO conterparts we give the offset between the X-ray and USNO-A2.0 positions in arcseconds.

\item [Column 10] I-band magnitudes from our KPNO image (\S \ref{zband_subsec}). `Yes' indicates the star
 is saturated while `Conf' indicates a confused region where X-ray position lost in the glare
 of a nearby source or reflection nebula.

\item [Columns 11-12] Near-IR counterparts: source numbers from the lists of \citet[][ASR]{Aspin94}
and \citet[][LAL]{Lada96}. Bl = blended cases where ACIS resolved
a close double that is unresolved in the infrared data (Figure
\ref{figure_acis_image}).

\item [Columns 13-15] Near-IR properties: K band magnitude, J-H and H-K color magnitudes based on
the near-IR photometry of \citet{Lada96}.  For sources \#9, 35 and
51, we use the photometry of  \citet{Aspin94}.

\item [Column 16] Footnotes indicate other counterparts including mid-IR, far-IR, mm/submm and radio counterparts.
 ISOCAM counterparts are based on visual inspection of public domain ISO images.
`?' indicates a weak or tentative ISOCAM source; these are not
included in the analysis.

\end{description}

\begin{description}

\item [Table \ref{properties_tbl}, Columns 1-2] Source number and name from Table \ref{src_tbl}.

\item [Columns 3-6] Quantities associated with event extraction from
the full band ($0.5-8$ keV) image (\S \ref{extraction_subsec}):
extracted counts $C_{xtr}$ from radius $R_{xtr}$, background
counts $B_{xtr}$, and the fraction $f_{PSF}$ of the source's
events collected within a circle of radius $R_{xtr}$.

\item [Columns 7] Average source count rate $CR$ during the observation (\S \ref{extraction_subsec}).

\item [Column 8] Variability class (\S \ref{var_analysis_subsec}): `Flare' (flare), `Pos flare' (possible flare)
 and `Const' (constant).

\item [Columns 9-10] Spectral parameters from one and two-temperature plasma
models. $\log N_H$ (in cm$^{-2}$) is the equivalent hydrogen
column density of intervening interstellar material producing soft
X-ray absorption, and $kT$ (in keV) is the energy of the plasma.
Parentheses indicate 1$\sigma$ errors.

\item [Columns 11-14] X-ray luminosities of the source assuming a
distance of 318 pc: $L_s$ = soft-band ($0.5-2$ keV) luminosity; $L_h$ = hard-band ($2-8$ keV) luminosity;
 $L_t$ = total band ($0.5-8$ keV) luminosity; and $L_c$ = total band luminosity corrected for
  the estimated interstellar absorption. See F02 for details.

\item [Column 15] Table notes giving details of non-standard spectral fits.

\end{description}

\section{Sources without stellar counterparts} \label{nocounterpart_subsec}

Extragalactic sources, mainly quasars and other active galactic
nuclei, can be seen through molecular clouds in the hard X-ray
band.  Assuming a faint quasar power law spectrum with photon
index $\Gamma = 1.4$ and a typical total column density of $\log
N_H = 22.0$ cm$^{-2}$ across the field, our 13 photon (mean count
value of 30 unidentified sources) point source corresponds to a
limiting sensitivity of $5.5 \times 10^{-15}$ erg s$^{-1}$
cm$^{-2}$ in the hard ($2-8$ keV) band. The corresponding
extragalactic source density is about $400-500$ sources per square
degree \citep{Brandt01b}, or $\sim 32-40$ in our $17\arcmin \times
17\arcmin$ ACIS-I field. From source counts alone, all 29 of our
sources without stellar counterparts plus source \#108 may thus be
extragalactic. Our deep I-band photometric data support this idea
with $m_I \geq 23$ for all these sources.

This interpretation is generally supported by the X-ray properties
of the unidentified sources: they are relatively weak ($5 \leq
C_{xtr} \leq 32$ with a median of 14 counts), exhibit no flares
and are highly absorbed ($\log N_H \geq 22.0$ cm$^{-2}$). Their
spatial distribution is roughly uniform, although a north-south
asymmetry may be marginally present. We will hereafter consider
the unidentified sources to be unrelated to the NGC 1333 young
stellar cluster and will omit them from further
discussion\footnote{A similar conclusion concerning ACIS sources
without counterparts was reached in studies of the Perseus cloud
IC 348 cluster \citep{Preibisch01}, the Ophiuchus cloud core
\citep{Imanishi01}, and the central region of the Pleiades
\citep{Krishnamurthi01}. However, the ACIS image of the Orion
Nebula Cluster had a several-fold excess of unidentified sources
which were too bright for quasars and strongly clustered around
the molecular cloud cores (F02).  The Orion unidentified sources
were therefore inferred to be mostly previously unknown YSOs.}.

\section{The X-ray stellar population \label{stel_pop_subsection}}

Seventy-three of the 77 ACIS sources with YSO counterparts have
near-infrared photometric measurements. Their locations are
displayed as yellow and red circles in Figure
\ref{counterparts2_fig}; about 80\% of them lie in the central
concentrations of the double cluster \citep{Lada96}. Their average
K-band magnitude $K \simeq 11$ corresponds to masses between
$0.2-0.4$ M$_\odot$ for a cluster age of $\sim 1$ Myr
\citep{Lada96}. The infrared $J-H$ $vs.$ $H-K$ diagram in Figure
\ref{color_color_fig} shows that about half of the X-ray sources
are consistent with reddened photospheres, classified as Class III
YSOs or weak-lined T Tauri stars, while half show infrared
excesses characteristic of Class II YSOs or classical T Tauri
stars. Here we adopt the reddening relationship $E(J-H)/E(H-K) =
1.7$ \citep{Cohen81}.  A similar fraction of circumstellar disks
is found among the 16 L-band sources detected in X-rays
\citep{Aspin97}. These are the magnetically active low-mass T
Tauri stars found in all X-ray images of nearby star forming
regions \citep{Feigelson99}.

Figure \ref{nh_vs_av_fig} compares the line-of-sight absorption
estimated from the infrared color-color diagram with the soft
X-ray absorption derived from our spectral analysis (\S
\ref{spectral_analysis_subsec}). The visual absorption $A_V$ is
estimated by projecting each star along the reddening vector to
the main sequence or classical T Tauri star locus in the $J-H$
$vs.$ $H-K$ diagram.  Error bars are provided for selected $\log
N_H$ values to illustrate the internal errors of the X-ray
absorption measurements, evaluated from the XSPEC $\chi^2$
statistic. We find good agreement between the two absorption
measurements for the stronger sources with $A_V \leq 5$ ($\log N_H
\leq 21.9$ cm$^{-2}$), consistent with the relationship $N_H = 2.2
\times 10^{21} A_V$ cm$^{-2}$ obtained by \citet{Ryter96} for a
standard interstellar gas-to-dust ratio, but several objects show
$A_V$ considerably greater than expected from the $N_H$ value.  An
offset towards higher $A_V$ was also seen in several hundred Orion
stars (F02). We cannot distinguish here between several possible
causes of this discrepancy: unmodeled infrared emission from an
inner disk, incorrect modelling of the X-ray spectrum, a
non-standard dust-to-gas ratio associated with an extremely young
YSO, or spatial separation between the X-ray emitting region and
the photosphere.

There are 6 sources which are neither probable background objects
(\S \ref{nocounterpart_subsec}) nor K-band sources.  Two of these
(\#59 and \#65) are previously unresolved companions of bright
K-band cluster members (Figure \ref{figure_acis_image}c and d).
One (\#18) has KPNO/$ISO$ counterparts only. Two (\#5 and \#109)
have optical counterparts: source \#5 \citep[HJ 110,][]{Herbig83}
along with \#33 (BD +30$^\circ$547) are probably foreground stars
\citep{Preibisch97b} while \#109 is probably a cluster member.
Source \#109 is the only one having offset $> 3\arcsec$ from its
optical counterpart, but taking into account the fact that it lies
far off-axis, where the larger offsets are common due to
anisotropies in the {\it Chandra} mirror point spread function, it
is reasonable to assume this identification valid.

The remaining source (\#49) has only radio/submm counterparts with
properties of an extremely young object, probably a Class I
protostar. We consider this source, together with seven K-band
stars that produce Herbig-Haro outflows (\#14, 23, 41, 50, 79, 86
and 90), in \S \ref{protostars_subsec} below. ACIS source \#33,
the foreground star with nearly 3000 ACIS counts overwhelms any
possible ACIS signal from the nearby protostar IRAS 2B = VLA 10 =
SK 7 \citep{Rodriguez99, Sandell01}.

\section{Cluster membership} \label{cluster_membership_section}

X-ray surveys provide insights into the census of a young stellar
population that are distinct from traditional optical and infrared
surveys: X-rays select for elevated magnetic activity, while other
methods mostly select for circumstellar disks.  For example, a
$ROSAT$ study of the Chamaeleon I cloud nearly doubled the cluster
membership known at that time and established that most members
are weak-lined T Tauri stars without prominent disks and that many
stars lose their disks before they reach the birthline
\citep{Feigelson93}.  In clouds such as NGC 1333 where sensitive
and spatially complete K-band surveys locate virtually all young
stellar objects \citep{Lada96}, X-ray observations serve
principally to discriminate cluster members from background
Galactic stars rather than discover new members.

Figure \ref{N_vs_K_fig} summarizes the K-band survey of
\citet{Lada96} and the ACIS survey obtained here.  The histogram
from $7 < K < 16$ shows the distribution of K magnitudes in the
region covered by the ACIS field.  The dotted portion shows the
expected Galactic contamination obtained from nearby control
fields and adjusted for extinction due to the cloud by
\citet{Lada96}. The K-band luminosity function (KLF) is incomplete
below $\rm {K}=14$ magnitude and the counts in the bins $> 14.5$
in Figure \ref{N_vs_K_fig} are estimates based on control field
observations of \citet{Lada96}. For these brightness bins, the
number of cluster members contributing to the KLF is presently
uncertain but likely to be small. The hatched portion shows the K
stars that are also ACIS sources.

We detect all of the clusters members with $K < 12$. These 55
ACIS/K-band sources with $K < 12$ constitute a nearly `complete'
sample. From $K \simeq 12$ to $K \simeq 16$, the X-ray
observations discriminate a decreasing fraction of cluster members
from background stars.  The 18 candidate faint ACIS sources listed
in Table \ref{faint_src_tbl} have $10 < K < 15$ with a median of
$K \simeq 13$, as shown in the double-hatched portion of Figure
\ref{N_vs_K_fig}. The facts that these candidate sources have K
magnitudes fainter than the X-ray brighter sources and near the $K
\simeq 13$ peak of the cluster luminosity function give confidence
that they are real X-ray sources. If they were spurious, they
would be associated with random K-band sources, most of which are
background stars around $14 < K < 16$. There is thus good reason
to accept most of the identifications in Table \ref{faint_src_tbl}
as valid.  This indicates that ACIS sources as faint as 3 counts
(with counterparts in other wavebands) can be valid, and that a
future deeper exposure should successfully detect most of the
remaining cluster members.

As noted in \S \ref{data_selection_subsec}, no matches of
tentative sources to sub-mm protostars \citep{Chini01} were found.
A stacking analysis \citep{Brandt01} yielded $\sim$5 counts for
the 6 undetected protostars; one of these events should be
background based on the number of sources, the approximate PSF
size in this region, and the average background rate. An ACIS
exposure, several times deeper than the 37.8 ks studied here would
thus be likely to detect these protostars.

The rightmost bar in Figure \ref{N_vs_K_fig} shows the 32 X-ray
sources without $K < 16$ counterparts.  The grey portion is our
estimation of the extragalactic contamination to the X-ray survey
(\S \ref{nocounterpart_subsec}).  The ACIS source in the hatched
$\rm {K} > 16$ portion is X-ray source \#49 which is associated
with radio/millimeter and mid-IR emitting protostar that is too
deeply embedded to appear in the K-band survey (\S
\ref{protostars_subsec}).  Unlike the ACIS Orion Nebula study
which revealed several dozen X-ray discovered deeply embedded
cluster members (F02), there is no indication here that the
existing K-band plus radio/millimeter surveys are incomplete, at
least for stars with masses $M \geq 0.2$ M$_\odot$\footnote{Recall
that pre-main sequence X-ray emission depends strongly on stellar
mass, for unknown reasons, so that our X-ray detection rate of M
stars and brown dwarfs here is low \citep[][F02]{Preibisch01}.}.

\section{Global X-ray properties} \label{global_properties_subsec}

Figure \ref{x_global_prop_fig} shows univariate distributions of
six measured X-ray properties: extracted counts $C_{xtr}$,
observed total X-ray luminosity $L_t$ ($0.5-8$ keV),
absorption-corrected X-ray luminosity $L_c$, variability class,
plasma energy $kT$, and absorbing column density $N_H$.  In these
histograms, we distinguish between the unidentified background
sources and two foreground stars (clear regions), the population
of cluster members (hatched regions), and stars producing
large-scale jets (marked as circles).  We omit the 32 probable
non-members from our discussion here to concentrate on the NGC
1333 cluster members.

The distribution of source counts $C_{xtr}$ ranges from 5 to 1707
events with log mean and standard deviation of $<\log C_{xtr}> =
1.7 \pm 0.6$ cts and median value $\sim 1.5$ cts (Figure
\ref{x_global_prop_fig}a). Background counts are not included
here; they are usually negligible. The shape of the distribution
is roughly flat at lower count rates\footnote{The apparent drop in
sources between 5 and 10 counts likely does not reflect the true
distribution in the cluster, as sensitivity to the weakest sources
decreases off-axis and we have omitted the faint sources in Table
\ref{faint_src_tbl}.} and falls to a lower level at high count
rates. The distribution of observed X-ray luminosities $\log L_t$
has a similar shape with mean value of $<\log L_t> = 29.1~\pm~0.6$
erg s$^{-1}$ (Figure \ref{x_global_prop_fig}b). This is somewhat
below the comparable value of $29.4~\pm~0.7$ erg s$^{-1}$, found
in the study of the Orion Nebula Cluster with a similar limiting
sensitivity (F02), perhaps because the latter has more higher mass
stars than the NGC 1333 cluster. The distribution of luminosities
after correction for absorption shows a lognormal distribution
with $<\log L_c> = 29.7~\pm~0.5$ erg s$^{-1}$ (Figure
\ref{x_global_prop_fig}c). The effect of the absorption correction
is to push more sources into the luminous $10^{30}-10^{31.5}$ erg
s$^{-1}$ range. Of course, the mean luminosities of the full
cluster population are lower than these values, as the faintest
cluster members are not detected here.

The variability class distribution is shown in Figure
\ref{x_global_prop_fig}d. Of 77 known cluster members 26 show
temporal variability with 14 indicating a strong X-ray flare. The
remaining 51 members with `Constant' emission are dominated by
sources with $C_{xtr} < 50$ counts which are too weak to clearly
show flaring activity. If one considers only the stronger sources,
the distribution among the three variability classes becomes
roughly equal, similar to that seen in the Orion X-ray population
(F02).

Figures \ref{x_global_prop_fig}e and  \ref{x_global_prop_fig}f
show the distribution of plasma energies and the interstellar
column densities. Only sources with $> 30$ extracted counts are
included here. Plasma energies are distributed  with mean and
standard deviation of $<kT> = 2.0 \pm 1.1$ keV and median $\sim
1.7$ keV. 84\% of sources have $kT$ between $0.6-3$ keV, which is
a typical range for X-ray active T Tauri stars \citep{Feigelson99,
Preibisch97a}.  One source (\#30) has unusually high plasma
temperature up to $\sim 7$ keV. Such ultra-hot plasmas have been
found in {\it ASCA} studies during powerful T Tauri and protostar
flares \citep[e.g.][]{Tsuboi98}. Most of NGC 1333 stars have
plasmas hotter than seen in the Sun, even during its most powerful
contemporary flares \citep{Reale01}. There is no apparent
relationship between variability (`Flare' and `Possible flare'
variability classes) and spectral hardness. This implies that the
`quiescent' emission of YSOs arises from plasma as hot as `flare'
emission, supporting the idea that quiescent emission in
magnetically active stars arises from microflares rather than
coronal processes \citep{Cargill94, Guedel97a, Drake00}.

The interstellar column densities derived from X-ray spectral
fitting (Figure \ref{x_global_prop_fig}f) are not an intrinsic
property of the X-ray emission, but rather reflect the location of
each star in relation to molecular cloud material in NGC 1333.
Most of the $\log N_H$ values are normally distributed with mean
$<\log N_H> \simeq 21.7$ cm$^{-2}$, but 5 sources suffer no
detectable absorption with $\log N_H < 20.0$ cm$^{-2}$. Seven
stellar sources have $\log N_H \geq 22.2$ cm$^{-2}$ ($A_V \geq 10$
magnitudes).  We cannot distinguish here between those that happen
to lie deep within or on the far side of the cloud, and those that
are very young protostars with dense circumstellar envelopes.

A simple correlation $L_s \sim 10^{-4}L_{bol}$ for the soft X-ray
emission in several young stellar populations has been reported
(e.g. Feigelson et al. 1993 for the Chamaeleon I cloud and
Casanova et al. 1995 for the Ophiuchi cloud). The astrophysical
cause of this relation is unknown, as these stars typically have
X-ray emission $10^{1}$-fold below saturation levels where the
correlation $L_x \propto R_\star^2 \propto L_{bol}$ is expected.
It may reflect the dependence of bolometric and X-ray luminosities
on the mass, age or internal dynamo processes of pre-main sequence
stars. As $J$-band magnitudes are a good empirical measure of
$L_{bol}$ \citep{Greene94} and the extinction cross-sections are
the same in the soft X-ray and $J$ bands \citep{Casanova95}, $\log
L_s$ and $J$ values can be directly compared without absorption
correction. Figure \ref{L_vs_J_fig} shows such a correlation for
the $0.5-2$ keV luminosities of NGC 1333 YSOs.  A similar relation
is seen for the total band ($0.5-8$ keV) luminosities. Excluding
three outliers, BD +30$^\circ$549, SVS 3 (\S
\ref{intermediate_mass_section}),  and SVS 16 (\S
\ref{intersting_sources_subsec}), linear fits to the data give
$\log L_s =  - 0.48(\pm 0.07) \times (J - 12) + 29.25(\pm 0.06)$
and $\log L_t =  - 0.47(\pm 0.06) \times (J - 12 ) + 29.59(\pm
0.07)$ for the soft and total energy band,
respectively\footnote{These regressions are the ordinary least
square bisector lines which treat the variables symmetrically
\citep{Feigelson92}.}. \citet{Casanova95} found $\log L_{s} = -
0.3 \times (J  - 12) + 29$ for {\it ROSAT}-detected stars in the
Ophiuchi cloud.

NGC 1333 and IC 348 are the two most prominent young stellar
clusters in the Perseus molecular cloud where IC 348 is older with
the age of about $3-7$ Myr \citep{Lada95, Herbig98}. It is
interesting to investigate whether the X-ray properties of the
stars in the NGC 1333 are different from those of the IC 348 due
perhaps to an evolution of magnetic activity. We have compared the
absorption-corrected X-ray luminosity functions of 77 members of
NGC 1333 and 168 members of IC 348 \citep[Table 1
in][]{Preibisch02} using the two-sided two-sample
Kolmogorov-Smirnov test. These samples constitute $70-80$\% of the
entire populations in both cases. No difference in the X-ray
luminosity functions is present. This supports the suggestion of
\citet{Preibisch97b} that, during the PMS evolution between $1-7$
Myr, an increase in the X-ray surface flux from an increase in the
rotation rate may be compensated by a decrease in the stellar
surface area.

\section{X-ray emission from late B stars}
\label{intermediate_mass_section}

Theoretically, X-ray emission from late B stars is not expected as
they lack both the convection zones driving magnetic activity and
the strong radiation driven stellar winds with shocks responsible
for X-ray emission in O and early B stars \citep{Pallavicini89}.
We detect here X-ray emission from both stars that illuminate the
optical reflection nebulae:  BD +30$^\circ$549 (B9, ACIS source
\#83) and SVS 3 (B6, \#64).

These two sources differ in X-ray luminosity by a factor of $\sim
40$, after correction for reddening.  Only 39 photons are seen
from the lightly absorbed B9 star compared to 1161 photons from
the heavily absorbed B6 star; the corresponding X-ray luminosities
are $\log L_c = 29.1$ and $30.9$ erg s$^{-1}$ and emissivities are
$\log L_c/L_{bol} = -6.4$ and $-3.8$ respectively.  A similar wide
dispersion among intermediate-mass young stars is seen in the
Orion Nebula Cluster (see their Figure 12 in F02). These stars are
outliers on different sides of the $L_s-J$ diagram in Figure
\ref{L_vs_J_fig}; a similar effect is reported for intermediate
mass stars in the IC~348 cluster \citep[][ see their Figure
7]{Preibisch02}.

The simplest resolution to these discrepancies between
intermediate- and low-mass pre-main sequence X-ray emission is
that the observed emission originates from unresolved late-type
companions and not from the late-B stars themselves.  This has
been argued by various researchers \citep[e.g.][]{Schmitt85,
Simon95}.  The `Constant' variability class of SVS 3 may suggest
that its emission does not arise from magnetic flares. Such strong
constant emission is unusual, but not unique, among X-ray luminous
T Tauri stars. For example, JW 286 is a solar-mass Orion Nebula
star with constant emission at $\log L_x = 30.5$ erg s$^{-1}$ over
two $\sim 40$ ks observations separated by 6 months (F02).

\section{On the X-ray properties of classical vs. weak-lined T Tauri stars}
\label{ctt_vs_wtt_section}

There has been some debate concerning the X-ray properties of
classical and weak-lined T Tauri stars. In several young stellar
clusters associated with nearby star formation regions --
Chamaeleon I \citep{Feigelson93}, Ophiuchus \citep{Casanova95,
Grosso00}, Orion Nebula Cluster (F02) and IC~348
\citep{Preibisch01} -- no substantial difference in X-ray
properties are found between these two classes of T Tauri stars.
However, $ROSAT$ studies of the more extended Taurus-Auriga clouds
show WTTs are on average several times more X-ray luminous than
CTTs \citep{Nuhaeuser95, Stelzer01}. \citet{Preibisch02} argue
that this arises from a combination of the $L_x - L_{bol}$
correlation (\S \ref{global_properties_subsec}) and incompleteness
in the sample of faint Taurus-Auriga WTTs.

For the NGC 1333 cluster, we examine the subsample of $K \le 12$
stars which has 24 CTTs and 22 WTTs (Figure \ref{N_vs_K_fig}).
This subsample is complete in two senses: it includes all cluster
members above a mass $M \simeq 0.2M_\odot$, and all of these stars
are X-ray detected here. Figure \ref{L_vs_K_fig} plots the total
band ($0.5-8$ keV) X-ray luminosities derived from our spectral
fitting versus observed $K$-magnitude. The X-ray luminosity
distributions show no systematic differences between the two
classes: the means and standard deviations are $<\log L_t> =
29.4~\pm~0.6$ erg s$^{- 1}$ for CTTs and $<\log L_t> =
29.2~\pm~0.6$ erg s$^{-1}$ for WTTs. Similar results are found for
absorption-corrected X-ray luminosities: $<\log L_c> =
29.8~\pm~0.6$ erg s$^{- 1}$ for CTTs and $<\log L_c> =
29.6~\pm~0.5$ erg s$^{-1}$ for WTTs.

The NGC 1333 results thus confirm the findings from most other
young stellar clusters, which suggest that there is no difference
in the astrophysical mechanism of WTT and CTT X-ray emission. We
support the interpretation of \citet{Preibisch02} that the
Taurus-Auriga measurement of higher X-ray luminosities in WTTs
compared to CTTs is a sample selection bias rather than an
astrophysical effect.

\section{X-ray spectra} \label{spectra2_section}

Spectroscopy can address important questions concerning the
astrophysical origins of young stellar X-ray emission.  The solar
flare paradigm predicts a statistical scaling between plasma
temperature and X-ray emission, roughly as $L_x \propto T^{3 \pm
1}$ \citep[e.g.][]{Schmitt90}. This is predicted by simple models
of plasma heated in magnetic coronal loops \citep{Rosner78}, with
or without additional heating from multiple weak flares
\citep{Drake00}, and is seen in $ROSAT$ studies of magnetically
late-type active stars \citep[e.g.][]{Preibisch97a, Guedel97}. The
$L_x-T$ diagram for NGC 1333 sources (not shown) between $29 <
\log L_c < 31$ erg s$^{-1}$ and $6.8 < \log T < 7.6$ $^\circ$K
shows a broad trend consistent with previous results, but the
random scatter is comparable to the correlation.  One source
(\#30) shows spectrum dominated by plasma temperature above 60 MK
($kT > 5$ keV).  Such super-hot temperatures are seen only during
rare powerful flares on the Sun, but were seen with {\it Chandra}
in $\sim 20\%$ of the Orion Nebula Cluster sources (F02). It is
possible that the prevalence of super-hot plasmas requires
non-solar-like magnetic reconnection, perhaps involving star-disk
magnetic interaction \citep[e.g.][]{Hayashi96, Montmerle00}, but
the evidence for this model is not yet convincing.

A second issue concerns the complexity of individual spectra.  The
Sun and magnetically active stars exhibit a distribution of
emission measures over a wide range of temperatures
\citep[e.g.][]{Brinkman01}. Time-variable abundance anomalies are
present in some solar flares, and were recently found to be very
prominent in powerful flares of magnetically active stars. The NGC
1333 sources are generally too faint to reveal such effects, but
some hints emerge. Figure \ref{spectra_fig}d shows that a
one-temperature plasma model gives a poor fit in the $2-4$ keV
range. This could either be due to a several-fold excess of sulfur
and argon compared to solar abundances or, more probably, to a
multi-component plasma with temperatures from $0.8-3$ keV.

Figure \ref{79.spectrum_fig} shows the spectrum of the CTT star
LkH$\alpha$ 270 (source \#79 with 1707 counts) which clearly needs
a two-temperature model. The top panel shows a two-temperature
model with components at 0.5 and 3.1 keV. Note the discrepancies
around $1.0-1.2$ keV; the spectrum shows a narrow line from the
Lyman-$\alpha$ transition of hydrogenic neon at 1.02 keV, while
the model has a weaker neon line with a broad shoulder towards
higher energies due to L-shell emission lines from iron-group
elements.  The bottom panel shows a two-temperature model where
the abundance of neon is enhanced 2.5 times over solar levels and
iron-like elements are reduced to 0.4 times solar levels.  This
reproduces the observed line shape nicely, and the ratio
[Ne]/[Fe]$\simeq 6$ is similar to the `Inverse First Ionization
Potential effect' recently seen in high-resolution X-ray spectra
of stellar flares in older magnetically active stars
\citep{Brinkman01, Audard01, Huenemoerder01} and TW Hya, the
nearest classical T Tauri star \citep{Kastner01}.  Although the
signal-to-noise ratio in the NGC 1333 spectra is not sufficiently
high to measure abundance effects with statistical significance,
Figure \ref{79.spectrum_fig} lends confidence that such studies
can be conducted with non-dispersed {\it Chandra} ACIS CCD
spectra. If the high [Ne]/[Fe] ratio is confirmed, it would
further demonstrate that the X-ray emission mechanism of classical
T Tauri stars is similar to that of magnetically active stars
without circumstellar disks.

\section{X-ray emission from the sources of HH outflows and embedded object} \label{protostars_subsec}

Protostellar jets in NGC 1333 are detected by their optical line
emission \citep[][see Figure \ref{counterparts2_fig}]{Bally96,
Bally01}, infrared H$_{2}$ line emission \citep{Garden90}, radio
continuum emission \citep{Rodriguez99}, and CO emission
\citep{Knee00}.  Many jets are highly collimated, consisting of a
series of knots and bow shocks, extending as far as several
parsecs out from the driving stellar system. Jet power appears to
decline from the youngest protostellar Class 0 stage through the
classical T Tauri Class II stage, presumably due to a decline in
accretion rate from the disk onto the star. These jets are powered
by the gravitational energy liberated in the accretion disk and
are supplied by both disk and shocked ambient material. Magnetic
fields are critical for converting planar Keplerian motions in the
disk to polar ejection. \citet{Reipurth01}  and \citet{Hartigan00}
provide a more extensive background on these issues.  Coupling
between fields and the disk material requires a modest ionization
fraction; YSO X-ray emission may be an important source of this
ionization \citep{Glassgold01}.

NGC 1333 has a number of relatively lightly absorbed YSOs driving
jets or outflows.  The C-shaped symmetry and external irradiation
of these jets suggests that their host stars have been dynamically
ejected from the cloud core \citep{Bally01}. Seven of these YSOs
are detected in our {\it Chandra} observation.  Their locations
are shown in red in Figure \ref{counterparts2_fig} and their X-ray
properties are denoted as filled circles in Figure
\ref{x_global_prop_fig}.  We describe them individually here:
\begin{description}

\item[ACIS source \#14 = LkH$\alpha$ 351] This object has a knotty H$\alpha$ jet \citep[HH~495;][]{Bally01}.
The X-ray source has an average luminosity of $\log L_t = 29.5$
erg s$^{-1}$, a relatively soft and unabsorbed spectrum ($kT
\simeq 1.3^{+0.1}_{-0.1}$ keV, $\log N_H \simeq
21.0^{+0.3}_{-0.4}$ cm$^{-2}$), and a `Constant' variability
class.  It is curious that its $JHK$ colors do not indicate an
infrared excess.

\item[\#23 = HJ 109] This source drives a bent [SII]-bright jet \citep[HH~498;][]{Bally01}.
Its X-ray properties are typical for cluster members: $\log L_t =
29.5$ erg s$^{-1}$, $\log N_H \simeq 21.3^{+0.1}_{-0.2}$
cm$^{-2}$, and $kT \simeq 2.0^{+0.4}_{-0.3}$ keV. This source
exhibits an X-ray flare.

\item[\#41 = HJ 8] This source drives a bipolar, asymmetrical and knotty jet \citep[HH~334;][]{Bally01}.
 X-ray source is weak and absorbed with $\log L_t = 28.7$ erg s$^{-1}$.

\item[\#50 = HJ 12] This source is the originator of bright bent jets emerging
from both sides of the star ending in terminal shocks
\citep[HH~499;][]{Bally01}. The X-ray source is very weak (9
counts) and heavily absorbed with $\log L_t = 28.5$ erg s$^{-1}$.

\item[\#79 = LkH$\alpha$ 270] This source lies in the middle of the NGC 1333 reflection nebula and drives a monopolar,
 knotty and bent jet 1$\arcmin$ long \citep[HH~335;][]{Bally96,Bally01}.  The X-ray luminosity is above average
  for the cluster at $\log L_t = 30.6$ erg s$^{-1}$, and a flare was seen during the observation.
    Its spectral properties are ordinary ($kT \simeq 2.2^{+0.2}_{-0.2}$ keV, $\log N_H \simeq 21.5^{+0.04}_{-0.03}$ cm$^{-2}$),
     but non-solar elemental abundances may be present (see \S \ref{spectra2_section} and Figure \ref{79.spectrum_fig}) .

\item[\#86 = IRAS f] This source drives a C-shaped monopolar jet \citep[HH~17;][]{Bally96}.
The X-ray spectrum extracted from only 16 counts indicates high
absorption.  The X-ray luminosity is below average at $\log L_t =
28.9$ erg s$^{-1}$.

\item[\#90 = LkH$\alpha$ 271 = SVS 20] The flow from this YSO is seen close to the star as a small and
bright H$\alpha$ bow shock (HH~346) and can be traced to a bright
arc (HH~345) 3\arcmin\/ away \citep{Bally96}. Faint wisps of
H$\alpha$ emission may trace a counterflow. As in the case of
source \#86,  the X-ray source is weak and highly absorbed with
$\log L_t = 29.2$ erg s$^{-1}$.

\end{description}

\noindent From their location in the distributions shown in Figure
\ref{x_global_prop_fig} and from the details given above, we do
not find any significant differences between the X-ray properties
of the jet-driving YSOs and the rest of the young stellar
population in NGC 1333.

More than 20 YSOs in the NGC 1333 star forming region produce jets
or molecular outflows \citep{Knee00, Bally96, Bally01}. The ACIS
image does not reveal X-rays from some of the youngest, deeply
embedded Class 0 and Class I protostars \citep[][and references
therein] {Sandell01, Rudolph01, Reipurth02}.  The $\simeq 1$
M$_\odot$ dense molecular ridge at the base of the high-velocity
HH 7-11 outflow has several YSOs including SVS 13 (IRAS 3 and
optically visible), SVS 13B (Class 0), VLA 2 (also a water maser),
VLA 3, and VLA 20. The far-IR/submillimeter sources IRAS 2A (Class
0), 2B and 2C drive at least two molecular bipolar flows.  IRAS 7
drives the HH 6 jet near the center of the ACIS field; it is also
a complicated system with several submillimeter/radio components
and multiple jets. The protostellar multiple source IRAS 4A-C also
drive two bipolar flows. None of these well-studied protostars
appear in the ACIS-I image.

Our failure to detect the youngest Class 0 protostars in NGC 1333
is consistent with them having the same luminosity distribution as
the Class I and T Tauri stars with typical $\log L_x \simeq
29.0-30.0$ erg s$^{-1}$ but with substantially higher absorptions.
However, it stands in contrast to the reported ACIS-I detection of
MMS 2 and MMS 3 in the Orion OMC-3 cloud cores with absorptions of
$\log N_H \simeq 23.0-23.5$ cm$^{-2}$ and luminosities $L_x \simeq
30.0-30.3$ erg s$^{-1}$ \citep{Tsuboi01}.  Our observation has the
same level of sensitivity as the OMC-2/3 observation and we would
have detected emission at that level. The levels seen in these
Orion protostars may be drawn from a larger population of
protostars and represent the high end of the Class 0 X-ray
luminosity function. It may also be the case that the protostars
in the OMC observation, with an exposure time more than twice that
of our observation, have been detected while they were in a high
variability state.

\citet{Sekimoto97} suggested that protostars may be preferentially
detected in X-rays when viewed along the jet axis, where local
obscuration may be reduced.  Our results are roughly consistent
with this scenario; for example,  we detect two sources (SVS 20,
which ejects HH 345-346, and LkH$\alpha$ 270, which ejects HH 335)
with high-radial velocity jets and optically visible sources and
do not detect objects (sources of HH 333, 336) with low-radial
velocity jets \citep{Bally96, Bally01}.  But the small number of
objects, the difficulty in establishing jet orientations, and wide
differences in obscuration prevent us from establishing this idea
with confidence.

Notably, we detected X-ray emission from a deeply embedded object
(\#49), probably a Class I protostar located in the vicinity of HH
7-11, that is not detected in optical or near-IR observations but
has mm/submm/radio counterparts (see Table
\ref{src_tbl})\footnote{Source \#49 = MMS5 had been mistakenly
classified as a T Tauri star \citep{Chini01} because of the
incorrect identification of this object with the near-IR source
ASR 7.}. The X-ray luminosity of about $10^{29.9}$ erg~s$^{-1}$
(see Table \ref{properties_tbl}) is similar to those of the
embedded young stellar objects in other regions
\citep{Feigelson99, Preibisch01}. Interestingly,  \#49 is also a
variable and circularly polarized gyrosynchrotron radio source VLA
19 \citep{Rodriguez99}, and joins CrA-IRS 5 \citep{Feigelson98} as
the only known X-ray/radio flaring protostar.

\section{Other interesting sources} \label{intersting_sources_subsec}

\begin{figure}
\centering
  \begin{minipage}[t]{1.0\textwidth}
  \centering
  \includegraphics[scale=0.6]{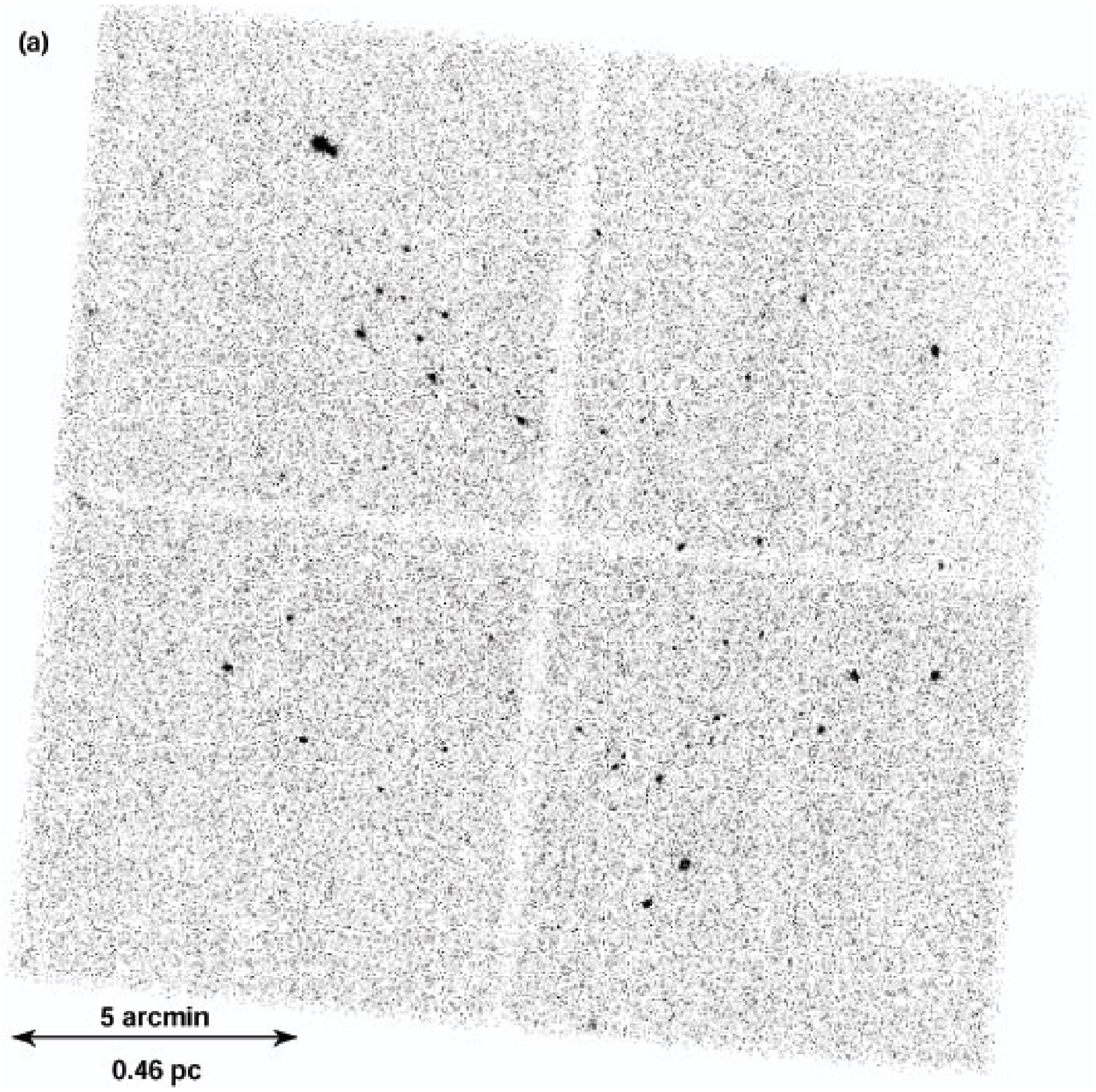}
\end{minipage} \\ [0.4in]
  \begin{minipage}[t]{1.0\textwidth}
  \centering
  \includegraphics[scale=0.15]{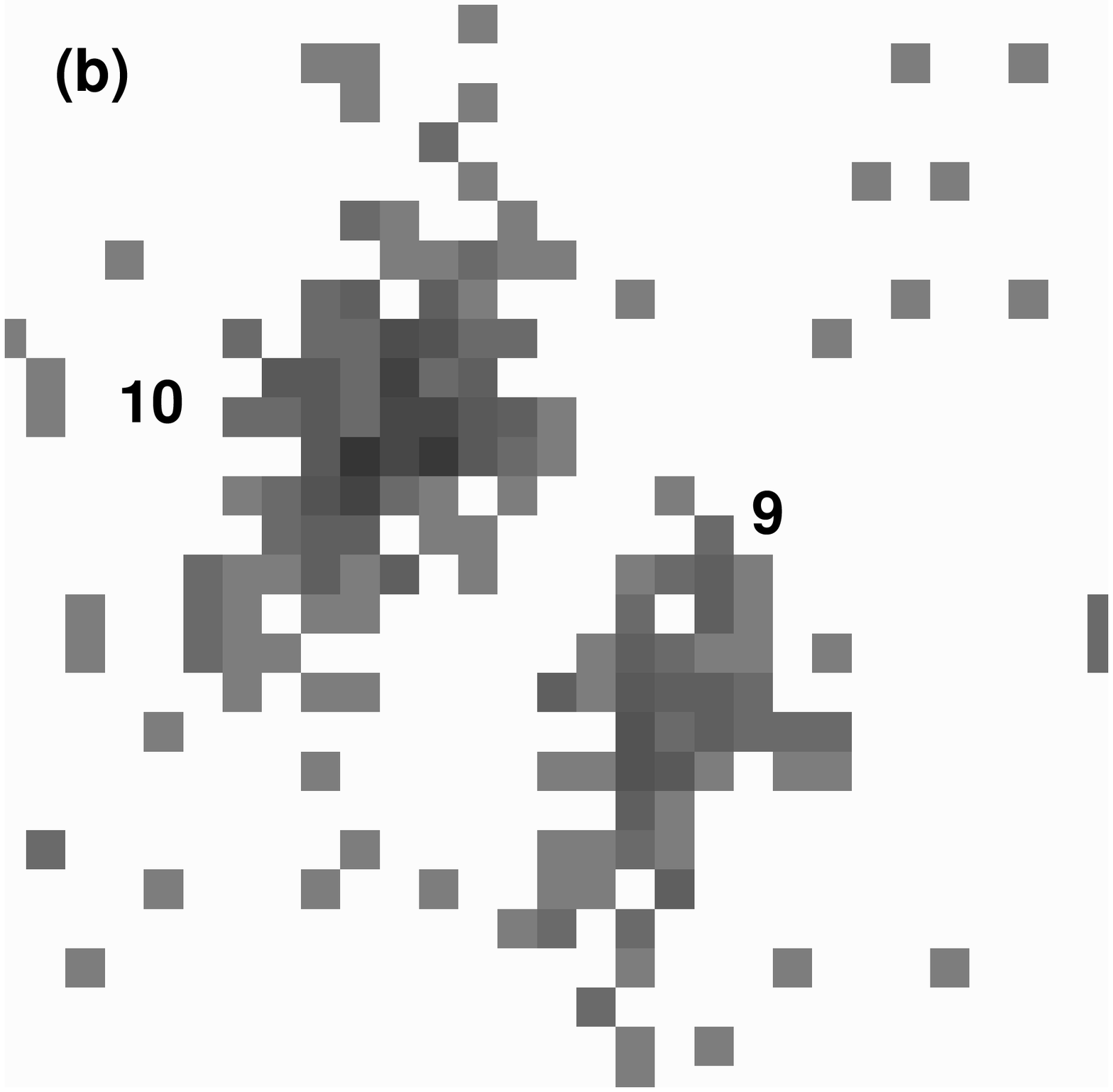} \hspace{0.1in}
\includegraphics[scale=0.15]{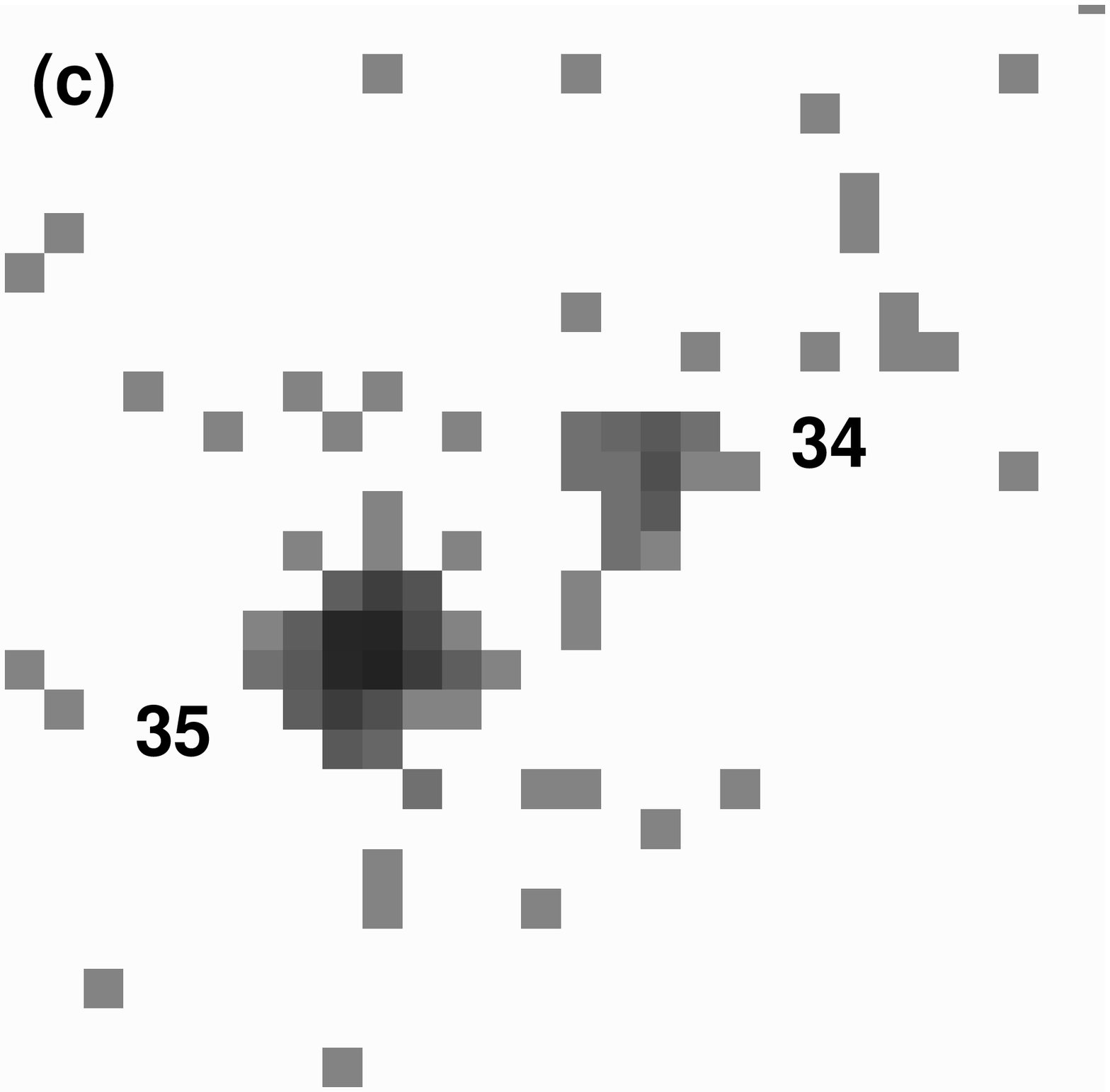} \hspace{0.1in}
\includegraphics[scale=0.15]{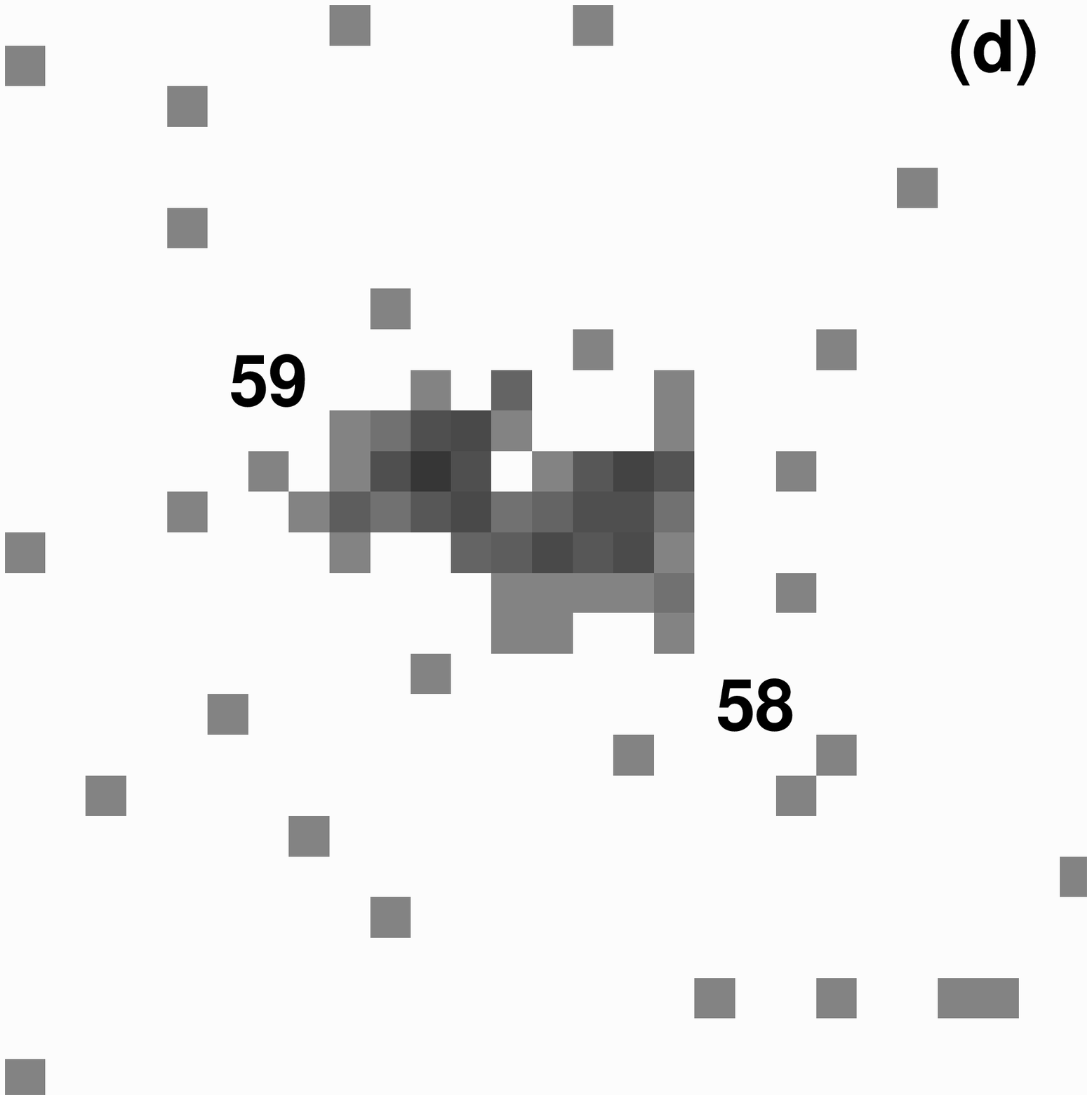} \hspace{0.1in}
\includegraphics[scale=0.15]{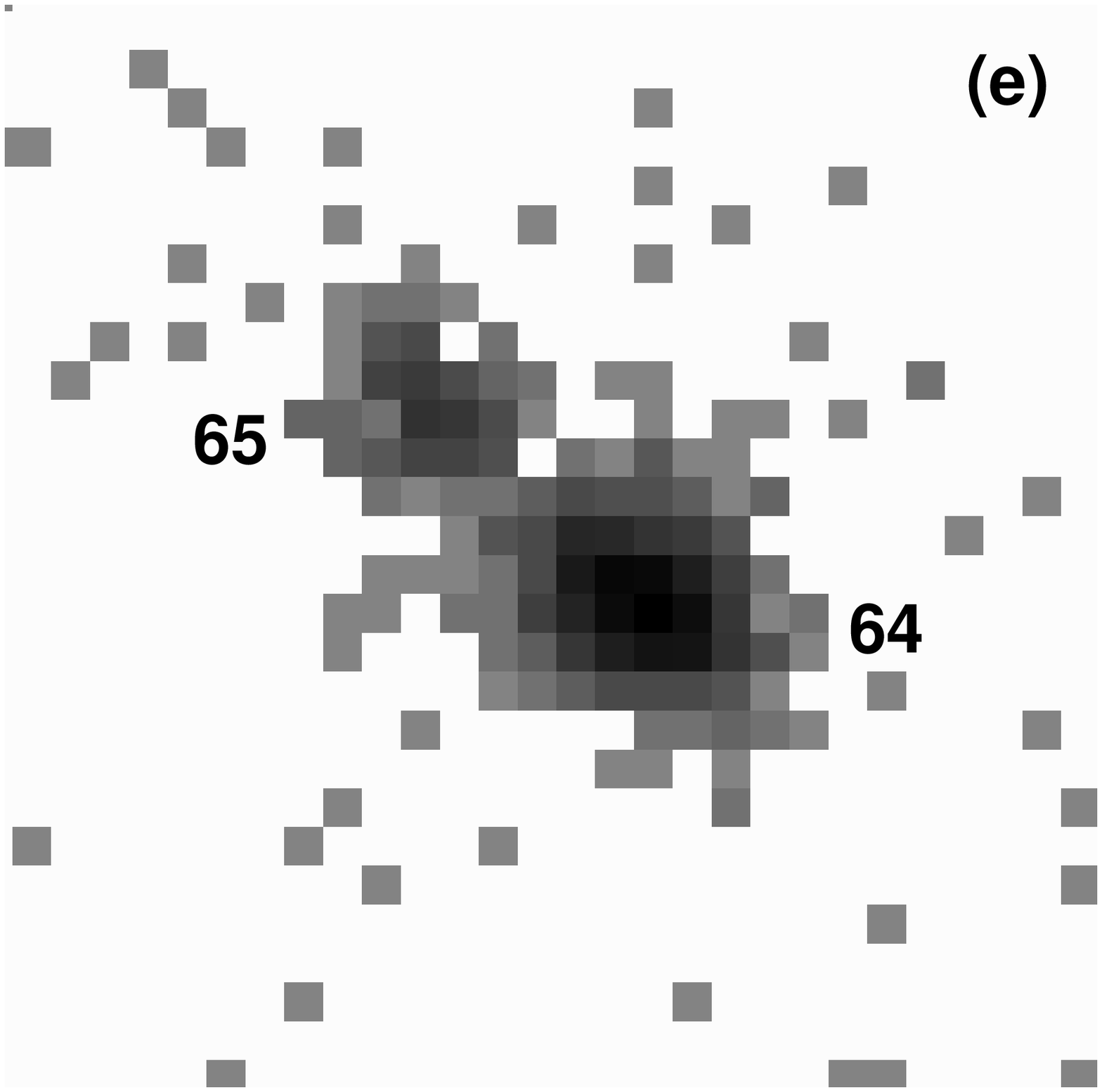}
\caption{(a) Full $17\arcmin \times 17\arcmin$ ACIS-I image of NGC
1333 shown at reduced resolution (2\arcsec/pixel). The small
panels are $14\arcsec \times 14\arcsec\/$ each and show detailed
views of close double sources at full resolution
(0.5\arcsec/pixel): (b) sources 9 and 10; (c) 34 and 35; (d) 58
and 59; and (e) 64 and 65. Note that sources $9+10$ are elongated
due to the {\it Chandra} PSF at $\sim 5\arcmin$ off-axis.
\label{figure_acis_image}}
\end{minipage}
\end{figure}

\begin{figure}
\centering
\includegraphics[width=6.5in]{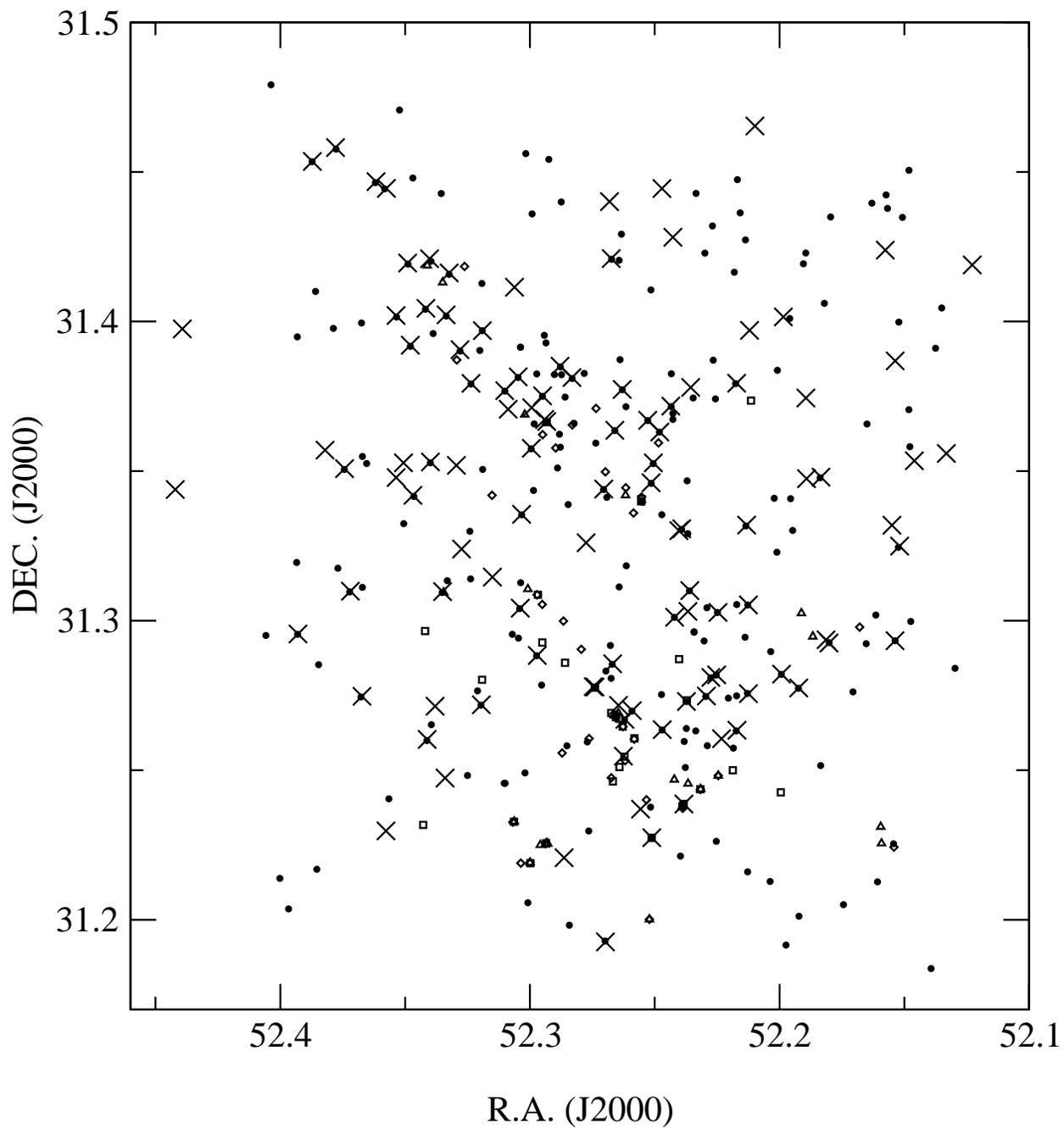}
\caption{Diagram showing the ACIS sources ($\times$) together with
near-infrared sources with $K<15$ ($\bullet$) \citep{Lada96},
far-IR ($\triangle$) \citep[IRAS catalog of Point Sources, IRAS
CPC observations by][]{Jennings87}, submm/mm ($\diamond$)
\citep{Sandell01, Chini01} and radio sources ($\Box$)
\citep{Rodriguez99}.   \label{counterparts1_fig}}
\end{figure}

\begin{figure}
\centering
\includegraphics[width=6.5in]{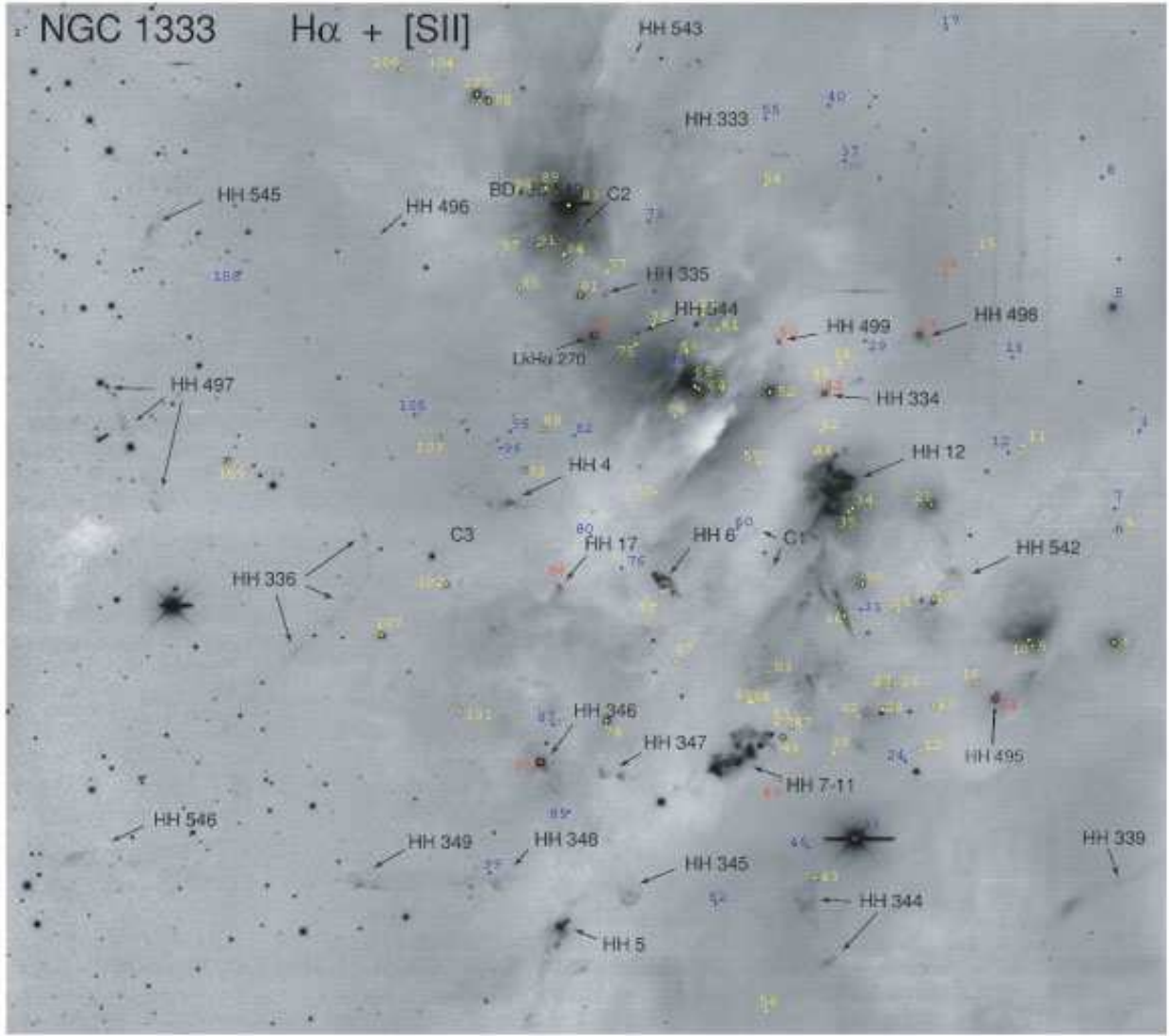}
\caption{ACIS sources and members of the young stellar cluster.
X-ray sources, shown as dots with source number labels, superposed
on an optical H$\alpha$ + [SII] image indicating Herbig-Haro
outflow sources \citep{Bally01}.  Here the X-ray sources are color
coded: outflow/protostellar sources (red), cluster T Tauri sources
(yellow), background sources without stellar counterparts and two
foreground sources (blue).  \label{counterparts2_fig}}
\end{figure}

\begin{figure}
\centering
  \begin{minipage}[t]{1.0\textwidth}
  \centering
  \includegraphics[angle=90.,scale=0.3]{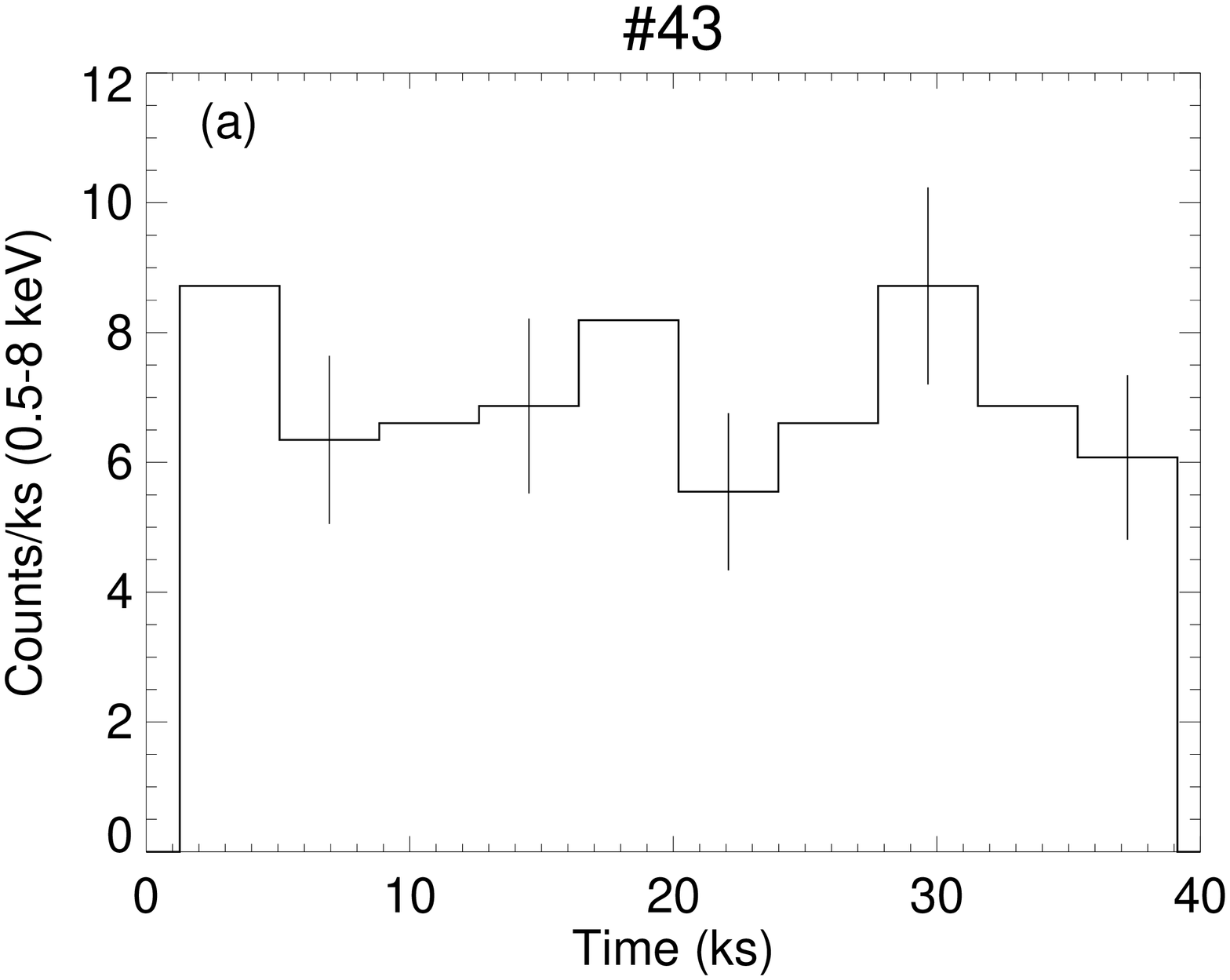} \hspace{0.1in}
  \includegraphics[angle=90.,scale=0.3]{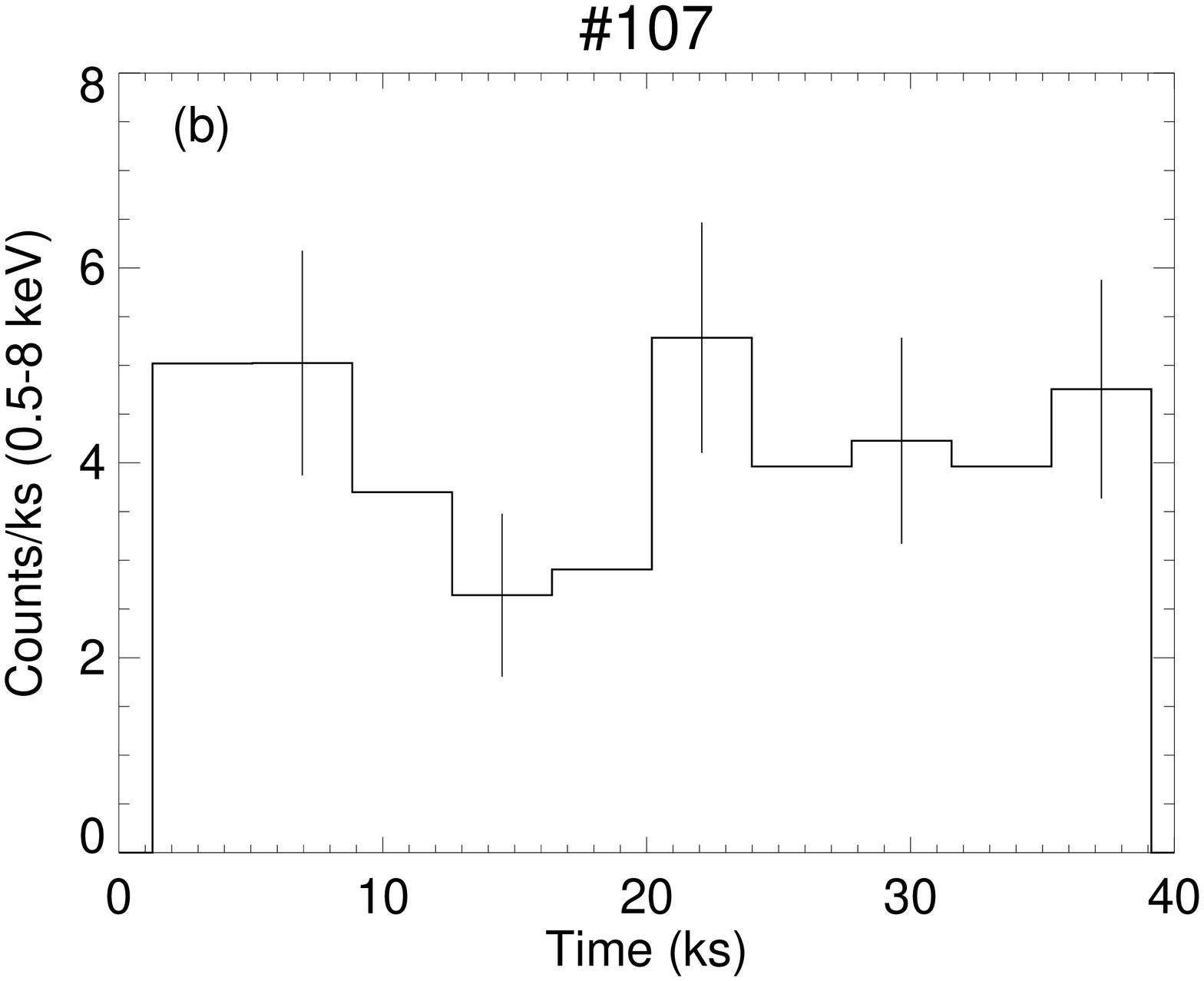}
\end{minipage} \\ [0.2in]

 \begin{minipage}[t]{1.0\textwidth}
  \centering
  \includegraphics[angle=90.,scale=0.3]{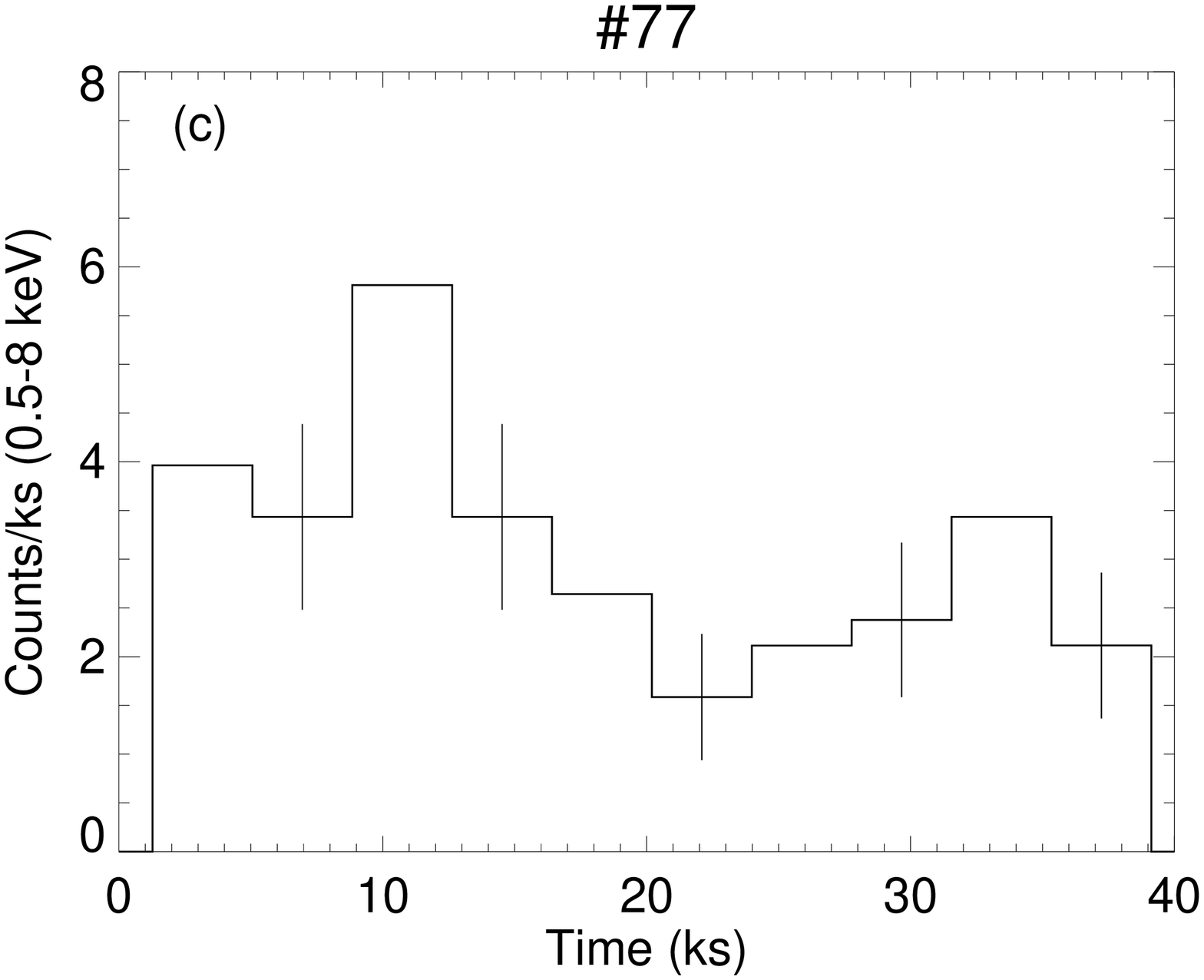} \hspace{0.1in}
  \includegraphics[angle=90.,scale=0.3]{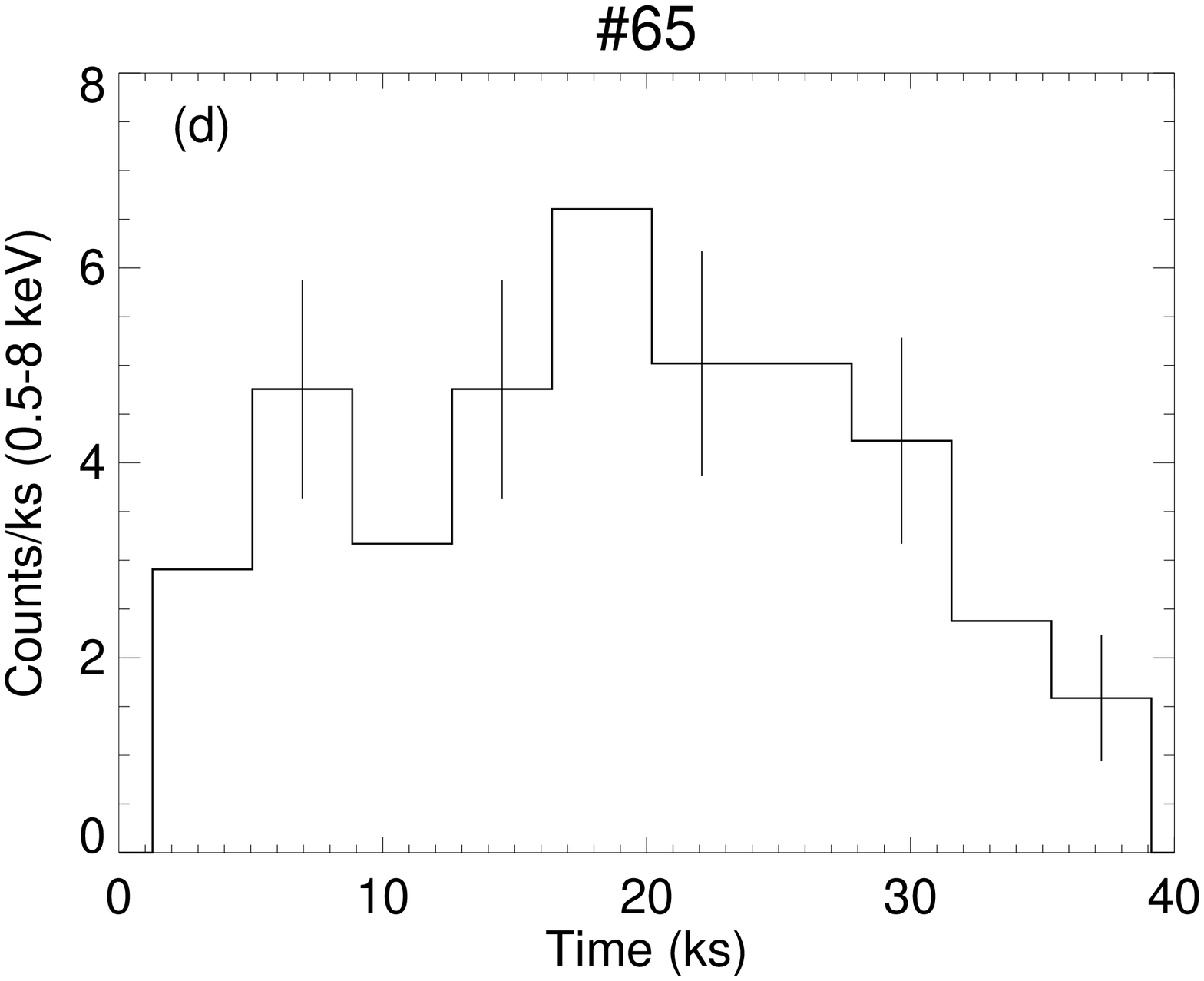}
\end{minipage} \\ [0.2in]

\begin{minipage}[t]{1.0\textwidth}
  \centering
  \includegraphics[angle=90.,scale=0.3]{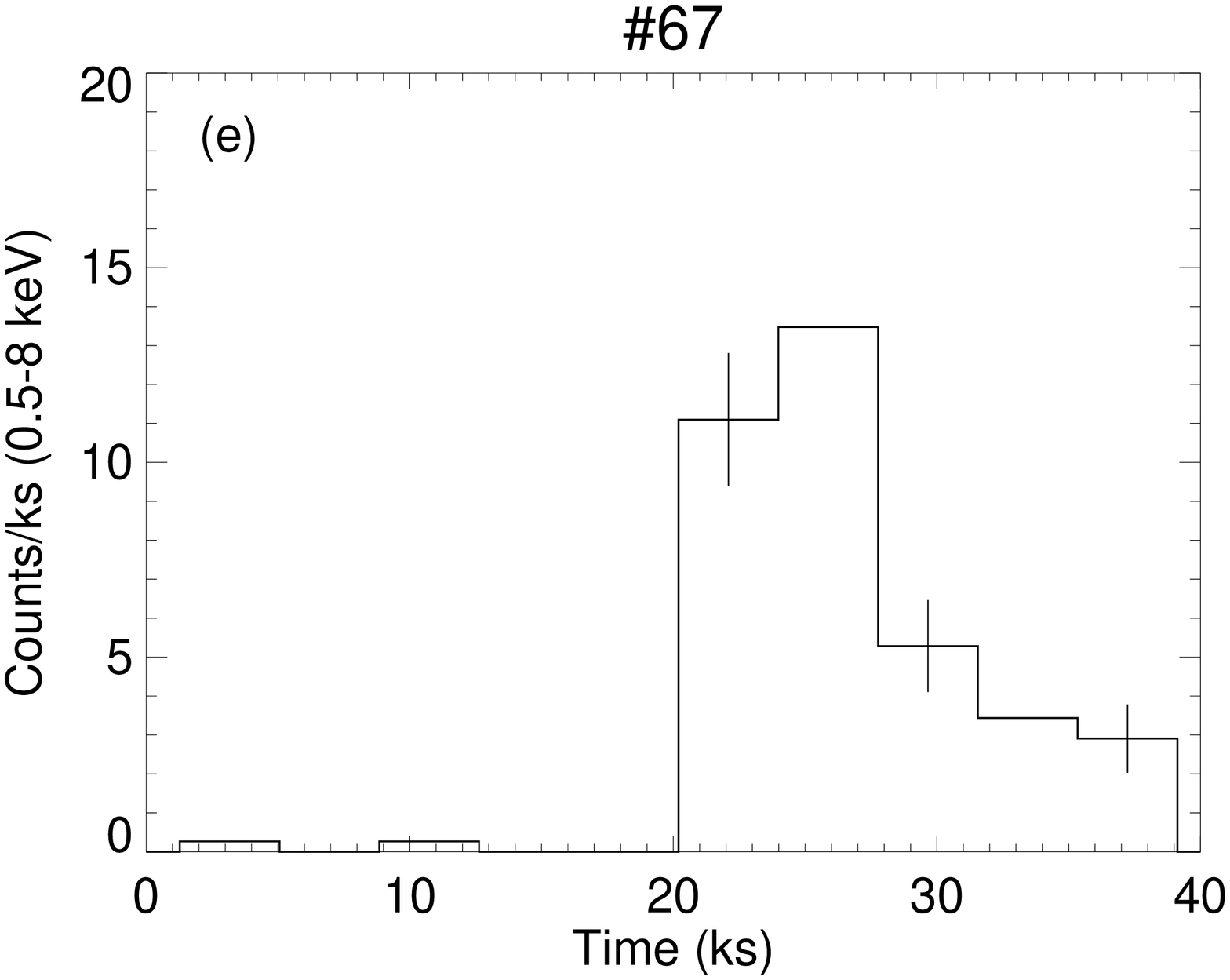} \hspace{0.1in}
  \includegraphics[angle=90.,scale=0.3]{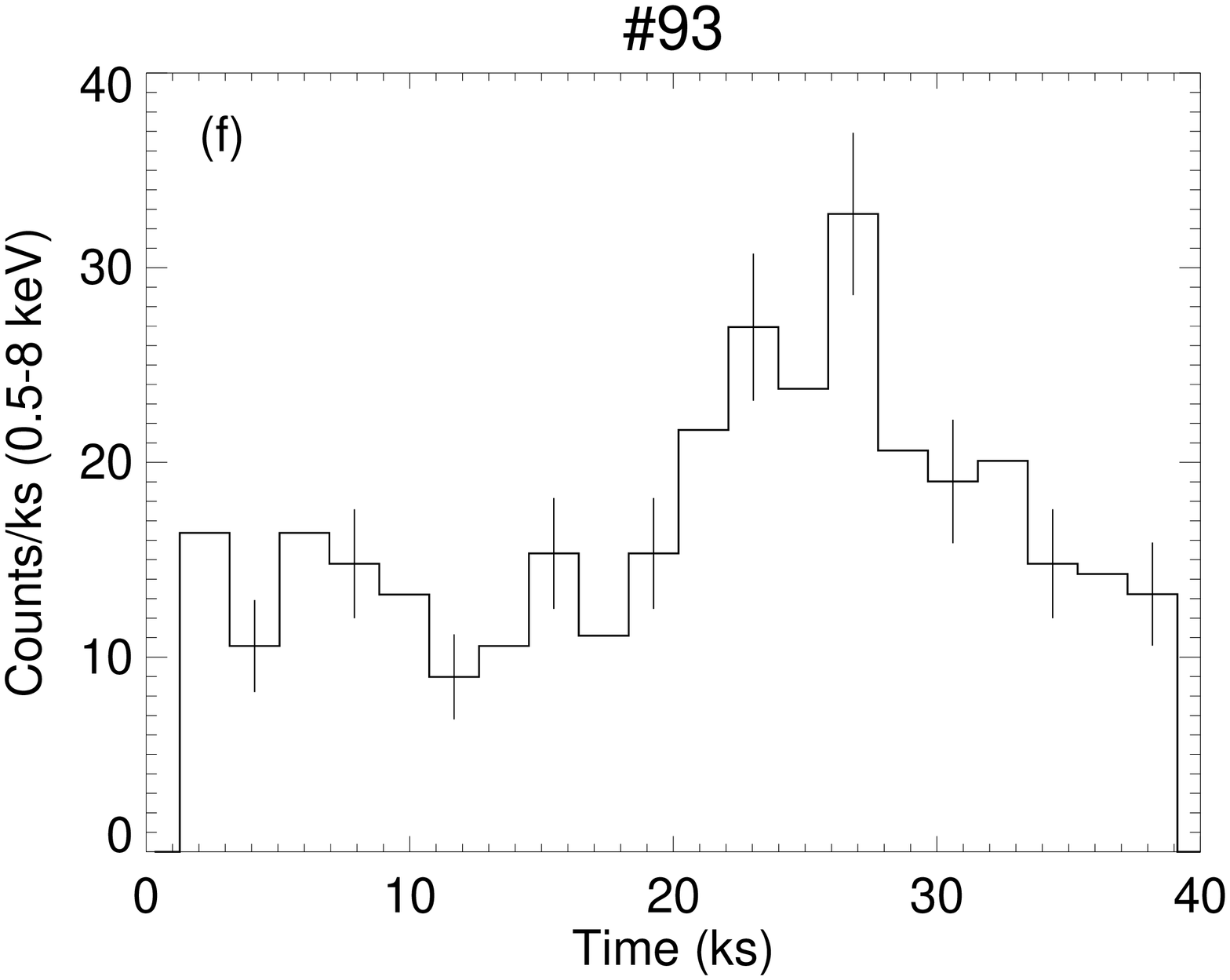}
\caption{Examples of ACIS NGC 1333 young stellar population
variability classes: (a), (b) Constant; (c), (d) Possible Flare;
(e), (f) Flare. The ordinate gives counts ks$^{-1}$ in the total
($0.5-8.0$ keV) band. Error bars show $\sqrt{N}$ uncertainties.
The abscissa gives time in ks, with 10 (20) bins per observation
for weaker (stronger) sources. \label{lightcurve_fig}}
\end{minipage}
\end{figure}

\begin{figure}
\centering
  \begin{minipage}[t]{1.0\textwidth}
  \centering
  \includegraphics[angle=-90.,scale=0.4]{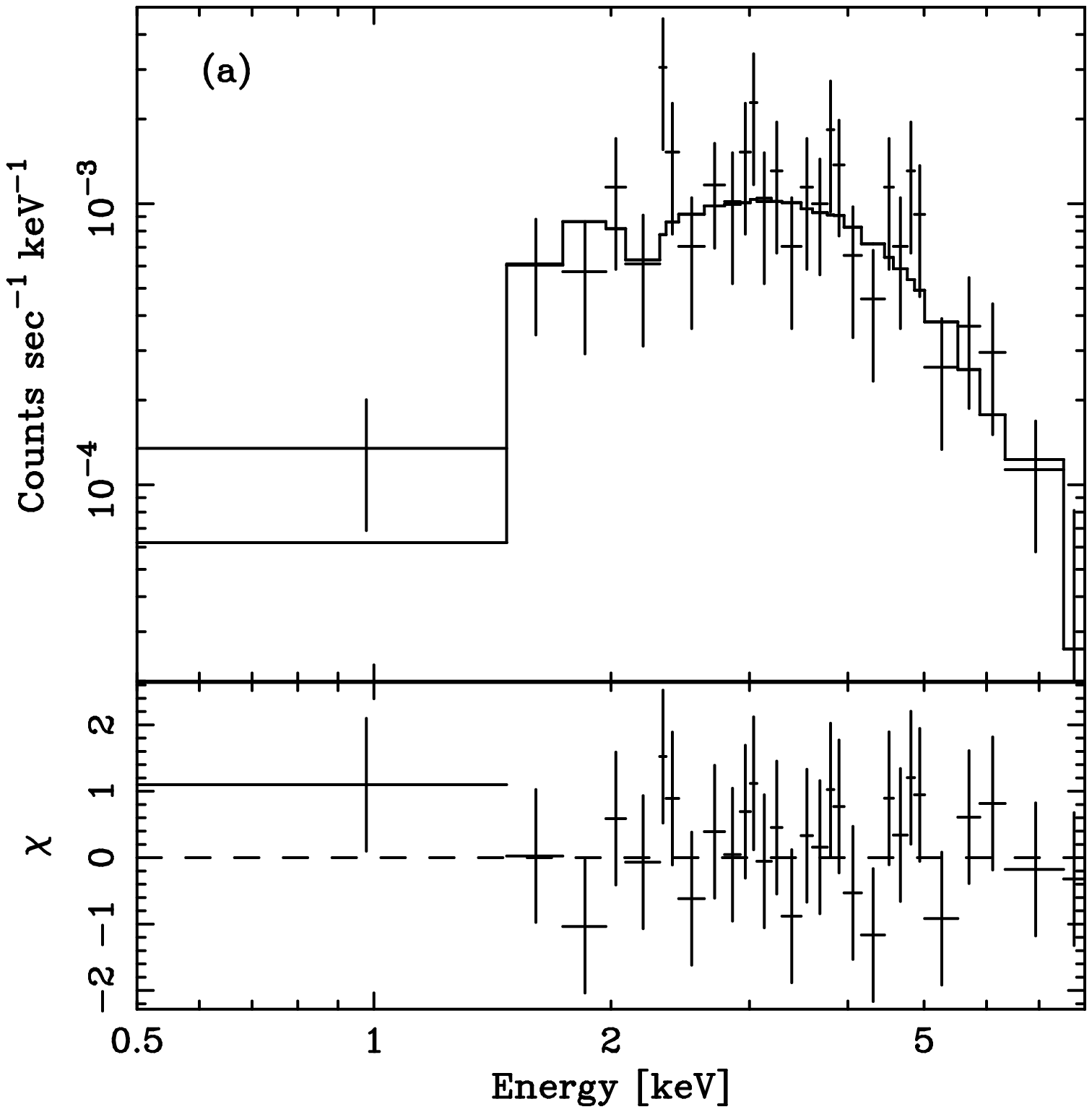} \hspace{0.1in}
  \includegraphics[angle=-90.,scale=0.4]{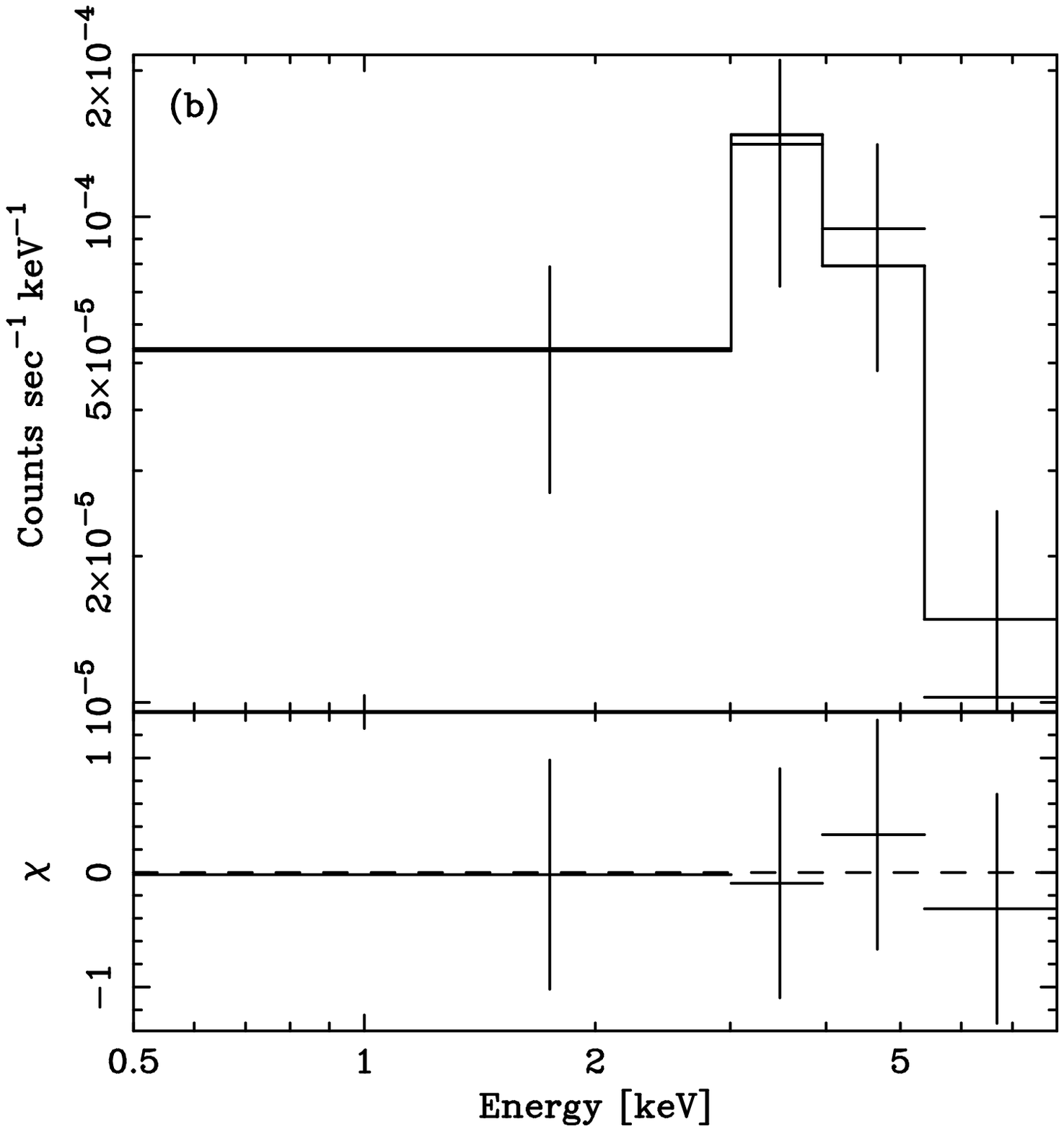}
\end{minipage} \\ [0.2in]

 \begin{minipage}[t]{1.0\textwidth}
  \centering
  \includegraphics[angle=-90.,scale=0.4]{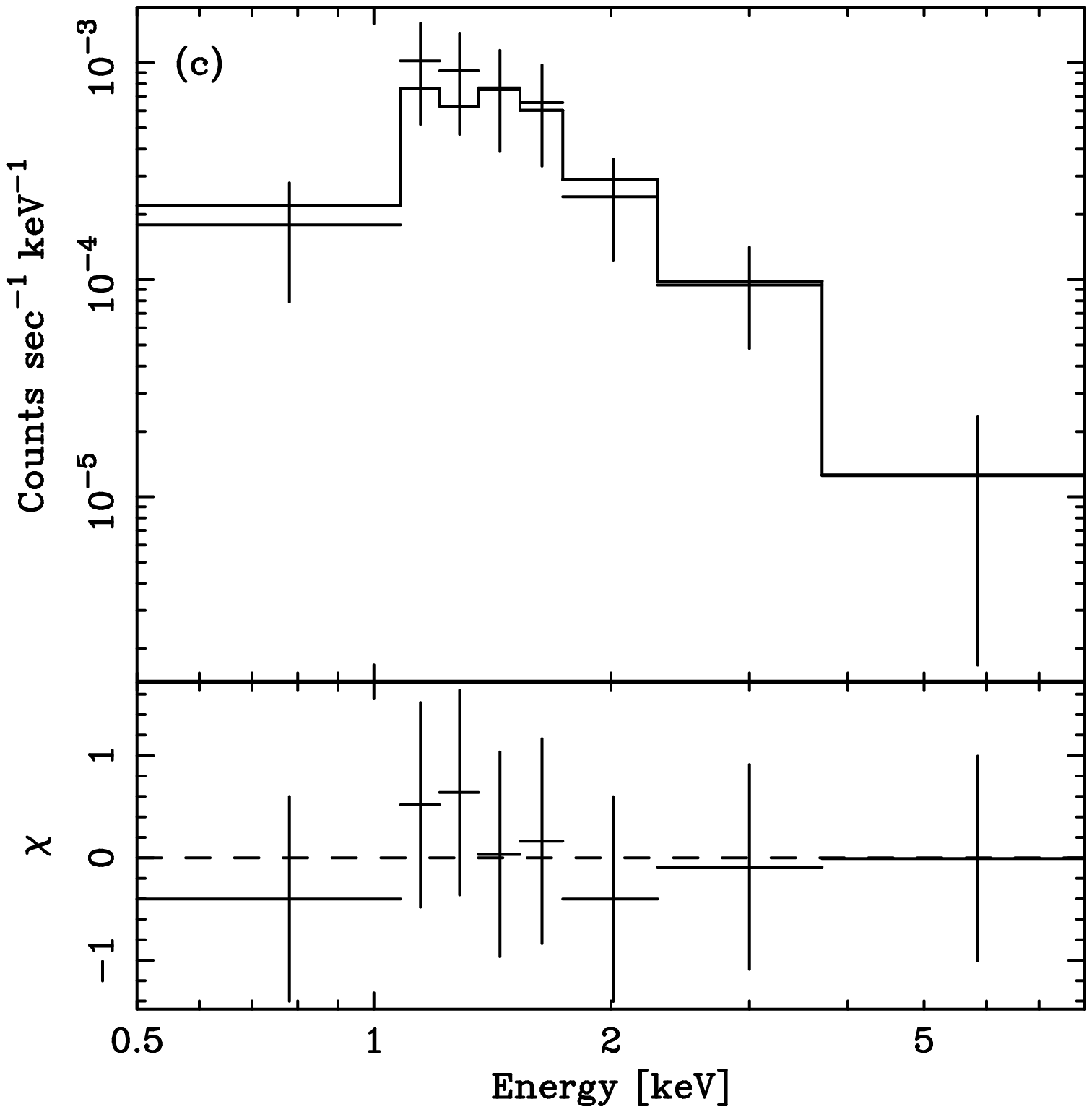} \hspace{0.1in}
  \includegraphics[angle=-90.,scale=0.4]{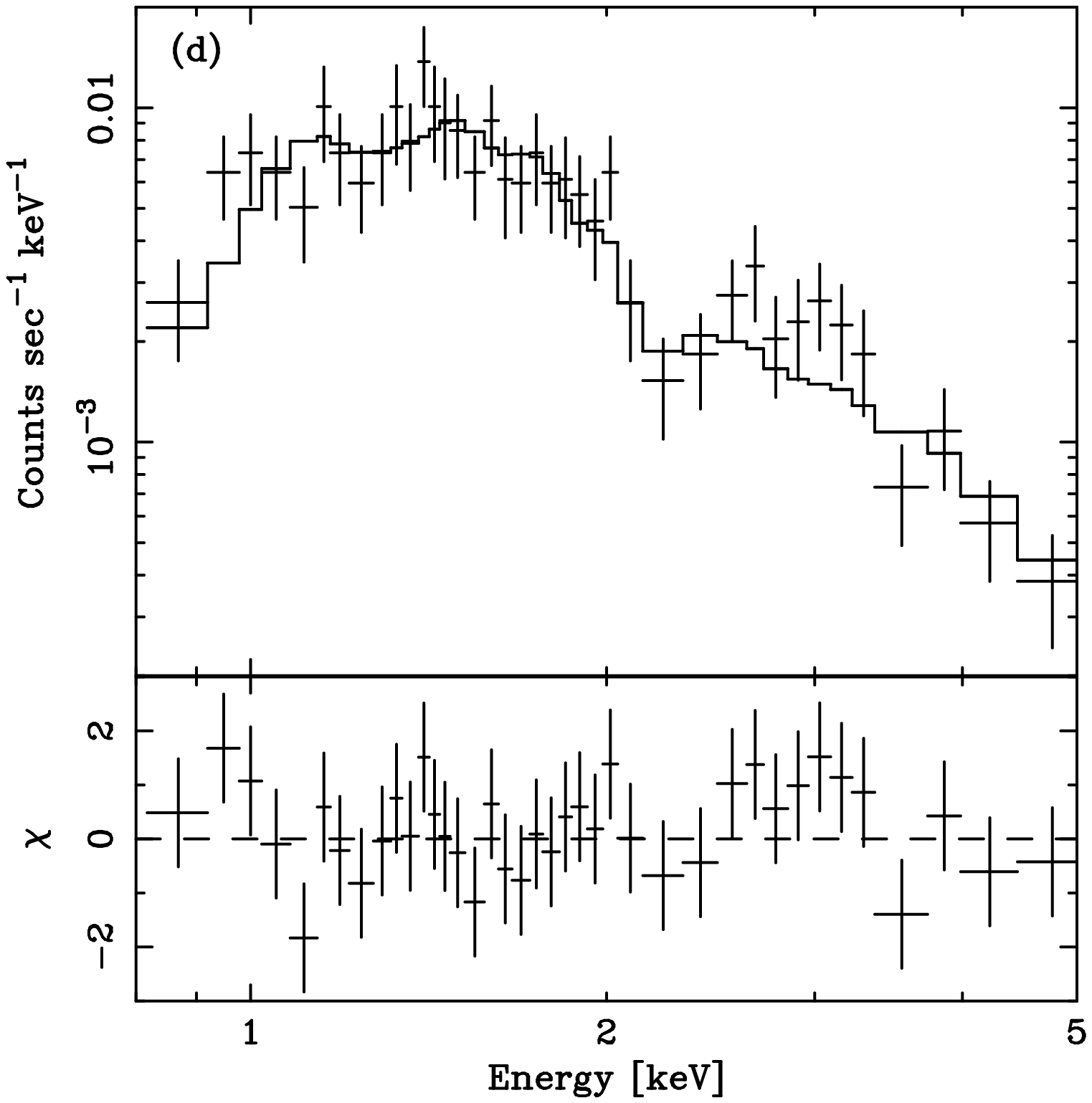}
\end{minipage} \\ [0.2in]

\begin{minipage}[t]{1.0\textwidth}
  \centering
  \includegraphics[angle=-90.,scale=0.4]{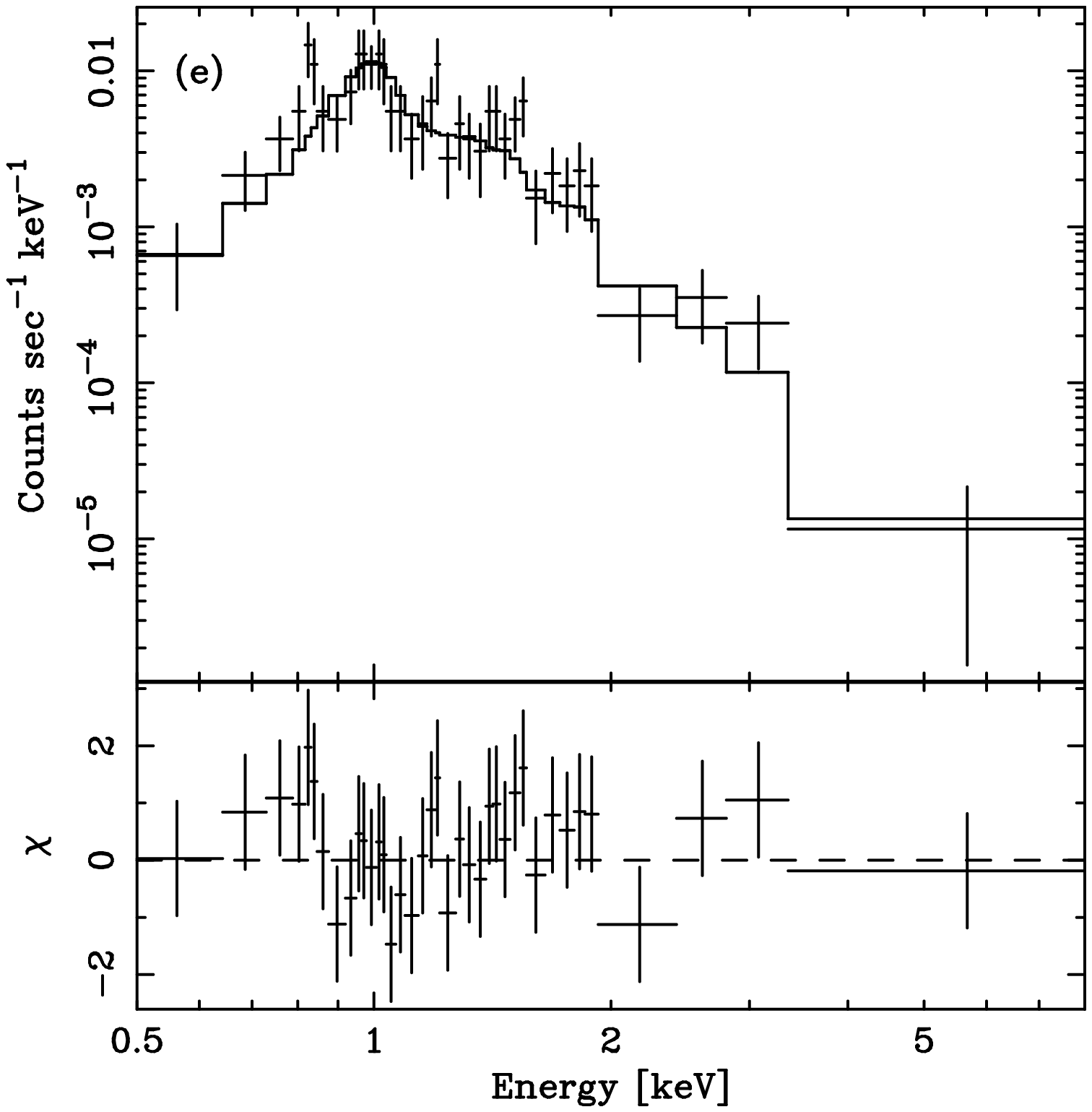} \hspace{0.1in}
  \includegraphics[angle=-90.,scale=0.4]{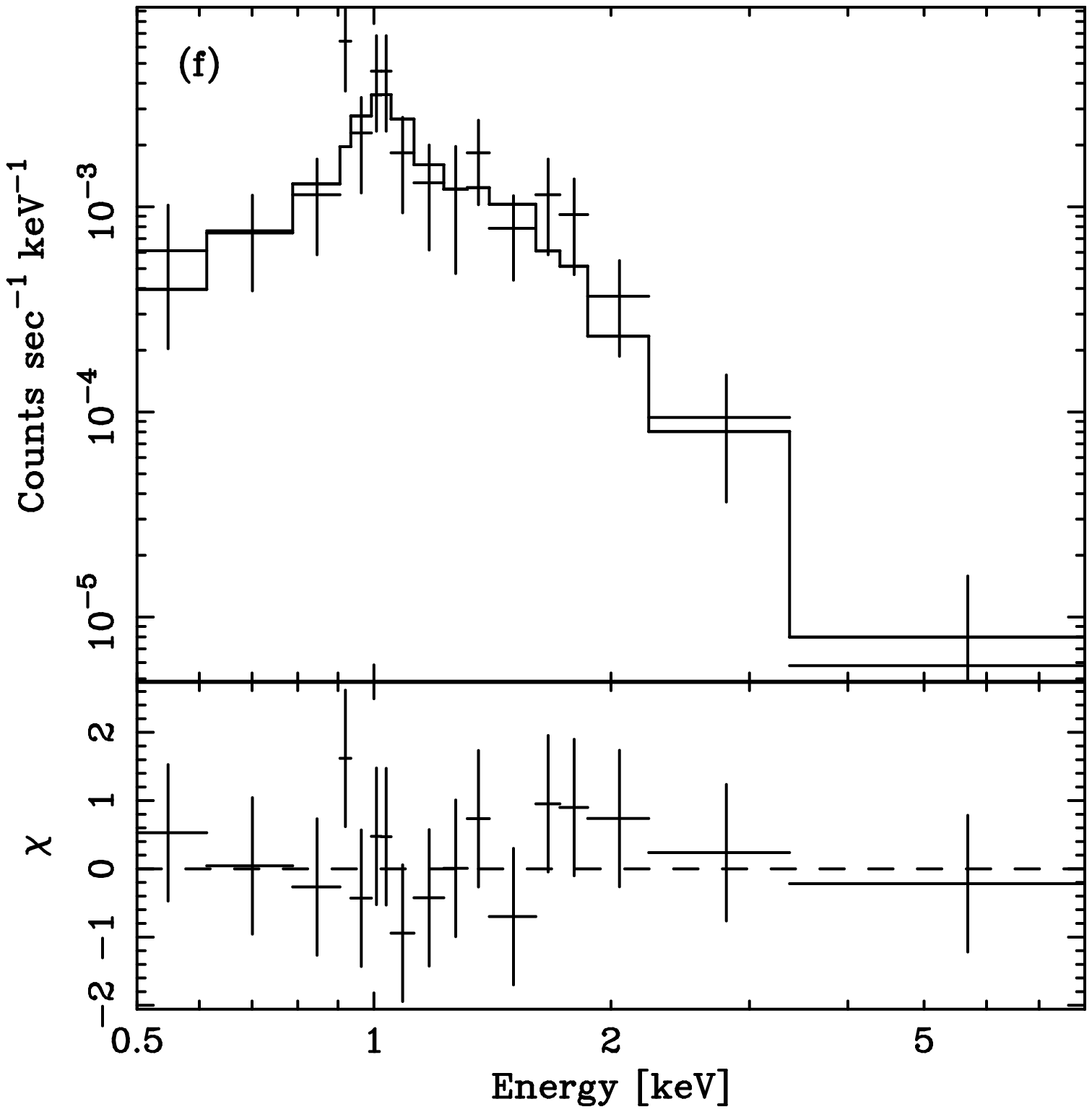}
\caption{Examples of X-ray spectra of young stellar objects in NGC
1333. In the top panels of each plot the points with error bars
represent the grouped observed spectra; the continuous solid lines
represent the best fit 1-temperature models. The residuals in
units of contribution to $\chi^2$ are plotted on the bottom
panels.  See (\S \ref{spectral_analysis_subsec}) for more details.
\label{spectra_fig}}
\end{minipage}
\end{figure}

\begin{figure}
\centering
\includegraphics[width=5.5in]{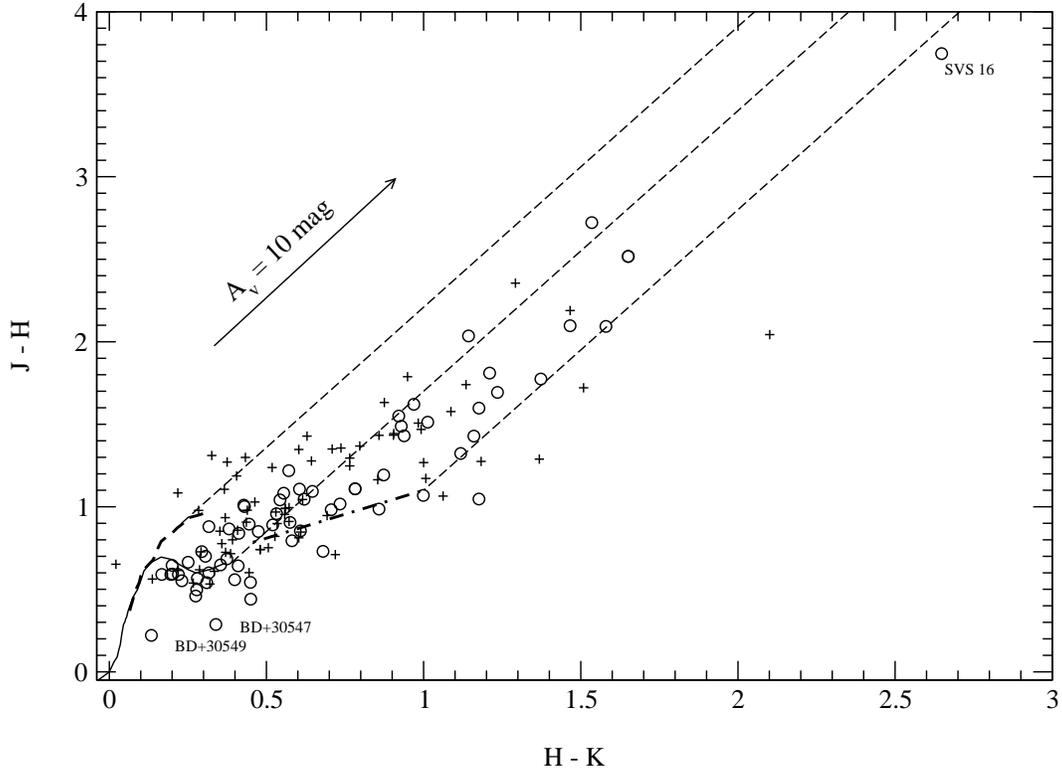}
\caption{Near-IR color-color diagram of the objects in the NGC
1333 field based on the JHK photometry data from \citet{Lada96}.
X-ray detected objects are plotted as circles, non-detected
objects as crosses. The solid line is the locus of points
corresponding to the unreddened main sequence; thick dashed line
-- the locus of positions of giant stars \citep{Bessel88}; thick
dot-dashed line -- classical T Tauri star locus \citep{Meyer97}.
The two leftmost parallel dashed lines define the reddening band
for main sequence stars; they are parallel to the reddening
vector. The arrow shows the reddening vector for $A_V = 10$ mag.
\label{color_color_fig}}
\end{figure}

\begin{figure}
\centering
\includegraphics[width=5.5in]{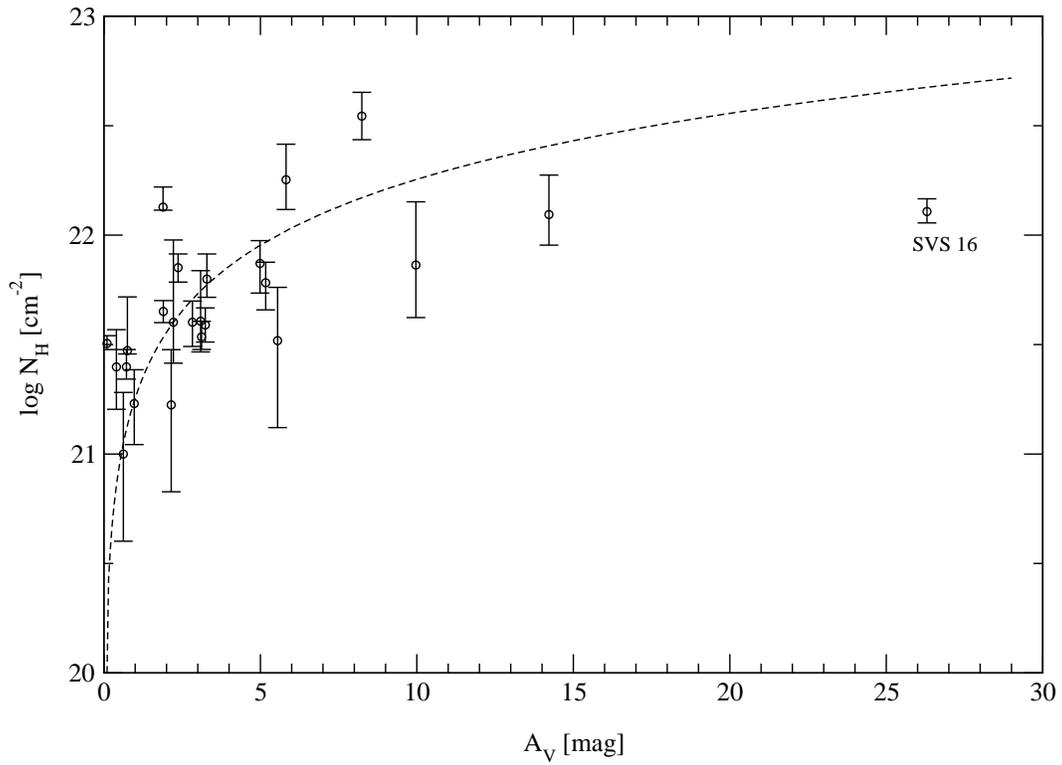}
\caption{Plot of intervening column density derived from X-ray
spectral $vs.$ visual absorption, estimated for 24 X-ray sources
with $>30$ counts. The line gives the expected relationship for a
standard gas-to-dust ratio. \label{nh_vs_av_fig}}
\end{figure}

\begin{figure}
\centering
\includegraphics[width=5.5in]{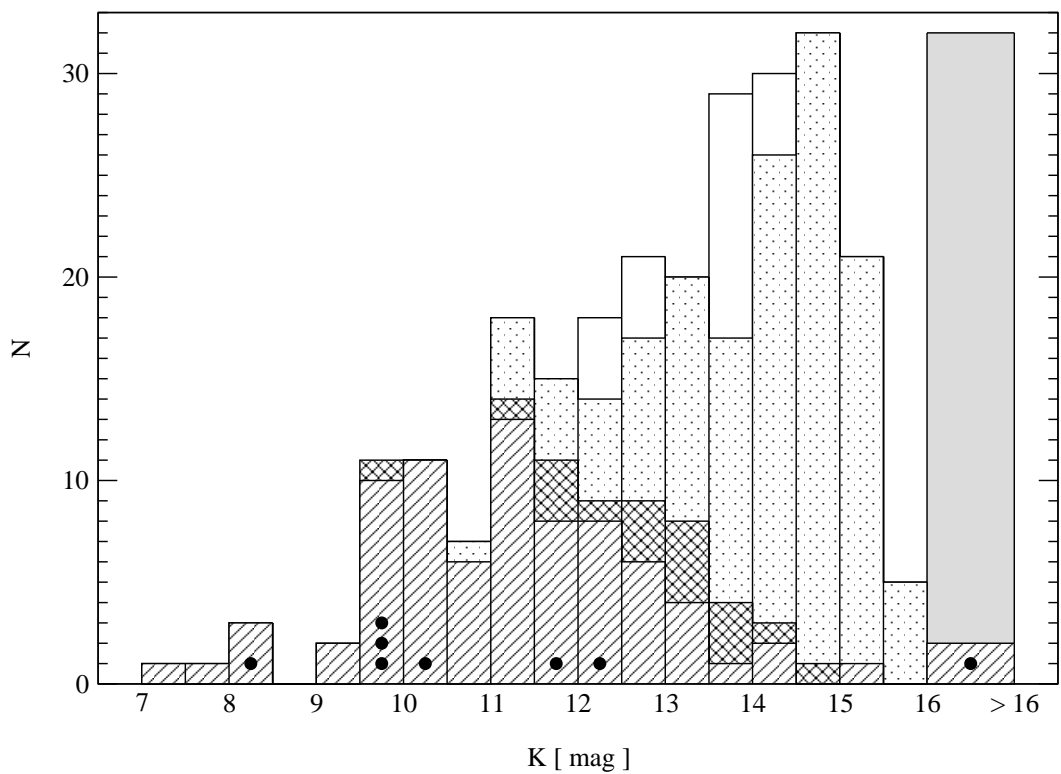}
\caption{The K luminosity function for all sources observed toward
the NGC 1333 region within ACIS-I field of view \citep{Lada96}.
The single hatched region indicates the 79 X-ray sources from
Table \ref{src_tbl}; the double-hatched region shows 18 additional
faint candidate X-ray sources from Table \ref{faint_src_tbl}; the
dotted region gives the expected Galactic background
contamination. The white region shows the remaining cluster
members undetected in X-rays. The bar at the right indicates the
X-ray sources that do not have $K \leq 16$ counterparts.  The grey
portion is the expected extragalactic contamination.  Seven YSOs
driving Herbig-Haro outflows and one deeply embedded protostar are
marked as solid dots. \label{N_vs_K_fig}}
\end{figure}

\begin{figure}
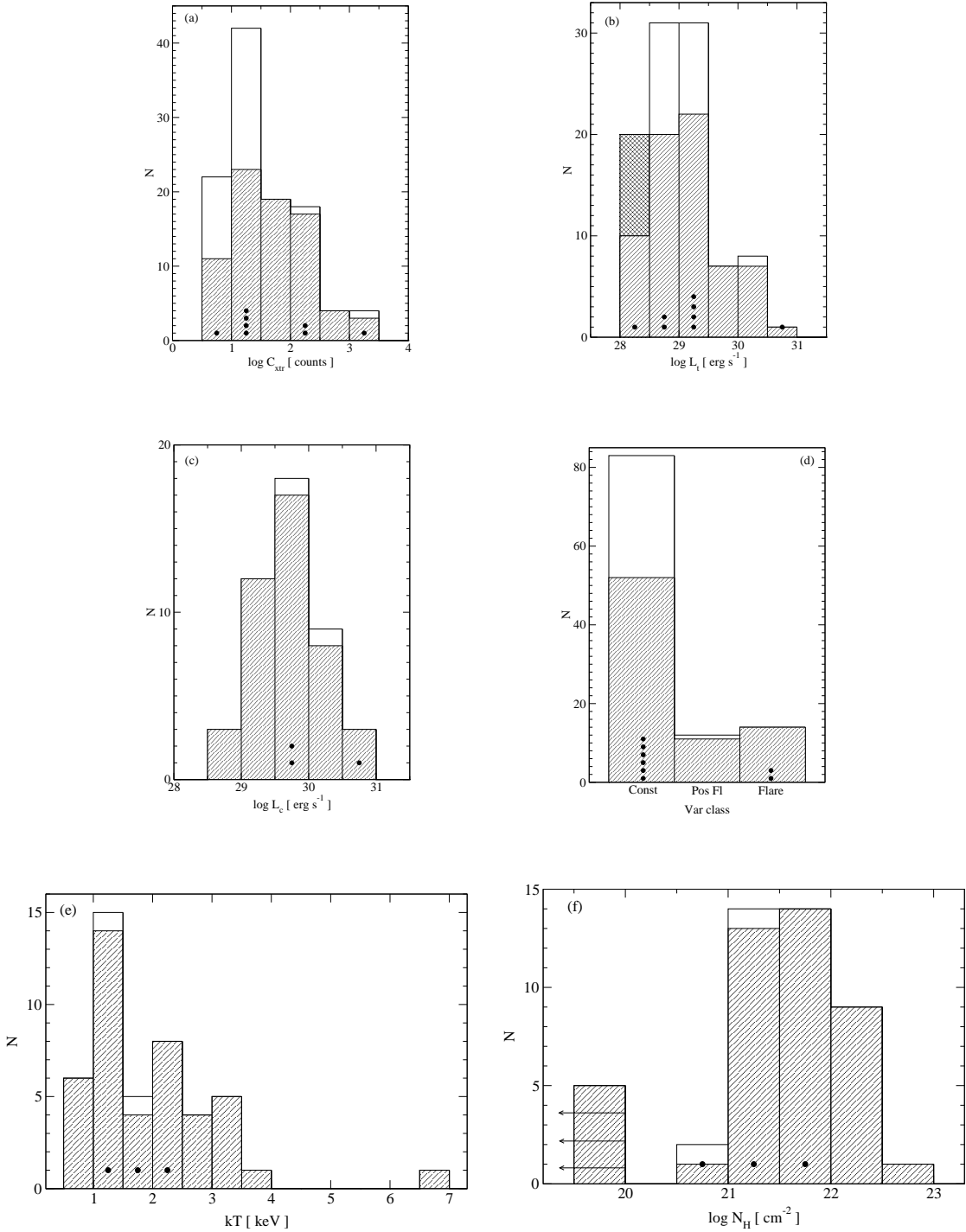

\centering
  \begin{minipage}[t]{1.0\textwidth}
  \centering
  \includegraphics[scale=0.24]{f10a.eps} \hspace{0.8in}
  \includegraphics[scale=0.24]{f10b.eps}
\end{minipage} \\ [0.4in]

\begin{minipage}[t]{1.0\textwidth}
  \centering
  \includegraphics[scale=0.24]{f10c.eps} \hspace{0.8in}
  \includegraphics[scale=0.24]{f10d.eps}
\end{minipage} \\ [0.4in]

  \begin{minipage}[t]{1.0\textwidth}
  \centering
\includegraphics[scale=0.3]{f10e.eps} \hspace{0.1in}
\includegraphics[scale=0.3]{f10f.eps}
\caption{Distributions of X-ray properties of the ACIS sources:
(a) extracted counts; (b) total band luminosity; (c) total band
luminosity corrected for absorption; (d) variability classes; (e)
plasma energy; and (f) line-of-sight column densities. Figures
(c), (e) and (f) include only sources with $> 30$ counts. The bin
with arrows indicates sources with very low column densities. The
hatching area indicates the 77 identified members of NGC 1333
population, while the white area indicates the properties of 30
unidentified sources and 2 foreground stars. The double-hatched
area indicates 10 members of NGC 1333 footnoted as `d' in Table
\ref{properties_tbl}. The 8 solid dots show seven YSOs driving
Herbig-Haro outflows and a deeply embedded protostar.
\label{x_global_prop_fig}}
\end{minipage}
\end{figure}

\begin{figure}
\centering
 \includegraphics[width=5.5in]{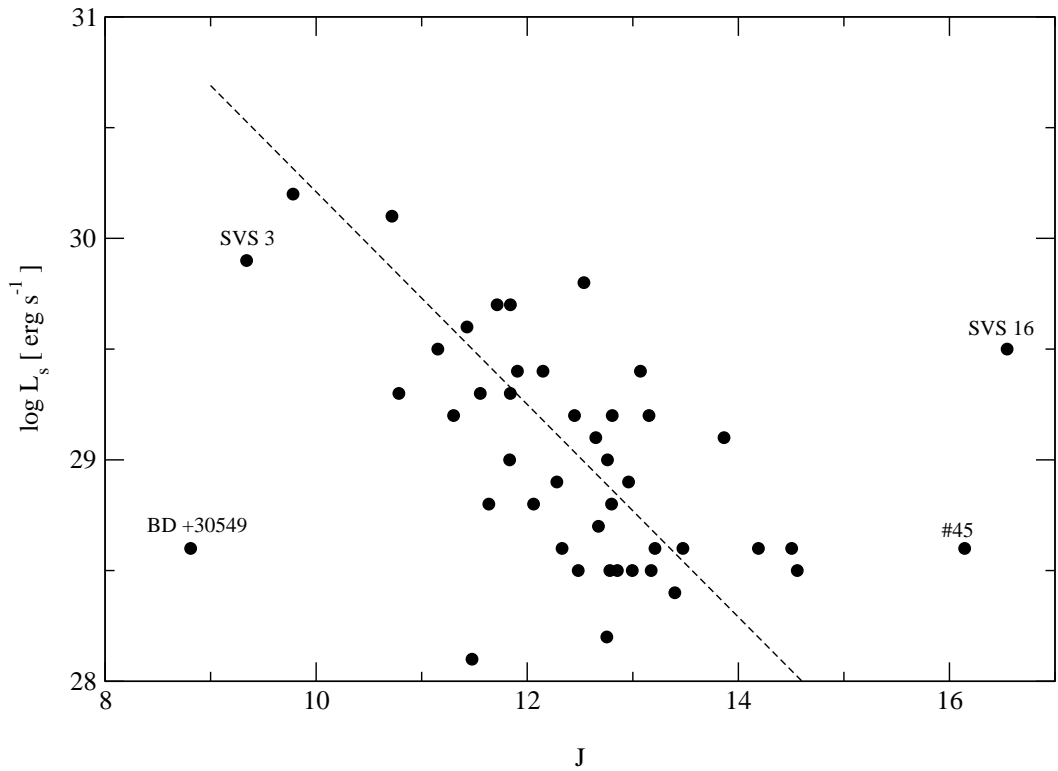}
 \caption{Soft band ($0.5-2$ keV) X-ray luminosity {\it vs.} the observed $J$-magnitude, both uncorrected for extinction.
 The regression line (dashed line) is obtained excluding BD +30$^\circ$549, SVS 3, and SVS 16. \label{L_vs_J_fig}}
\end{figure}

\begin{figure}
\centering
\includegraphics[width=5.5in]{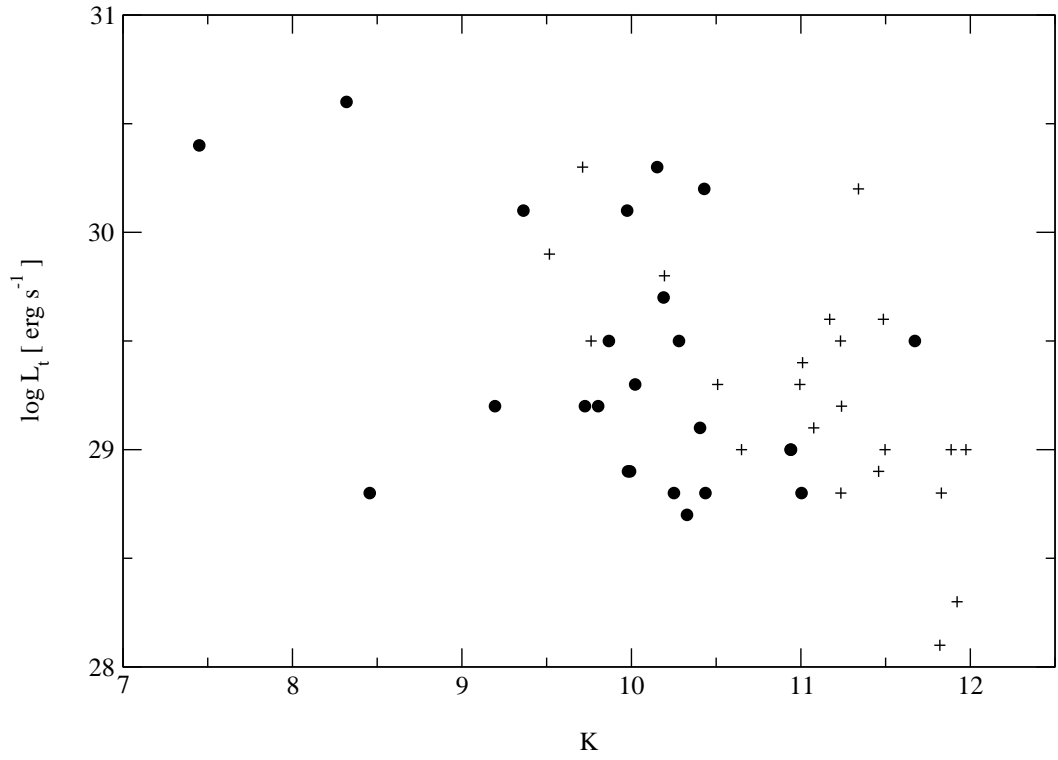}
\caption{Total band ($0.5-8$ keV) X-ray luminosities of the
complete subsamples of T Tauri stars {\it vs.} the observed
$K$-magnitude. CTTs are plotted as filled circles, WTTs as
crosses. \label{L_vs_K_fig}}
\end{figure}

\begin{figure}
\centering
  \begin{minipage}[t]{1.0\textwidth}
  \centering
  \includegraphics[scale=0.5]{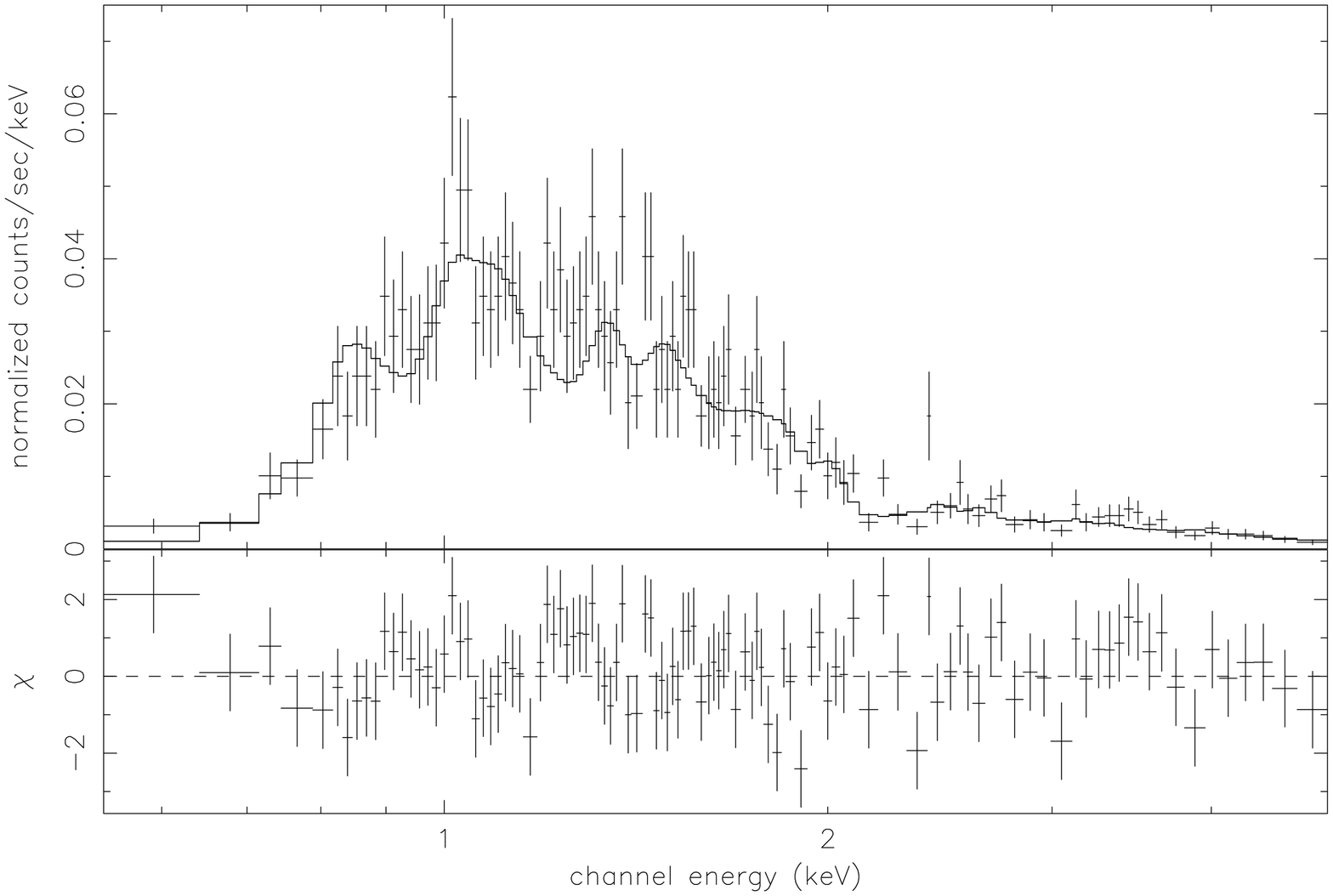} \hspace{0.1in}
\end{minipage} \\ [0.4in]
  \begin{minipage}[t]{1.0\textwidth}
  \centering
  \includegraphics[scale=0.5]{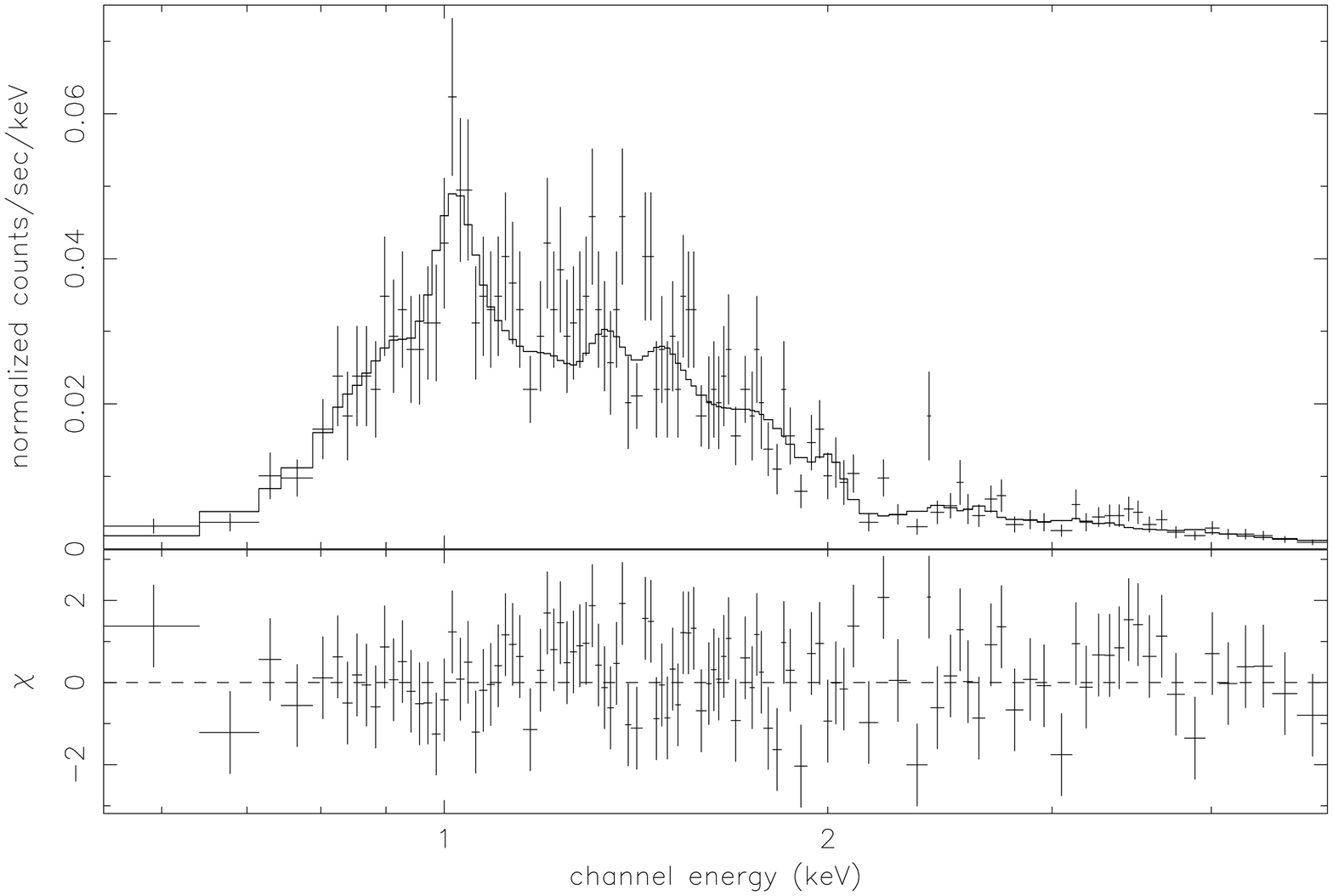} \hspace{0.1in}
\caption{ACIS spectrum of the X-ray bright classical T Tauri star
LkH$\alpha$ 270, driving an optically visible jet: (top)
two-temperature plasma model with standard elemental abundances;
and (bottom) two-temperature model with enhanced neon and reduced
iron. The residuals in units of contribution to $\chi^2$ are
plotted on the bottom panels.\label{79.spectrum_fig}}
  \end{minipage}
\end{figure}

\begin{figure}
\centering
\includegraphics[angle=-90.,scale=0.7]{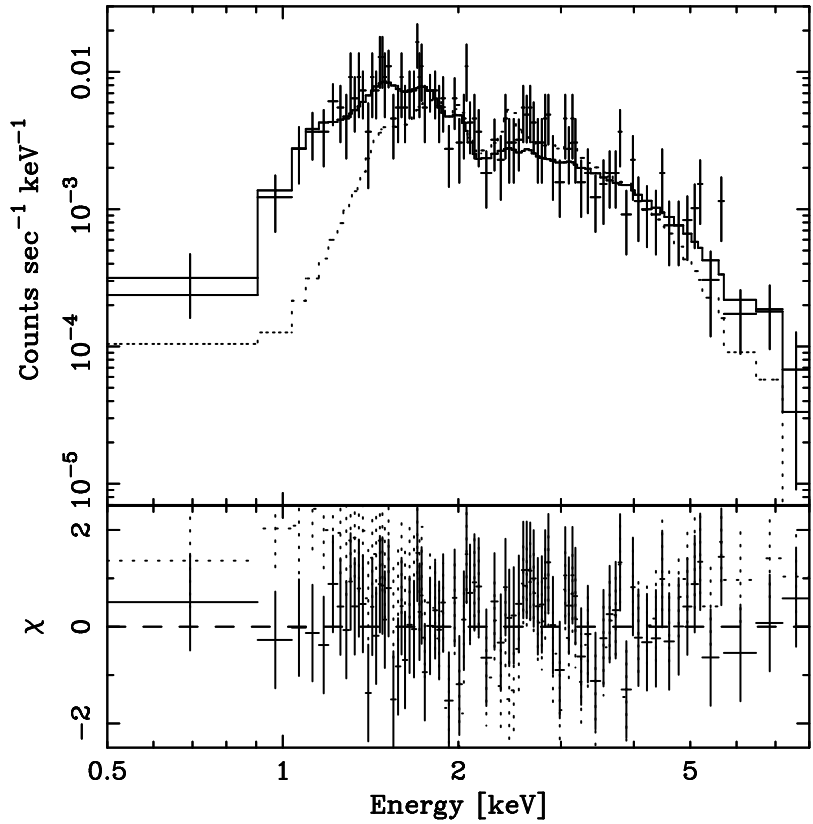}
\caption{{\it Chandra} X-ray spectrum of SVS 16. The continuous
solid line represents the best fit model with $\log N_H = 22.1$
cm$^{-2}$ ($A_V \simeq 7$); the dotted line represents the best
fit model for the fixed column density $\log N_H = 22.6$ cm$^{-2}$
($A_V \simeq 20$). The residuals in units of contribution to
$\chi^2$ are plotted on the bottom panel.\label{39_spectrum_fig}}
\end{figure}

\begin{description}

\item[\#39, 1$\arcmin\/$ northwest from HH 7-11 region]
Infrared photometry \citep[e.g.][]{Aspin94, Aspin97, Lada96} shows
that SVS 16 is highly obscured ($A_V \simeq 26$).
\citet{Preibisch98} report this is a binary system consisting of
two very young M-type PMS stars without significant amount of
circumstellar matter with a separation of 1$\arcsec$. {\it ROSAT}
data \citep{Preibisch97b, Preibisch98} indicated SVS 16 has an
extremely high constant X-ray luminosity $\log L_x \simeq 32.3$
erg s$^{-1}$ in the $0.1-2.4$ keV, based on the extinction of $A_V
= 26.2$ mag from near-IR data. In contrast to that result, {\it
Chandra} data indicate an evident flare showing a much lower
column density $\log N_H = 22.1$ cm$^{-2}$ or $A_V \simeq 7$
(Figure \ref{39_spectrum_fig}). The presence of soft X-rays
indicates that the source is not very deeply embedded into the
cloud. The lower column density reduces the inferred corrected
X-ray luminosity to $\log L_c = 30.6$ erg s$^{-1}$. Both IR and
X-ray data are quite reliable and suggest that the plasma emitting
the X-rays does not originate in the same region as the IR
emission from this source.

\item[\#58 and \#59, Figure \ref{figure_acis_image}d]
Source \#58 is a known classical T-Tauri star exhibiting strong
variation and large circular polarization in the radio continuum
\citep{Rodriguez99}, which indicates the presence of
gyrosynchrotron emission from particles accelerated in situ by
magnetic reconnection flares. This demonstrates the similarity in
the production mechanism of the X-ray emission to that of
magnetically active WTTs \citep{Feigelson85}. The only known
optical counterpart (from our KPNO image) has $m_I \sim 22.0$,
which means that the source is deeply embedded into the molecular
cloud. {\it Chandra} source \#59 is an X-ray discovered companion
to the \#58. It has a projected separation of $2\arcsec (\simeq
640$ AU) from \#58 (Figure \ref{figure_acis_image}d). Both
components have nearly identical spectra with $\log N_H = 22.2$
cm$^{-2}$, suggesting it may be a physical binary. With 71 counts,
\#59 exhibits an X-ray flare.

\item[\#65 and \#64, Figure \ref{figure_acis_image}e]
At optical and infrared wavelengths it is hard to detect any
objects in the vicinity of SVS 3 (\#64) because of its extreme
brightness. The difference between X-ray brightnesses is not so
dramatic, even with the moderately high luminosity $\log L_t =
30.4$ erg s$^{-1}$ for SVS 3, so we are able to discover a flaring
young T-Tauri star (\#65) with a projected separation of
3$\arcsec\/$ from SVS 3 (Figure \ref{figure_acis_image}e). An
X-ray spectrum extracted from 153 counts shows about the same
absorption column density $\log N_H \simeq 22.0^{+0.1}_{-0.1}$
cm$^{-2}$ for source \#65 as for SVS 3 and the plasma energy of
$2.8^{+1.0}_{-0.7}$ keV.

\end{description}


\section{Conclusions} \label{conclusion_sec}
We present a 37.8 ks {\it Chandra} ACIS observation of the NGC
1333 star formation region within the Perseus molecular cloud
complex. The main conclusions of our study are as follows:
\begin{enumerate}

\item We detect 127 X-ray sources with a limiting luminosity of $\sim 10^{28.0}$ erg~s$^{-1}$.
Among them, 30 are probably extragalactic background objects, and
2 are foreground stars. The remaining 95 (77 bright and 18 faint
tentative ACIS sources) are identified with cluster members. Two T
Tauri stars are discovered in the ACIS images as previously
unknown components of visual binaries.

\item The X-ray luminosity function of the sample of 77 bright cluster members is
approximately lognormal with mean and standard deviation $<\log
L_c> = 29.7~\pm~0.5$ erg s$^{- 1}$ after correction for
interstellar absorption. Most of the sources have plasma energies
between $0.6-3$ keV, a typical range for the X-ray active T Tauri
stars. The trend between increased variability and increased
hardness ratio can not be discerned.

\item A complete K-band luminosity function distribution of the observed region including Table \ref{faint_src_tbl} indicates that
 we detect all cluster members
 with $K < 12$ and over half of members with $K > 12$. Their average K-band magnitude of
$K \simeq 11$ corresponds to masses between $0.2-0.4$ M$_\odot$
for a cluster age of $\sim 1$ Myr.

\item We find a good correlation $L_x \propto J$ which confirms a well known
relation $L_x \propto L_{bol}$ found in many star forming regions.

\item The observed X-ray emission from two late-B stars is consistent with
emission originating from unresolved late-type companions.

\item We detect seven X-ray emitting YSOs which drive optically visible jets
 as well as one deeply embedded object that has not been detected in near-IR
 observations. The last one is presumably Class I protostar.

\item We also find no systematic differences in X-ray luminosity distributions between two
complete subsamples of CTTs and WTTs. That suggests that there is
no difference in the astrophysical mechanism of WTT and CTT X-ray
emission production. The presence or absence of an outflows does
not appear to produce any difference in X-ray properties of YSOs.

\item We find that the X-ray counterpart of SVS 16 has the column density much lower
than that expected from near-IR photometry and thus its X-ray
luminosity is not anomalously high, as has been previously
suggested.

\end{enumerate}

\acknowledgments We thank Patrick Broos and Yohko Tsuboi (Penn
State) for expert assistance with data analysis; Joao Alves (ESO)
for useful discussions and assistance with near-IR data; Thomas
Preibisch (Max-Planck-Institut f{\" u}r Radioastronomie) for
communication of unpublished {\it Chandra} findings on IC 348; and
the anonymous referee for helpful comments. This work was
supported by the SAO grant GO0-1092X.

\newpage

\bibliography{aj-jour}



\clearpage
\newpage

\clearpage
\newpage

\clearpage
\newpage

\clearpage
\newpage

\clearpage
\newpage

\clearpage
\newpage


\clearpage
\newpage

\clearpage
\newpage

\clearpage
\newpage

\clearpage
\newpage


\clearpage
\newpage


\clearpage
\newpage


\clearpage
\newpage



\begin{deluxetable}{rrrrrcccccccccrrrcc}
\tabletypesize{\scriptsize}
\tablewidth{0pt} \tablecolumns{19} \tablecaption{ACIS NGC 1333
sources and stellar identifications \label{src_tbl}} \rotate
\tablehead{ \multicolumn{6}{c}{X-ray source} &&
\multicolumn{4}{c}{Optical counterpart} &&
\multicolumn{5}{c}{Near-IR counterpart} &&
\multicolumn{1}{c}{Notes}
\\ \cline{1-6} \cline{8-11} \cline{13-17}
\colhead{Src}& \colhead{CXONGC1333 J} & \colhead{R.A.} &
\colhead{Dec.} &
\colhead{$\theta$} & \colhead{Pr} && \colhead{HJ} & \colhead{BR} &
\colhead{USNO} & \colhead{KPNO-I} && \colhead{ASR} & \colhead{LAL}
& \colhead{K} & \colhead{J-H} & \colhead{H-K} &
\\
\colhead{\#}& \colhead{} & \multicolumn{2}{c}{(J2000)} &
\colhead{\arcmin} & \colhead{\#} && \colhead{\#} & \colhead{} &
\colhead{\arcsec} & \colhead{(mag)} && \colhead{\#} & \colhead{\#}
& \colhead{(mag)} & \colhead{(mag)} & \colhead{(mag)} &
}
\startdata

1 & 032829.4+312508 & 03 28 29.47 & +31 25 08.1 & 9.7 & \nodata & & \nodata & \nodata & \nodata & \nodata & & \nodata & \nodata & \nodata & \nodata & \nodata & & \nodata \\
2 & 032831.9+312121 & 03 28 31.96 & +31 21 21.3 & 7.6 & \nodata & & \nodata & \nodata & \nodata & \nodata & & \nodata & \nodata & \nodata & \nodata & \nodata & & \nodata \\
3 & 032835.0+312112 & 03 28 35.04 & +31 21 12.5 & 6.9 & \nodata & & \nodata & \nodata & \nodata & \nodata & & \nodata & \nodata & \nodata & \nodata & \nodata & & \nodata \\
4 & 032836.4+311929 & 03 28 36.46 & +31 19 29.7 & 6.4 & \nodata & & 29 & yes & 0.5 & 14.5 & & \nodata & 55 & 11.83 & 0.57 & 0.28 & & \nodata \\
5 & 032836.8+312312 & 03 28 36.88 & +31 23 12.6 & 7.3 & 4 & & 110 & yes & 0.7 & 14.2 & & \nodata & \nodata & \nodata & \nodata & \nodata & & $\ast$ \\
6 & 032836.8+311735 & 03 28 36.89 & +31 17 35.9 & 6.7 & 3 & & 28 & yes & 0.3 & 15.1 & & \nodata & 56 & 10.20 & 1.11 & 0.60 & & $\ast$ \\
7 & 032837.1+311954 & 03 28 37.19 & +31 19 54.9 & 6.3 & \nodata & & \nodata & \nodata & \nodata & \nodata & & \nodata & \nodata & \nodata & \nodata & \nodata & & \nodata \\
8 & 032837.8+312525 & 03 28 37.83 & +31 25 25.9 & 8.5 & \nodata & & \nodata & \nodata & \nodata & \nodata & & \nodata & \nodata & \nodata & \nodata & \nodata & & \nodata \\
9 & 032843.2+311733 & 03 28 43.24 & +31 17 33.1 & 5.4 & \nodata & & 27 & yes & \nodata & 15.9 & & 126 & Bl & 10.19 & 1.81 & 1.21 & & $\ast$ \\
10 & 032843.5+311737 & 03 28 43.52 & +31 17 37.1 & 5.3 & \nodata & & 26 & yes & 1.7 & 15.6 & & 127 & 79 & 9.87 & 0.98 & 0.71 & & $\ast$ \\
11 & 032844.0+312052 & 03 28 44.01 & +31 20 52.8 & 5.0 & \nodata & & \nodata & yes & \nodata & 17.6 & & \nodata & 83 & 12.74 & 0.79 & 0.58 & & $\ast$ \\
12 & 032845.4+312050 & 03 28 45.40 & +31 20 50.9 & 4.7 & \nodata & & \nodata & \nodata & \nodata & \nodata & & \nodata & \nodata & \nodata & \nodata & \nodata & & \nodata \\
13 & 032845.4+312227 & 03 28 45.47 & +31 22 28.0 & 5.3 & \nodata & & \nodata & \nodata & \nodata & \nodata & & \nodata & \nodata & \nodata & \nodata & \nodata & & \nodata \\
14 & 032846.1+311639 & 03 28 46.15 & +31 16 39.0 & 5.3 & 5 & & 30 & yes & 0.4 & 14.2 & & 128 & 89 & 9.76 & 0.73 & 0.29 & & $\ast$ \\
15 & 032847.6+312405 & 03 28 47.63 & +31 24 05.5 & 6.1 & \nodata & & \nodata & yes & \nodata & 17.9 & & \nodata & 93 & 11.24 & 2.04 & 1.14 & & \nodata \\
16 & 032847.8+311655 & 03 28 47.80 & +31 16 55.7 & 4.8 & \nodata & & 32 & yes & \nodata & 16.3 & & 111 & 97 & 10.94 & 0.99 & 0.86 & & $\ast$ \\
17 & 032850.3+312755 & 03 28 50.37 & +31 27 55.2 & 9.1 & \nodata & & \nodata & \nodata & \nodata & \nodata & & \nodata & \nodata & \nodata & \nodata & \nodata & & \nodata \\
18 & 032850.8+312349 & 03 28 50.88 & +31 23 49.3 & 5.4 & \nodata & & \nodata & \nodata & \nodata & 19.8 & & \nodata & \nodata & \nodata & \nodata & \nodata & & $\ast$ \\
19 & 032850.9+311818 & 03 28 50.99 & +31 18 18.9 & 3.5 & 6 & & \nodata & yes & 0.2 & 14.0 & & 122 & 106 & 9.20 & 1.32 & 1.12 & & $\ast$ \\
20 & 032850.9+311632 & 03 28 51.00 & +31 16 32.2 & 4.5 & \nodata & & 33 & yes & 1.0 & 15.5 & & 44 & 107 & 12.22 & 0.68 & 0.37 & & $\ast$ \\
21 & 032851.1+311955 & 03 28 51.17 & +31 19 55.1 & 3.3 & 7 & & 9 & yes & 0.1 & 14.2 & & 125 & 110 & 9.97 & 1.09 & 0.65 & & $\ast$ \\
22 & 032852.0+311548 & 03 28 52.09 & +31 15 48.1 & 4.9 & \nodata & & 42 & yes & 0.5 & 16.1 & & 45 & 125 & 12.08 & 0.54 & 0.45 & & $\ast$ \\
23 & 032852.1+312245 & 03 28 52.12 & +31 22 45.7 & 4.4 & \nodata & & 109 & yes & 0.3 & 14.1 & & \nodata & 120 & 11.67 & 0.91 & 0.57 & & $\ast$ \\
24 & 032853.5+311537 & 03 28 53.58 & +31 15 38.0 & 4.9 & \nodata & & \nodata & \nodata & \nodata & \nodata & & \nodata & \nodata & \nodata & \nodata & \nodata & & \nodata \\
25 & 032853.9+311809 & 03 28 53.92 & +31 18 10.0 & 3.0 & \nodata & & \nodata & \nodata & \nodata & 21.7 & & 40 & 129 & 10.94 & 2.10 & 1.47 & & $\ast$ \\
26 & 032854.0+311654 & 03 28 54.04 & +31 16 54.8 & 3.8 & \nodata & & 37 & yes & \nodata & 15.8 & & 42 & 131 & 11.71 & 0.85 & 0.47 & & \nodata \\
27 & 032854.6+311651 & 03 28 54.61 & +31 16 51.6 & 3.8 & \nodata & & 38 & yes & \nodata & 16.2 & & 43 & 136 & 10.28 & 1.43 & 0.94 & & $\ast$ \\
28 & 032855.0+311629 & 03 28 55.03 & +31 16 29.3 & 4.0 & \nodata & & 35 & yes & \nodata & 17.7 & & 107 & 141 & 10.44 & 1.49 & 0.93 & & $\ast$ \\
29 & 032856.5+312240 & 03 28 56.54 & +31 22 40.7 & 3.7 & \nodata & & \nodata & \nodata & \nodata & \nodata & & \nodata & \nodata & \nodata & \nodata & \nodata & & \nodata \\
30 & 032856.6+311836 & 03 28 56.62 & +31 18 36.2 & 2.3 & 8 & & 21 & yes & 0.1 & 15.5 & & 120 & 150 & 9.80 & 1.51 & 1.01 & & $\ast$ \\
31 & 032856.8+311811 & 03 28 56.81 & +31 18 11.2 & 2.5 & \nodata & & \nodata & \nodata & \nodata & \nodata & & \nodata & \nodata & \nodata & \nodata & \nodata & & \nodata \\
32 & 032856.9+311622 & 03 28 56.95 & +31 16 22.8 & 3.9 & \nodata & & 39 & yes & \nodata & 17.9 & & 118 & 154 & 10.25 & 1.77 & 1.37 & & $\ast$ \\
33 & 032857.1+311419 & 03 28 57.19 & +31 14 19.4 & 5.7 & 10 & & \nodata & yes & 0.5 & 18.1 & & 130 & 157 & 7.84 & 0.29 & 0.34 & & $\ast$ \\
34 & 032857.4+311950 & 03 28 57.41 & +31 19 50.6 & 1.9 & 9 & & 19 & yes & \nodata & 14.5 & & 115 & 158 & 10.65 & 0.89 & 0.52 & & $\ast$ \\
35 & 032857.6+311948 & 03 28 57.67 & +31 19 48.3 & 1.8 & \nodata & & 20 & yes & 1.3 & 15.7 & & 113 & Bl & 10.43 & 0.73 & 0.68 & & $\ast$ \\
36 & 032858.0+311804 & 03 28 58.08 & +31 18 04.2 & 2.4 & \nodata & & 7 & yes & 0.3 & 14.8 & & 36 & 162 & 11.46 & 0.90 & 0.45 & & $\ast$ \\
37 & 032858.2+312541 & 03 28 58.25 & +31 25 41.4 & 6.4 & \nodata & & \nodata & \nodata & \nodata & \nodata & & \nodata & \nodata & \nodata & \nodata & \nodata & & \nodata \\
38 & 032858.4+312218 & 03 28 58.45 & +31 22 18.2 & 3.2 & \nodata & & \nodata & \nodata & \nodata & \nodata & & \nodata & 166 & 11.35 & \nodata & \nodata & & $\ast$ \\
39 & 032859.2+311548 & 03 28 59.28 & +31 15 48.9 & 4.2 & 11 & & \nodata & \nodata & \nodata & \nodata & & 106 & 171 & 10.15 & 3.75 & 2.65 & & $\ast$ \\
40 & 032859.3+312640 & 03 28 59.31 & +31 26 40.1 & 7.3 & \nodata & & \nodata & \nodata & \nodata & \nodata & & \nodata & \nodata & \nodata & \nodata & \nodata & & \nodata \\
41 & 032859.5+312146 & 03 28 59.51 & +31 21 47.0 & 2.6 & \nodata & & 8 & yes & 0.4 & 14.5 & & \nodata & 173 & 10.33 & 1.07 & 1.00 & & $\ast$ \\
42 & 032900.1+312109 & 03 29 00.13 & +31 21 10.0 & 2.0 & \nodata & & \nodata & \nodata & \nodata & 22.5 & & \nodata & 175 & 13.29 & 1.69 & 1.24 & & \nodata \\
43 & 032900.2+311338 & 03 29 00.28 & +31 13 38.8 & 6.2 & 12 & & 4 & yes & 0.3 & 15.8 & & 51 & 178 & 11.49 & 1.04 & 0.54 & & $\ast$ \\
44 & 032900.3+312045 & 03 29 00.35 & +31 20 46.0 & 1.7 & \nodata & & 18 & yes & \nodata & 16.6 & & \nodata & 177 & 11.92 & 0.96 & 0.53 & & \nodata \\
45 & 032900.6+312201 & 03 29 00.67 & +31 22 01.1 & 2.7 & \nodata & & \nodata & \nodata & \nodata & 22.5 & & \nodata & 180 & 11.89 & 2.72 & 1.53 & & \nodata \\
46 & 032901.3+311413 & 03 29 01.36 & +31 14 13.2 & 5.6 & \nodata & & \nodata & \nodata & \nodata & \nodata & & \nodata & \nodata & \nodata & \nodata & \nodata & & \nodata \\
47 & 032902.1+311611 & 03 29 02.16 & +31 16 11.7 & 3.6 & \nodata & & 40 & yes & 0.4 & 15.9 & & 3 & 184 & 13.70 & 0.56 & 0.20 & & \nodata \\
48 & 032902.8+311601 & 03 29 02.87 & +31 16 01.3 & 3.7 & 13 & & 6 & yes & 0.5 & 14.2 & & 2 & 188 & 11.97 & 0.59 & 0.20 & & \nodata \\
49 & 032903.0+311516 & 03 29 03.02 & +31 15 17.0 & 4.5 & \nodata & & \nodata & \nodata & \nodata & \nodata & & \nodata & \nodata & \nodata & \nodata & \nodata & & $\ast$ \\
50 & 032903.1+312238 & 03 29 03.13 & +31 22 38.3 & 3.1 & \nodata & & 12 & yes & 0.7 & 17.0 & & \nodata & 189 & 12.25 & 1.05 & 1.18 & & $\ast$ \\
51 & 032903.4+311618 & 03 29 03.48 & +31 16 18.2 & 3.4 & \nodata & & \nodata & \nodata & \nodata & \nodata & & 132 & \nodata & 15.14 & \nodata & \nodata & & \nodata \\
52 & 032903.8+312149 & 03 29 03.84 & +31 21 49.1 & 2.3 & 14 & & 11 & yes & 0.4 & 14.3 & & \nodata & 195 & 9.36 & 1.19 & 0.87 & & $\ast$ \\
53 & 032904.0+311708 & 03 29 04.05 & +31 17 08.1 & 2.6 & \nodata & & 23 & yes & 0.3 & 15.5 & & 8 & 199 & 12.49 & 0.50 & 0.28 & & \nodata \\
54 & 032904.1+312515 & 03 29 04.19 & +31 25 15.8 & 5.7 & \nodata & & \nodata & yes & \nodata & 17.5 & & \nodata & 198 & 11.00 & 1.55 & 0.92 & & \nodata \\
55 & 032904.3+312624 & 03 29 04.35 & +31 26 24.3 & 6.9 & \nodata & & \nodata & \nodata & \nodata & \nodata & & \nodata & \nodata & \nodata & \nodata & \nodata & & \nodata \\
56 & 032904.7+311133 & 03 29 04.74 & +31 11 33.5 & 8.2 & \nodata & & \nodata & \nodata & \nodata & \nodata & & 99 & 205 & 14.21 & \nodata & 2.64 & & $\ast$ \\
57 & 032904.9+312038 & 03 29 04.90 & +31 20 38.4 & 1.1 & \nodata & & \nodata & \nodata & \nodata & \nodata & & \nodata & 204 & 12.67 & \nodata & \nodata & & $\ast$ \\
58 & 032905.7+311640 & 03 29 05.74 & +31 16 40.1 & 3.0 & \nodata & & \nodata & \nodata & \nodata & 22.0 & & 7 & 207 & 10.02 & 2.52 & 1.65 & & $\ast$ \\
59 & 032905.9+311640 & 03 29 05.90 & +31 16 40.6 & 3.0 & \nodata & & \nodata & \nodata & \nodata & Conf & & \nodata & Bl & \nodata & \nodata & \nodata & & $\ast$ \\
60 & 032906.6+311933 & 03 29 06.60 & +31 19 33.8 & 0.1 & \nodata & & \nodata & \nodata & \nodata & \nodata & & \nodata & \nodata & \nodata & \nodata & \nodata & & \nodata \\
61 & 032907.9+312252 & 03 29 07.96 & +31 22 52.5 & 3.3 & \nodata & & 13 & yes & \nodata & 16.9 & & \nodata & 215 & 10.40 & 1.62 & 0.97 & & \nodata \\
62 & 032908.7+311314 & 03 29 08.72 & +31 13 14.7 & 6.5 & \nodata & & \nodata & \nodata & \nodata & \nodata & & \nodata & \nodata & \nodata & \nodata & \nodata & & \nodata \\
63 & 032909.1+312306 & 03 29 09.11 & +31 23 06.3 & 3.6 & \nodata & & \nodata & yes & \nodata & 19.7 & & \nodata & 222 & 11.68 & 1.43 & 1.16 & & \nodata \\
64 & 032910.3+312159 & 03 29 10.38 & +31 21 59.7 & 2.6 & 15 & & \nodata & yes & \nodata & 15.7 & & \nodata & 230 & 7.45 & 1.11 & 0.78 & & $\ast$ \\
65 & 032910.5+312202 & 03 29 10.59 & +31 22 02.0 & 2.6 & 15 & & \nodata & \nodata & \nodata & Conf & & \nodata & Bl & \nodata & \nodata & \nodata & & \nodata \\
66 & 032910.7+312230 & 03 29 10.79 & +31 22 30.6 & 3.1 & \nodata & & \nodata & yes & \nodata & 18.6 & & \nodata & 233 & 12.89 & 1.05 & 0.62 & & \nodata \\
67 & 032911.2+311718 & 03 29 11.29 & +31 17 18.1 & 2.6 & \nodata & & \nodata & yes & \nodata & 16.8 & & 24 & 243 & 12.97 & 0.44 & 0.45 & & \nodata \\
68 & 032911.8+312216 & 03 29 11.83 & +31 22 16.3 & 3.0 & \nodata & & \nodata & \nodata & \nodata & \nodata & & \nodata & \nodata & \nodata & \nodata & \nodata & & \nodata \\
69 & 032911.8+312127 & 03 29 11.87 & +31 21 27.4 & 2.2 & \nodata & & \nodata & \nodata & \nodata & \nodata & & \nodata & 248 & 12.51 & \nodata & \nodata & & \nodata \\
70 & 032912.8+312008 & 03 29 12.80 & +31 20 08.0 & 1.5 & \nodata & & \nodata & yes & \nodata & 17.5 & & 83 & 257 & 13.35 & 0.64 & 0.41 & & \nodata \\
71 & 032912.9+311815 & 03 29 12.93 & +31 18 15.1 & 2.0 & \nodata & & \nodata & \nodata & \nodata & \nodata & & 30 & 261 & 14.23 & \nodata & \nodata & & $\ast$ \\
72 & 032913.1+312253 & 03 29 13.13 & +31 22 53.4 & 3.7 & \nodata & & 15 & yes & \nodata & 16.1 & & \nodata & 262 & 9.98 & 1.60 & 1.18 & & \nodata \\
73 & 032913.4+312441 & 03 29 13.48 & +31 24 41.3 & 5.4 & \nodata & & \nodata & \nodata & \nodata & \nodata & & \nodata & \nodata & \nodata & \nodata & \nodata & & \nodata \\
74 & 032914.0+312214 & 03 29 14.09 & +31 22 14.4 & 3.2 & \nodata & & \nodata & \nodata & \nodata & \nodata & & \nodata & \nodata & \nodata & \nodata & \nodata & & \nodata \\
75 & 032914.4+312236 & 03 29 14.43 & +31 22 36.8 & 3.5 & \nodata & & \nodata & yes & \nodata & 17.5 & & \nodata & 269 & 13.11 & \nodata & \nodata & & \nodata \\
76 & 032915.6+311852 & 03 29 15.61 & +31 18 52.4 & 2.2 & \nodata & & \nodata & \nodata & \nodata & \nodata & & \nodata & \nodata & \nodata & \nodata & \nodata & & \nodata \\
77 & 032916.5+312348 & 03 29 16.57 & +31 23 49.0 & 4.8 & \nodata & & \nodata & yes & \nodata & 16.5 & & \nodata & 276 & 11.17 & 1.22 & 0.57 & & \nodata \\
78 & 032916.6+311618 & 03 29 16.67 & +31 16 18.9 & 4.1 & 16 & & 3 & yes & 0.3 & 14.3 & & 121 & 280 & 10.51 & 0.59 & 0.20 & & $\ast$ \\
79 & 032917.6+312245 & 03 29 17.67 & +31 22 45.7 & 4.1 & 17 & & 103 & yes & 0.6 & 14.3 & & \nodata & 283 & 8.32 & 0.85 & 0.61 & & $\ast$ \\
80 & 032918.5+311926 & 03 29 18.58 & +31 19 26.2 & 2.7 & \nodata & & \nodata & \nodata & \nodata & \nodata & & \nodata & \nodata & \nodata & \nodata & \nodata & & \nodata \\
81 & 032918.7+312326 & 03 29 18.72 & +31 23 26.1 & 4.7 & 18 & & 102 & yes & 0.3 & 14.4 & & \nodata & 293 & 11.23 & 0.66 & 0.25 & & $\ast$ \\
82 & 032919.0+312106 & 03 29 19.07 & +31 21 06.8 & 3.2 & \nodata & & \nodata & \nodata & \nodata & \nodata & & \nodata & \nodata & \nodata & \nodata & \nodata & & \nodata \\
83 & 032919.7+312458 & 03 29 19.76 & +31 24 58.4 & 6.2 & \nodata & & \nodata & yes & 1.4 & yes & & \nodata & 295 & 8.46 & 0.22 & 0.13 & & $\ast$ \\
84 & 032920.0+312408 & 03 29 20.06 & +31 24 08.0 & 5.5 & \nodata & & \nodata & \nodata & \nodata & \nodata & & \nodata & 296 & 12.47 & \nodata & \nodata & & \nodata \\
85 & 032920.1+311450 & 03 29 20.16 & +31 14 50.6 & 5.7 & \nodata & & \nodata & \nodata & \nodata & \nodata & & \nodata & \nodata & \nodata & \nodata & \nodata & & \nodata \\
86 & 032920.4+311834 & 03 29 20.40 & +31 18 35.0 & 3.3 & \nodata & & \nodata & yes & \nodata & 18.9 & & 112 & 300 & 9.99 & 2.09 & 1.58 & & $\ast$ \\
87 & 032921.1+311617 & 03 29 21.12 & +31 16 17.1 & 4.7 & \nodata & & \nodata & \nodata & \nodata & \nodata & & \nodata & \nodata & \nodata & \nodata & \nodata & & \nodata \\
88 & 032921.5+312111 & 03 29 21.57 & +31 21 11.1 & 3.7 & \nodata & & 17 & yes & 0.5 & 14.5 & & \nodata & 304 & 11.50 & 0.55 & 0.23 & & $\ast$ \\
89 & 032921.6+312515 & 03 29 21.65 & +31 25 15.6 & 6.6 & \nodata & & \nodata & yes & \nodata & 16.3 & & \nodata & 303 & 11.82 & 0.84 & 0.41 & & \nodata \\
90 & 032921.8+311537 & 03 29 21.88 & +31 15 37.1 & 5.3 & \nodata & & 46 & yes & 0.4 & 14.3 & & 123 & 307 & 9.73 & 1.02 & 0.73 & & $\ast$ \\
91 & 032922.0+312415 & 03 29 22.00 & +31 24 15.9 & 5.8 & \nodata & & 100 & yes & 0.3 & 14.2 & & \nodata & 306 & 11.08 & 0.59 & 0.17 & & \nodata \\
92 & 032923.2+312030 & 03 29 23.25 & +31 20 30.9 & 3.8 & \nodata & & 2 & yes & 1.5 & 14.4 & & 78 & 310 & 11.25 & 0.60 & 0.32 & & $\ast$ \\
93 & 032923.5+312331 & 03 29 23.50 & +31 23 31.5 & 5.5 & \nodata & & 101 & yes & \nodata & 15.2 & & \nodata & 311 & 11.34 & 0.88 & 0.32 & & \nodata \\
94 & 032923.7+312510 & 03 29 23.76 & +31 25 10.4 & 6.8 & \nodata & & \nodata & yes & \nodata & 16.9 & & \nodata & 312 & 12.43 & 1.00 & 0.43 & & \nodata \\
95 & 032924.1+312109 & 03 29 24.16 & +31 21 09.9 & 4.2 & \nodata & & \nodata & \nodata & \nodata & \nodata & & \nodata & \nodata & \nodata & \nodata & \nodata & & \nodata \\
96 & 032924.8+312052 & 03 29 24.87 & +31 20 52.2 & 4.3 & \nodata & & \nodata & \nodata & \nodata & \nodata & & \nodata & \nodata & \nodata & \nodata & \nodata & & \nodata \\
97 & 032924.8+312407 & 03 29 24.88 & +31 24 07.2 & 6.1 & \nodata & & \nodata & yes & \nodata & 16.6 & & \nodata & 315 & 13.41 & 0.54 & 0.31 & & \nodata \\
98 & 032925.8+312640 & 03 29 25.80 & +31 26 40.2 & 8.3 & \nodata & & \nodata & yes & 1.4 & 13.6 & & \nodata & 318 & 9.52 & 1.08 & 0.56 & & $\ast$ \\
99 & 032925.8+311347 & 03 29 25.84 & +31 13 47.0 & 7.3 & \nodata & & \nodata & \nodata & \nodata & \nodata & & \nodata & \nodata & \nodata & \nodata & \nodata & & \nodata \\
100 & 032926.7+312648 & 03 29 26.80 & +31 26 48.3 & 8.6 & 19 & & \nodata & yes & 0.5 & 13.9 & & \nodata & 321 & 9.71 & 0.70 & 0.31 & & $\ast$ \\
101 & 032928.1+311628 & 03 29 28.15 & +31 16 29.0 & 5.8 & \nodata & & 48 & yes & 0.2 & 15.7 & & \nodata & 328 & 12.07 & 0.46 & 0.28 & & \nodata \\
102 & 032929.2+311835 & 03 29 29.26 & +31 18 35.5 & 5.1 & \nodata & & 1 & yes & 0.2 & 14.9 & & \nodata & 331 & 11.01 & 1.01 & 0.43 & & $\ast$ \\
103 & 032929.8+312102 & 03 29 29.83 & +31 21 03.0 & 5.3 & \nodata & & 53 & yes & 1.0 & 15.1 & & \nodata & 333 & 11.24 & 0.87 & 0.38 & & $\ast$ \\
104 & 032930.6+312729 & 03 29 30.70 & +31 27 29.4 & 9.6 & \nodata & & \nodata & yes & \nodata & 16.4 & & \nodata & 335 & 12.60 & 0.65 & 0.35 & & \nodata \\
105 & 032931.6+312125 & 03 29 31.70 & +31 21 25.2 & 5.8 & \nodata & & \nodata & \nodata & \nodata & \nodata & & \nodata & \nodata & \nodata & \nodata & \nodata & & \nodata \\
106 & 032932.9+312712 & 03 29 32.92 & +31 27 12.6 & 9.6 & \nodata & & \nodata & yes & 1.1 & 15.2 & & \nodata & 344 & 12.22 & 0.56 & 0.40 & & \nodata \\
107 & 032934.3+311743 & 03 29 34.30 & +31 17 44.0 & 6.4 & 20 & & 51 & yes & 0.4 & 14.1 & & \nodata & 347 & 10.99 & 0.65 & 0.20 & & $\ast$ \\
108 & 032945.4+312351 & 03 29 45.43 & +31 23 51.1 & 9.6 & \nodata & & \nodata & \nodata & \nodata & 22.7 & & \nodata & \nodata & \nodata & \nodata & \nodata & & \nodata \\
109 & 032946.1+312037 & 03 29 46.11 & +31 20 37.8 & 8.7 & \nodata & & \nodata & yes & \nodata & \nodata & & \nodata & \nodata & \nodata & \nodata & \nodata & & \nodata \\

\enddata
\begin{scriptsize}
 \raggedright
~\\
Notes to Table 1: \\
\#\#6, 11, 18, 20, 21, 22, 23, 38, 41, 50, 56, 78~~ISOCAM
($6.7\mu$m) \\
\#\#25, 27, 28, 34, 35, 57, 59, 71, 88, 102, 103~~ISOCAM ($6.7\mu$m, $14.3\mu$m) \\
\#5~~ISOCAM ($6.7\mu$m); VSS 32 \citep{Vrba76} \\
\#9~~ISOCAM ($6.7\mu$m, $14.3\mu$m);
HBC 340 \citep{Herbig88}; IRAS 03256+3107 \\
\#10~~ISOCAM ($6.7\mu$m, $14.3\mu$m); HBC
341 \citep{Herbig88} \\
\#14~~ISOCAM
($6.7\mu$m, $14.3\mu$m); VSS 24 \citep{Vrba76}; LkH$\alpha$ 351
\citep{Herbig56}; SVS 21 \citep{Strom76} \\
\#16~~ISOCAM ($6.7\mu$m, $14.3\mu$m); SVS 17
\citep{Strom76} \\
\#19~~ISOCAM ($6.7\mu$m,
$14.3\mu$m); LkH$\alpha$ 352A \citep{Herbig56}; SVS 10
\citep{Strom76} \\
\#30~~ISOCAM
($6.7\mu$m, $14.3\mu$m); HBC 344 \citep{Herbig88}; SVS 11
\citep{Strom76}; M1 star (SIMBAD) \\
\#32~~ISOCAM
($6.7\mu$m, $14.3\mu$m); SVS 15 \citep{Strom76}; VLA 8
\citep{Rodriguez99} \\
\#33~~ISOCAM ($6.7\mu$m,
$14.3\mu$m, $16.6\mu$m); BD +30$^\circ$547; SVS 19
\citep{Strom76}; VSS 10 \citep{Vrba76}; VLA 9 \citep{Rodriguez99};
K0 star (SIMBAD) \\
\#36~~ISOCAM ($6.7\mu$m,
$14.3\mu$m); HBC 345 \citep{Herbig88} \\
\#39~~ISOCAM
($6.7\mu$m, $14.3\mu$m, $16.6\mu$m); SVS 16 \citep{Strom76} \\
\#43~~ISOCAM? ($6.7\mu$m); VLA 13
\citep{Rodriguez99} \\
\#49~~ISOCAM? ($6.7\mu$m); MMS
5 \citep{Chini01}; VLA 16 \citep{Rodriguez99} \\
\#52~~ISOCAM ($6.7\mu$m); SVS 8 \citep{Strom76} \\
\#58~~ISOCAM ($6.7\mu$m, $14.3\mu$m); SVS 14
\citep{Strom76}; VLA 22 \citep{Rodriguez99} \\
\#64~~IRAS 03260+3111 ; SiO maser \citep{Harju98};
SVS 3 \citep{Strom76}; VLA 44 \citep{Rodriguez99}; B6 star
(SIMBAD) \\
\#79~~VSS 4 \citep{Vrba76}; LkH$\alpha$
270 \citep{Herbig56}; SVS 2 \citep{Strom76}; K2 star (SIMBAD) \\
\#81~~ISOCAM ($15.01\mu$m) \\
\#83~~BD
+30$^\circ$549; IRAS 9 \citep{Jennings87}; SVS 1 \citep{Strom76};
VSS 3 \citep{Vrba76}; B9 star (SIMBAD) \\
\#86~~ISOCAM
($6.7\mu$m, $14.3\mu$m); IRAS 03262+3108; HH 17; HH 17A; SVS 5
\citep{Strom76} \\
\#90~~VSS 27 \citep{Vrba76};
LkH$\alpha$ 271 \citep{Herbig56}; SVS 20 \citep{Strom76}; HBC 13
\citep{Herbig88}; G star (SIMBAD) \\
\#92~~ISOCAM
($6.7\mu$m, $14.3\mu$m); LkH$\alpha$ 355 \citep{Herbig56} \\
\#98~~VSS 2 \citep{Vrba76} \\
\#100~~VSS
1 \citep{Vrba76}; SVS 6 \citep{Strom76} \\
\#107~~VSS
30 \citep{Vrba76} \\
\end{scriptsize}

\end{deluxetable}

\newpage

\begin{deluxetable}{rcrrrrccccrrcrrrrc}
\tabletypesize{\scriptsize}
\tablewidth{0pt}
\tablecolumns{18}
\tablecaption{X-ray properties of NGC 1333 sources \label{properties_tbl}}
\rotate
\tablehead{
\multicolumn{6}{c}{X-ray extraction} && \multicolumn{2}{c}{Variability} &&
\multicolumn{2}{c}{Spectrum} && \multicolumn{4}{c}{Luminosity} & \multicolumn{1}{c}{Notes}
\\ \cline{1-6} \cline{8-9} \cline{11-12} \cline{14-17}
\colhead{Src}& \colhead{CXONGC1333 J} & \colhead{$C_{xtr}$} &
\colhead{$B_{xtr}$} & \colhead{$R_{xtr}$} & \colhead{$f_{PSF}$} &&
\colhead{$CR$} & \colhead{Var Cl} && \colhead{log$N_H$} &
\colhead{$kT$} && \colhead{log$L_s$} & \colhead{log$L_h$} &
\colhead{log$L_t$} & \colhead{log$L_c$} &

 \\
\colhead{} & \colhead{} & \multicolumn{2}{c}{(counts)} &
\colhead{(\arcsec)} & \colhead{} && \colhead{(ct ks$^{-1}$)} &
\colhead{} && \colhead{(cm$^{-2}$)} & \colhead{(keV)} &&
\multicolumn{4}{c}{(erg s$^{-1}$)} & \colhead{} }
\startdata

1 & 032829.4+312508 & 20 & 4 & 17.6 & 0.88 & &  0.5 & Const & & 22.0 (21.2-22.5) & 3.0 (1.5-16.6) & & \nodata & \nodata & 28.9 & \nodata & a \\
2 & 032831.9+312121 & 14 & 2 & 11.3 & 0.87 & &  0.4 & Const & & 22.8  & $>10$  & & \nodata & \nodata & 29.1 & \nodata & a,b \\
3 & 032835.0+312112 & 13 & 3 &  9.5 & 0.86 & &  0.3 & Const & & 22.9 (22.3-23.4) & 2.1  & & \nodata & \nodata & 28.9 & \nodata & a \\
4 & 032836.4+311929 & 32 & 1 &  8.6 & 0.89 & &  0.9 & Const & & 21.4 (20.9-21.8) & 0.6 (0.3-0.7) & &  28.7  & $<28.0$ & 28.8 & 29.1 & \nodata \\
5 & 032836.8+312312 & 260 & 7 & 12.4 & 0.94 & &  7.1 & Const & & 21.5 (21.4-21.7) & 1.5 (1.4-1.6) & & $<29.4$  & $<29.3$ & $<29.7$ & $<29.9$ & c \\
6 & 032836.8+311735 & 272 & 2 & 10.9 & 0.93 & &  7.7 & Flare & & 21.6 (21.5-21.7) & 2.1 (1.8-2.4) & &  29.4  & 29.5 & 29.8 & 30.0 & \nodata \\
7 & 032837.1+311954 & 11 & 1 &  6.1 & 0.74 & &  0.4 & Const & & 20.8 & $>10$  & & \nodata &\nodata&28.6&\nodata& a,b \\
8 & 032837.8+312525 & 16 & 4 & 13.6 & 0.88 & &  0.4 & Const & & 22.9  & $>10$  & & \nodata &\nodata& 29.1 &\nodata& a,b \\
9 & 032843.2+311733 & 88 & 2 &  6.1 & 0.87 & &  2.6 & Const & & 22.5 (22.4-22.6) & 2.3 (2.0-2.8) & &  28.6  & 29.6 & 29.7 & 30.2 & \nodata \\
10 & 032843.5+311737 & 193 & 6 &  6.6 & 0.91 & &  5.4 & Const & & 21.5 (21.5-21.7) & 1.5 (1.4-1.6) & &  29.3  & 29.2 & 29.5 & 29.7 & \nodata \\
11 & 032844.0+312052 & 8 & 2 &  5.6 & 0.90 & &  0.2 & Const & &\nodata  &\nodata & & \nodata &\nodata& \nodata &\nodata& d \\
12 & 032845.4+312050 & 13 & 1 &  5.2 & 0.87 & &  0.4 & Const & & 22.3  & $>10$  & & \nodata &\nodata& 29.1 &\nodata& a,b \\
13 & 032845.4+312227 & 8 & 1 &  6.2 & 0.90 & &  0.2 & Const & &\nodata  &\nodata & & \nodata &\nodata&\nodata& \nodata & d \\
14 & 032846.1+311639 & 212 & 1 &  7.7 & 0.95 & &  5.9 & Const & & 21.0 (20.6-21.3) & 1.3 (1.2-1.4) & &  29.3  & 28.9 & 29.5 & 29.6 & \nodata \\
15 & 032847.6+312405 & 48 & 4 &  9.2 & 0.94 & &  1.2 & Flare & & 21.9 (21.7-22.0) & 3.5 (2.2-12.4) & &  28.6  & 29.1 & 29.2 & 29.4 & \nodata \\
16 & 032847.8+311655 & 28 & 1 &  5.4 & 0.90 & &  0.8 & Const & & 21.1 (20.0-21.5) & $>10$  & &  28.5  & 28.9 & 29.0 & 29.1 & a,b \\
17 & 032850.3+312755 & 29 & 7 & 15.5 & 0.77 & &  0.8 & Const & & 22.0 (21.6-22.4) & $>10$  & &  28.2  & 29.1 & 29.2 & 29.3 & a,b \\
18 & 032850.8+312349 & 12 & 2 &  6.3 & 0.89 & &  0.3 & Const & & 22.0 & 1.5 & & \nodata &\nodata& 28.4 & \nodata & a \\
19 & 032850.9+311818 & 66 & 1 &  4.9 & 0.95 & &  1.8 & Pos fl & & 21.2 (20.8-21.5) & 3.2 (2.2-5.3) & &  28.8  & 29.0 & 29.2 & 29.3 & \nodata \\
20 & 032850.9+311632 & 11 & 1 &  8.0 & 0.96 & &  0.3 & Const & & 20.9 & 1.5 & &  \nodata  & \nodata & 28.3& \nodata & a \\
21 & 032851.1+311955 & 521 & 3 &  4.8 & 0.95 & & 14.4 & Flare & & 21.5 (21.5-21.6) & 2.3 (2.1-2.5) & &  29.7  & 29.8 & 30.1 & 30.2 & \nodata \\
22 & 032852.0+311548 & 7 & 1 &  5.6 & 0.88 & &  0.2 & Const & &\nodata &\nodata & &  \nodata  & \nodata & \nodata & \nodata & d \\
23 & 032852.1+312245 & 157 & 2 &  5.9 & 0.95 & &  4.3 & Flare & & 21.3 (21.0-21.4) & 2.0 (1.7-2.5) & &  29.2  & 29.1 & 29.5 & 29.6 & \nodata \\
24 & 032853.5+311537 & 5 & 1 &  3.2 & 0.69 & &  0.2 & Const & & \nodata  &\nodata & &  \nodata  & \nodata & \nodata & \nodata & d \\
25 & 032853.9+311809 & 35 & 1 &  3.0 & 0.90 & &  1.0 & Const & & 21.9 (21.6-22.1) & 1.7 (1.4-2.4) & &  28.6  & 28.7 & 29.0 & 29.3 & \nodata \\
26 & 032854.0+311654 & 7 & 1 &  4.0 & 0.92 & &  0.2 & Const & &\nodata  &\nodata & &  \nodata  & \nodata & \nodata & \nodata & d \\
27 & 032854.6+311651 & 143 & 1 &  5.3 & 0.95 & &  4.0 & Const & & 21.8 (21.7-21.9) & 2.3 (1.8-2.8) & &  29.1  & 29.3 & 29.5 & 29.8 & \nodata \\
28 & 032855.0+311629 & 32 & 1 &  4.0 & 0.89 & &  0.9 & Const & & 22.3 (22.1-22.4) & 0.9 (0.6-1.3) & &  28.5  & 28.6 & 28.8 & 29.8 & \nodata \\
29 & 032856.5+312240 & 6 & 2 &  3.9 & 0.87 & &  0.1 & Const & &\nodata  &\nodata & & \nodata &\nodata&\nodata&\nodata& d \\
30 & 032856.6+311836 & 56 & 2 &  3.8 & 0.94 & &  1.5 & Pos fl & & 21.5 (21.1-21.8) & 7.0 (3.3-55.2) & &  28.6  & 29.1 & 29.2 & 29.3 & \nodata \\
31 & 032856.8+311811 & 5 & 1 &  2.8 & 0.88 & &  0.1 & Const & &\nodata &\nodata & & \nodata  &\nodata &\nodata &\nodata & d \\
32 & 032856.9+311622 & 29 & 1 &  4.0 & 0.91 & &  0.8 & Const & & 21.8 (21.4-22.1) & 0.8 (0.2-11.2) & &  28.4  & 28.5 & 28.8 & 28.8 & a  \\
33 & 032857.1+311419 & 2896 & 13 & 15.0 & 0.99 & & 77.0 & Pos fl & & 20.8 & 1.2 & &  30.1  & 29.5 & 30.2 & 30.3& c \\
34 & 032857.4+311950 & 34 & 1 &  2.5 & 0.92 & &  0.9 & Const & & 21.7 (21.4-22.0) & 1.1 (0.8-1.5) & &  28.8  & 28.5 & 29.0 & 29.3 & \nodata \\
35 & 032857.6+311948 & 243 & 3 &  3.8 & 0.95 & &  6.7 & Flare & & 21.6 (21.5-21.7) & 3.6 (2.7-4.2) & &  29.7  & 30.0 & 30.2 & 30.3 & \nodata \\
36 & 032858.0+311804 & 49 & 1 &  3.8 & 0.95 & &  1.3 & Const & & 21.5 (21.1-22.1) & 1.3 (0.5-4.0) & &  28.8  & 28.3 & 28.9 & 29.5 & \nodata \\
37 & 032858.2+312541 & 7 & 1 &  8.4 & 0.88 & &  0.2 & Const & & \nodata  &\nodata & &  \nodata  & \nodata & \nodata & \nodata & d \\
38 & 032858.4+312218 & 7 & 1 &  3.2 & 0.86 & &  0.2 & Const & & \nodata  &\nodata & &  \nodata  & \nodata & \nodata & \nodata & d \\
39 & 032859.2+311548 & 515 & 7 &  5.5 & 0.94 & & 14.3 & Flare & & 22.1 (22.1-22.2) & 2.8 (2.4-3.4) & &  29.5  & 30.2 & 30.3 & 30.6 & \nodata \\
40 & 032859.3+312640 & 16 & 1 & 10.4 & 0.87 & &  0.5 & Const & & 22.3 (22.0-22.6) & $>10$  & &  $<28.0$  & 29.2 & 29.2 & 29.4 & a,b \\
41 & 032859.5+312146 & 14 & 1 &  2.8 & 0.88 & &  0.4 & Const & & 22.1  & 2.7  & &  \nodata  & \nodata & 28.7 & 29.1 & a \\
42 & 032900.1+312109 & 5 & 1 &  2.5 & 0.90 & &  0.1 & Const & & \nodata  & \nodata & &  \nodata  & \nodata & \nodata & \nodata & d \\
43 & 032900.2+311338 & 267 & 1 &  9.8 & 0.94 & &  7.5 & Const & & 21.6 (21.5-21.7) & 1.5 (1.3-1.7) & &  29.4  & 29.3 & 29.6 & 29.9 & \nodata \\
44 & 032900.3+312045 & 11 & 1 &  2.5 & 0.93 & &  0.3 & Const & & 21.1  & 1.2  & &  \nodata  & \nodata & 28.3 & \nodata & a \\
45 & 032900.6+312201 & 39 & 1 &  2.8 & 0.89 & &  1.1 & Pos fl & & 22.2 (21.9-22.3) & 1.3 (0.9-2.0) & &  28.6  & 28.8 & 29.0 & 29.6 & \nodata \\
46 & 032901.3+311413 & 7 & 1 &  6.6 & 0.86 & &  0.2 & Const & & \nodata  & \nodata & &  \nodata  & \nodata & \nodata & \nodata & d \\
47 & 032902.1+311611 & 20 & 1 &  3.7 & 0.91 & &  0.6 & Const & & 21.7 (21.1-22.1) & 0.3 (0.1-1.2) & &  \nodata  & \nodata & 28.4 & \nodata & a \\
48 & 032902.8+311601 & 85 & 2 &  5.2 & 0.96 & &  2.3 & Const & & $<20.0$ & 1.0 (0.9-1.1) & &  29.0  & 28.2 & 29.0 & 29.0& \nodata \\
49 & 032903.0+311516 & 16 & 1 &  4.7 & 0.87 & &  0.5 & Const & & 22.9 (22.5-23.3) & 1.9 (0.7-4.9) & &  $<28.0$  & 29.1 & 29.1 & 29.9 & a \\
50 & 032903.1+312238 & 9 & 1 &  3.2 & 0.88 & &  0.2 & Const & & 22.8  & 0.5  & &  \nodata  & \nodata & 28.5 & 30.9 & a \\
51 & 032903.4+311618 & 34 & 1 &  3.5 & 0.88 & &  1.0 & Flare & & 22.4 (22.2-22.6) & 2.1 (1.4-4.3) & &  28.2  & 29.1 & 29.2 & 29.7 & \nodata \\
52 & 032903.8+312149 & 478 & 3 &  3.8 & 0.94 & & 13.4 & Flare & & 21.8 (21.7-21.9) & 2.3 (2.1-2.8) & &  29.6  & 30.0 & 30.1 & 30.3 & \nodata \\
53 & 032904.0+311708 & 19 & 1 &  2.8 & 0.91 & &  0.5 & Const & & 20.9 (20.0-21.8) & 0.9 (0.5-1.3) & &  \nodata  & \nodata & 28.4 & \nodata & a \\
54 & 032904.1+312515 & 34 & 1 &  6.9 & 0.87 & &  1.0 & Const & & 22.2 (22.0-22.4) & 0.7 (0.4-1.1) & &  28.6  & 28.5 & 28.8& 29.9 & \nodata \\
55 & 032904.3+312624 & 6 & 1 &  9.4 & 0.86 & &  0.2 & Const & & \nodata  & \nodata & &  \nodata  & \nodata & \nodata & \nodata & d \\
56 & 032904.7+311133 & 30 & 4 & 12.9 & 0.87 & &  0.8 & Pos fl & & 22.5 (22.3-22.8) & 1.6 (0.9-6.7) & &  28.1  & 29.0 & 29.1 & 29.8 & \nodata \\
57 & 032904.9+312038 & 7 & 1 &  2.2 & 0.89 & &  0.2 & Const & & \nodata  & \nodata & &  \nodata  & \nodata & \nodata & \nodata & d \\
58 & 032905.7+311640 & 71 & 2 &  2.6 & 0.91 & &  2.0 & Const & & 22.1 (21.9-22.3) & 3.1 (2.1-7.1) & &  28.6  & 29.2 & 29.3 & 29.6 & \nodata \\
59 & 032905.9+311640 & 71 & 1 &  2.0 & 0.88 & &  2.1 & Pos fl & & 22.2 (22.1-22.3) & 2.2 (1.3-3.1) & &  28.7  & 29.3 & 29.4 & 29.8 & \nodata \\
60 & 032906.6+311933 & 15 & 1 &  1.8 & 0.84 & &  0.4 & Const & & 22.7  & 2.4  & &  \nodata  & \nodata & 29.0 & \nodata & a \\
61 & 032907.9+312252 & 24 & 1 &  3.4 & 0.88 & &  0.7 & Const & & 22.5 (22.3-22.8) & 0.9 (0.5-1.5) & &  28.5  & 28.9 & 29.1 & 30.3 & a \\
62 & 032908.7+311314 & 11 & 3 &  8.7 & 0.86 & &  0.2 & Const & & 22.4  & $>10$  & &  \nodata  & \nodata & 29.1 & \nodata & a,b \\
63 & 032909.1+312306 & 8 & 1 &  3.7 & 0.89 & &  0.2 & Const & & \nodata  & \nodata & &  \nodata  & \nodata & \nodata & \nodata & d \\
64 & 032910.3+312159 & 1161 & 9 &  4.7 & 0.95 & & 32.1 & Const & & 22.1 (22.1-22.2) & 1.5 (1.2-1.6) & &  29.9  & 30.3 & 30.4 & 30.9 & \nodata \\
65 & 032910.5+312202 & 153 & 3 &  2.7 & 0.87 & &  4.6 & Pos fl & & 22.0 (21.9-22.2) & 2.8 (2.1-3.8) & &  29.0  & 29.6 & 29.7 & 30.0 & \nodata \\
66 & 032910.7+312230 & 25 & 1 &  3.1 & 0.90 & &  0.7 & Const & & 21.6 (21.3-21.8) & 1.5 (0.4-2.0) & &  28.5  & 28.3 & 28.7 & 28.9 & a \\
67 & 032911.2+311718 & 139 & 1 &  4.0 & 0.95 & &  3.8 & Flare & & $<20.0$  & 3.0 (2.4-3.9) & &  29.1  & 29.1 & 29.4 & 29.4 & b \\
68 & 032911.8+312216 & 5 & 1 &  3.0 & 0.85 & &  0.1 & Const & & \nodata  & \nodata & &  \nodata  & \nodata & \nodata & \nodata & d \\
69 & 032911.8+312127 & 9 & 1 &  2.6 & 0.87 & &  0.2 & Const & & \nodata  & \nodata & &  \nodata  & \nodata & \nodata & \nodata & d \\
70 & 032912.8+312008 & 7 & 1 &  2.3 & 0.92 & &  0.2 & Const & & \nodata  & \nodata & &  \nodata  & \nodata & \nodata & \nodata & d \\
71 & 032912.9+311815 & 150 & 1 &  3.8 & 0.93 & &  4.2 & Pos fl & & 22.6 (22.5-22.7) & 3.3 (2.3-6.0) & &  28.5  & 29.9 & 30.0 & 30.4 & \nodata \\
72 & 032913.1+312253 & 20 & 1 &  3.8 & 0.89 & &  0.6 & Pos fl & & 22.2 (21.9-22.4) & 1.9 (1.2-3.5) & &  28.2  & 28.7 & 28.9 & 29.3 & a \\
73 & 032913.4+312441 & 16 & 5 &  6.3 & 0.88 & &  0.3 & Const & & 23.5  & 0.9  & &  \nodata  & \nodata & 29.0 & \nodata & a \\
74 & 032914.0+312214 & 14 & 1 &  3.2 & 0.87 & &  0.4 & Const & & 22.1  & $>10$  & &  \nodata  & \nodata & 28.9 & \nodata & a,b \\
75 & 032914.4+312236 & 28 & 1 &  3.6 & 0.90 & &  0.8 & Flare & & 21.9 (21.5-22.29) & 1.3 (0.6-2.5) & &  28.4  & 28.4 & 28.7 & 29.1 & a \\
76 & 032915.6+311852 & 16 & 1 &  2.6 & 0.87 & &  0.5 & Const & & 23.3  & 2.8  & &  \nodata  & \nodata & 29.2 & \nodata & a \\
77 & 032916.5+312348 & 117 & 1 &  6.8 & 0.94 & &  3.3 & Pos fl & & 21.9 (21.7-22.0) & 3.4 (2.2-5.2) & &  28.9  & 29.5 & 29.6 & 29.8 & \nodata \\
78 & 032916.6+311618 & 168 & 3 &  5.5 & 0.96 & &  4.5 & Flare & & $<20.0$  & 1.2 (0.9-1.5) & &  29.2  & 28.6 & 29.3 & 29.3 & b \\
79 & 032917.6+312245 & 1707 & 19 & 14.0 & 0.99 & & 45.1 & Flare & & 21.5 (21.5-21.5) & 2.2 (2.2-2.6) & &  30.2  & 30.4 & 30.6 & 30.7 & \nodata \\
80 & 032918.5+311926 & 25 & 1 &  2.8 & 0.87 & &  0.7 & Const & & 22.6 (22.4-22.9) & 1.8 (0.9-6.4) & &  $<28.0$  & 29.1 & 29.1 & 29.8 & a \\
81 & 032918.7+312326 & 234 & 1 &  6.6 & 0.95 & &  6.5 & Const & & 21.4 (21.2-21.6) & 1.0 (0.9-1.1) & &  29.4  & 28.8 & 29.5 & 29.7 & \nodata \\
82 & 032919.0+312106 & 19 & 1 &  3.2 & 0.88 & &  0.5 & Const & & 21.9 (21.4-22.2) & $>10$  & &  \nodata  & \nodata & 29.0 & \nodata & a,b \\
83 & 032919.7+312458 & 39 & 2 &  8.0 & 0.91 & &  1.1 & Const & & 21.2 (20.9-21.7) & 1.6 (1.1-2.2) & &  28.6  & 28.3 & 28.8 & 28.9 & \nodata \\
84 & 032920.0+312408 & 30 & 2 &  6.4 & 0.88 & &  0.8 & Pos fl & & 22.4 (22.1-22.6) & $>10$  & &  28.0  & 29.3 & 29.3 & 29.5 & a,b \\
85 & 032920.1+311450 & 8 & 1 &  6.9 & 0.88 & &  0.2 & Const & & \nodata  & \nodata  & &  \nodata  & \nodata & \nodata & \nodata & d \\
86 & 032920.4+311834 & 16 & 1 &  3.3 & 0.88 & &  0.5 & Const & & 22.7 (22.6-22.9) & 1.5 (1.0-2.1) & &  \nodata  & \nodata & 28.9 & 29.9 & a \\
87 & 032921.1+311617 & 10 & 1 &  5.3 & 0.88 & &  0.3 & Const & & 22.9  & $>10$  & &  \nodata  & \nodata & 29.0 & \nodata & a,b \\
88 & 032921.5+312111 & 69 & 1 &  5.2 & 0.95 & &  1.9 & Const & & 21.4 (21.1-21.7) & 1.2 (0.9-1.4) & &  28.9  & 28.4 & 29.0 & 29.2 & \nodata \\
89 & 032921.6+312515 & 11 & 2 &  9.0 & 0.90 & &  0.3 & Const & & 21.8  & 0.5  & &  \nodata  & \nodata & 28.1 & \nodata & a \\
90 & 032921.8+311537 & 23 & 1 &  6.2 & 0.89 & &  0.7 & Const & & 22.3 (22.2-22.5) & 5.4 (2.4-15.3) & &  28.1  & 29.2 & 29.2 & 29.5 & a \\
91 & 032922.0+312415 & 85 & 5 &  8.6 & 0.95 & &  2.2 & Const & & 21.1 (20.1-21.5) & 1.3 (1.1-1.6) & &  29.0  & 28.5 & 29.1 & 29.2 & \nodata \\
92 & 032923.2+312030 & 6 & 1 &  4.0 & 0.90 & &  0.1 & Const & & \nodata  & \nodata & &  \nodata  & \nodata & \nodata & \nodata & d \\
93 & 032923.5+312331 & 643 & 5 &  8.2 & 0.95 & & 17.8 & Flare & & 21.7 (21.6-21.7) & 2.6 (2.3-2.9) & &  29.8  & 30.0 & 30.2 & 30.4 & \nodata \\
94 & 032923.7+312510 & 14 & 2 &  9.2 & 0.89 & &  0.4 & Const & & $<20.0$  & 2.7  & &  \nodata  & \nodata & 28.5 & \nodata & a,b \\
95 & 032924.1+312109 & 11 & 1 &  4.4 & 0.88 & &  0.3 & Const & & 22.7  & 2.9  & &  \nodata  & \nodata & 28.9 & \nodata & a \\
96 & 032924.8+312052 & 6 & 1 &  4.5 & 0.87 & &  0.2 & Const & & \nodata  & \nodata & &  \nodata  & \nodata & \nodata & \nodata & d \\
97 & 032924.8+312407 & 12 & 1 &  7.8 & 0.91 & &  0.3 & Const & & 21.8  & $>0.2$  & &  \nodata  & \nodata & 28.3 & \nodata & a \\
98 & 032925.8+312640 & 231 & 10 & 13.8 & 0.91 & &  6.4 & Flare & & 21.8 (21.7-21.9) & 1.7 (1.2-2.1) & &  29.5  & 29.7 & 29.9 & 30.2 & \nodata \\
99 & 032925.8+311347 & 14 & 4 & 10.5 & 0.88 & &  0.3 & Const & & 22.1  & 4.0  & &  \nodata  & \nodata & 28.7 & \nodata & a \\
100 & 032926.7+312648 & 1055 & 16 & 16.0 & 0.95 & & 28.9 & Const & & 21.4 (21.3-21.5) & 1.5 (1.5-1.5) & &  30.1  & 29.9 & 30.3 & 30.5 & \nodata \\
101 & 032928.1+311628 & 153 & 5 &  8.5 & 0.95 & &  4.1 & Pos fl & & $<20.0$  & 1.3 (1.2-1.5) & &  29.2  & 28.7 & 29.3 & 29.3 & b \\
102 & 032929.2+311835 & 142 & 2 &  7.5 & 0.95 & &  3.9 & Const & & 21.6 (21.5-21.8) & 1.4 (1.1-1.5) & &  29.2  & 29.0 & 29.4 & 29.7 & \nodata \\
103 & 032929.8+312102 & 29 & 1 &  6.2 & 0.90 & &  0.8 & Const & & $<20.0$  & 2.8 (1.8-5.1) & &  28.5  & 28.5 & 28.8 & 28.8 & a \\
104 & 032930.6+312729 & 13 & 5 & 17.2 & 0.90 & &  0.2 & Const & & 21.4  & 0.5  & &  \nodata  & \nodata & 28.2 & \nodata & a \\
105 & 032931.6+312125 & 16 & 4 &  7.2 & 0.87 & &  0.4 & Const & & 21.7  & $>10$  & &  \nodata  & \nodata & 28.8 & \nodata & a,b \\
106 & 032932.9+312712 & 30 & 6 & 20.1 & 0.95 & &  0.7 & Const & & 21.5 (21.0-21.9) & 0.8 (0.4-1.1) & &  28.5  & $<28.0$ & 28.6 & 28.9 & a \\
107 & 032934.3+311743 & 157 & 3 & 10.3 & 0.96 & &  4.2 & Const & & 21.6 (21.4-22.0) & 0.6 (0.2-0.6) & &  29.3  & 28.2 & 29.3 & 29.8 & \nodata \\
108 & 032945.4+312351 & 23 & 11 & 10.0 & 0.42 & &  0.8 & Const & & 22.9  & $>10$  & &  $<28.0$  & 29.4 & 29.4 & 29.8 & a,b \\
109 & 032946.1+312037 & 36 & 9 & 14.6 & 0.81 & &  0.9 & Const & & $<20.0$  & 1.4 (1.1-1.6) & &  28.8  & 28.3 & 28.9 & 28.9 & b \\

\enddata

\tablenotetext{a}{{\it kT} and $\log N_H$ may be unreliable but
are reported here to allow the reproduction of the spline-like
fits to the event distribution for the calculation of broad-band
luminosities $\log L_t$.} \tablenotetext{b}{Fitted plasma energies
above $\sim 10$ keV and column densities below $\sim 20.0$
cm$^{-2}$ are not well-determined.} \tablenotetext{c}{Foreground
stars: \#5 (distance is less than 318 pc) and \#33 (distance is
192 pc, Preibisch97b).} \tablenotetext{d}{No spectral fit is done
as a result of a small number of counts. Most of these sources
have an approximate X-ray luminosity $\log L_t \simeq 10^{28}$ erg
s$^{-1}$.}

\end{deluxetable}
\begin{deluxetable}{rccr}
\tabletypesize{\scriptsize}
\tablewidth{0pt}
\tablecolumns{5}
\tablecaption{Tentative ACIS sources \label{faint_src_tbl}}

\tablehead{ \colhead{LAL} & \colhead{$C_{xtr}$} &
\colhead{$\theta$} &
\colhead{K} \\

\colhead{\#} &
\colhead{(counts)} & \colhead{(\arcmin)} &
\colhead{(mag)} \\
}
\startdata
34  & 6 & 8.0 & 14.20 \\
48  & 6 & 6.9 & 13.57 \\
50  & 3 & 7.1 & 11.91 \\
95  & 6 & 9.2 & 12.68 \\
99  & 5 & 3.9 & 14.58 \\
137 & 4 & 2.8 & 13.23 \\
147 & 3 & 3.6 & 11.80 \\
164 & 3 & 3.1 & 13.38 \\
165 & 5 & 3.8 & 13.79 \\
202 & 5 & 0.9 & 13.74 \\
206 & 3 & 2.0 & 13.46 \\
208 & 6 & 5.9 & 12.76 \\
224 & 4 & 2.2 & 12.32 \\
256 & 4 & 5.0 & 11.96 \\
265 & 3 & 2.5 & 13.12 \\
314 & 4 & 9.6 & 12.82 \\
336 & 4 & 5.3 & 11.09 \\
(a) & 6 & 4.0 & 10.00 \\
\enddata
\tablenotetext{(a)}{This X-ray source has the optical counterpart HJ 36
\citep{Herbig83}. $K$ magnitude is obtained using \citet{Aspin94} photometry for the source ASR 108, which they claim to be the counterpart of HJ 36, although it has a projected separation of 3\arcsec\/ from HJ 36.}

\end{deluxetable}


\clearpage
\newpage

\end{document}